\documentstyle[12pt]{article}
\def\be{\begin{equation}}
\def\ee{\end{equation}}
\def\ba{\begin{eqnarray}}
\def\ea{\end{eqnarray}}
\def\bq{\begin{quote}}
\def\eq{\end{quote}}

\newcommand{\reef}[1]{(\ref{#1})}
\newcommand{\vs}[1]{\vspace{#1 mm}}

\def\part{\partial}

\def\beq{\begin{equation}}
\def\eeq{\end{equation}}
\def\beqa{\begin{eqnarray}}
\def\eeqa{\end{eqnarray}}

\def\G{{\cal G}}
\def\A{{\cal A}}
\def\B{{\cal B}}

\def\ie{{\it i.e.,}\ }
\def\eg{{\it e.g.,}\ }

\begin{document}

\thispagestyle{empty}
\rightline{\small hep-th/9901045 \hfill SU-ITP-98-19}
\vs{-1.5}
\rightline{\small \hfill McGill/98-30}
\vs{-1.5}
\rightline{\small \hfill January 1999}
\vskip 2cm

\begin{center}
{\Large \bf The O(dd) Story of Massive Supergravity}
 \\
\vspace*{1cm}
Nemanja Kaloper\footnote{E-mail: kaloper@leland.stanford.edu}\\
\vspace*{0.2cm}
{\it Department of Physics, Stanford University}\\ {\it Stanford,
CA 94305-4060, USA}\\
\vspace*{0.4cm}
Robert C. Myers\footnote{E-mail: rcm@hep.physics.mcgill.ca} \\
\vspace*{0.2cm}
{\it Department of Physics, McGill University}\\ {\it Montr\'eal,
PQ, H3A 2T8, Canada}\\
\vspace{2cm}
ABSTRACT
\end{center}
The low energy effective action describing the standard
Kaluza-Klein reduction of heterotic string theory on a
$d$-torus possesses a manifest $O(d,d+16)$ symmetry. We consider
generalized Scherk-Schwarz
reductions of the heterotic string to construct
massive gauged supergravities. We show that the resulting action
can still be written in a manifestly $O(d,d+16)$ invariant
form, however, the U-duality transformations also act on
the mass parameters. The latter play the dual role of defining the scalar
potential and the nonabelian
structure constants. We conjecture that just as for the
standard reduction, a subgroup of this symmetry corresponds
to an exact duality symmetry of the heterotic string theory.
\vfill
\setcounter{page}{0}
\setcounter{footnote}{0}
\newpage

\section{Introduction}

Recently progress in understanding string dualities and the role of
$p$-branes has lead to interest in constructing massive supergravity
theories through ``unconventional" compactifications of massless
supergravities in higher dimensions \cite{cow,others,berg,flux}.
The seminal insight made in ref.~\cite{cow} was realizing that
given a theory containing an axion, that is a massless scalar with only
derivative couplings, a consistent compactification could be made
in which the axion was given a linear dependence on the internal coordinates.
While the reduced action remains independent of the internal coordinates,
the slope parameters of this linear dependence appear as mass parameters
in the reduced theory. As realized in \cite{cow}, these reductions are
actually a special class within the general framework developed by
Scherk and Schwarz\cite{schs} for producing masses from dimensional
reduction. The Scherk-Schwarz approach focuses on the global symmetries, \ie
the U-duality symmetries, of the action, and the dependence
on the internal coordinates takes the form of a U-duality transformation
that varies (in a specific way) over the internal space.
For the generalized axion reductions above, the relevant symmetry is the
shift symmetry of the scalar axion, and the fact that the corresponding
symmetry
generator is nilpotent yields the simple linear dependence mentioned above
\cite{kkm}.
Various applications and extensions of the generalized
reductions with axionic masses were explored
in \cite{others,berg,flux,kkm,cosmo,hull,ortin}.

However,
the picture arising from the exploratory investigations of \cite{cow}
was one of many disjoint massive supergravities in lower dimensions.
This situation contrasts with the prevailing theme in string theory of
recent years in which U-duality has played a central role in
unifying disparate (supergravity and superstring) theories as
various phases of a single U-theory\cite{witten,Uthe}.
In part, the fragmented picture for the massive supergravities
appeared because the linear ansatz
described above could only accommodate specific combinations of masses.
That is in many instances where the higher dimensional theory contained
a number of axions, all of the corresponding mass parameters could not
be simultaneously turned on within this scheme. This point was clarified
in \cite{kkm}, where it was shown that the problem arose because the
corresponding axionic symmetries did not commute. However, any
combination of masses was
easily accommodated within the Scherk-Schwarz framework, although
the reduction ansatz now involved a polynomial
(or even more general) dependence on the
internal coordinates. Given that there is no restriction on the
types of axionic masses, it was conjectured \cite{kkm} that various
distinct massive supergravities should all be a part of a single
U-duality invariant massive theory.

In the present paper, we demonstrate how the preceding conjecture
is realized for generalized toroidal compactifications of
heterotic string theory. In this case, the standard Kaluza-Klein
reduction on a $d$-dimensional torus from 10 to $10-d$ dimensions
produces a theory with global $O(d,d+16)$ symmetry and with a
$U(1)^{2d+16}$ gauge group. As shown by ref.~\cite{actor}, the
effective action can be organized to make the former U-duality
symmetry manifest. Essentially the $2d+16$ gauge fields may be
assembled as a vector under this symmetry, while there are
$d(d+16)$ moduli scalars transforming as a traceless symmetric
tensor. We will show that this global symmetry is retained in the
massive theories produced by generalized Scherk-Schwarz
reductions. The bosonic part of effective action may be written
as:
\begin{eqnarray}
S &=& \int d^{D} x \sqrt{{-g}} e^{-\phi} \Bigl\{{R}
 + ({\nabla} \phi)^2 + \frac{1}{8}
L_{ab} {\cal D}_\mu M^{bc} L_{cd} {\cal D}^\mu M^{da}
 \nonumber \\
&&~~~~~~~~~~~~~~~~~~~~~ - \frac{1}{4} F^a_{\mu\nu}
 L_{ab} M^{bc} L_{cd} F^{d\mu\nu}
- \frac{1}{12} {H}^2_{\mu\nu\lambda} - W(M) \Bigr\}
\label{result}
\end{eqnarray}
where the scalar potential takes the simple form:
\begin{equation}
W(M)=
\frac{1}{12} M^{ad}
M^{be} M^{cf} f_{abc} f_{def}
- \frac{1}{4} M^{ad} L^{be}L^{cf} f_{abc}f_{def}\ .
\label{potcovariant}
\end{equation}
(The reader is invited to read the main text for a full explanation
of this result.) The essential point is that the various mass parameters
introduced by the generalized reduction can be organized as a
completely antisymmetric three-index tensor $f_{abc}$ under the $O(d,d+16)$
transformations. These parameters play a dual role in the reduced
theory: first, as mass parameters defining the scalar potential
\reef{potcovariant}, and second as structure constants in the {\it non-abelian}
gauge group of this theory, implicitly appearing in $F^a_{\mu\nu}$ and
${\cal D}^\mu M^{ab}$. That is the generalized reduction has
produced a gauged supergravity with a nontrivial non-abelian symmetry.

A simple intuition which explains (at least in part) the emergence of this
nonabelian symmetry is as follows: The Scherk-Schwarz reduction introduces
an axionic shift which depends on internal coordinates. Now a part of
gauge symmetry in the reduced theory can be thought of as local shifts
of the internal coordinates. These Kaluza-Klein gauge transformations
are inherited from the diffeomorphism invariance of original
ten-dimensional theory. Hence consistency of this symmetry in the generalized
reduction requires that these gauge transformations be accompanied
by a {\it local} axionic shift. That is the latter symmetries, which
are ``ordinarily'' only a part of the global U-duality group, have now
been incorporated as a part of the local gauge group. Further given that
these axionic symmetries in general do not commute \cite{kkm}, it must be that
the gauge group is modified to become non-abelian.

For the standard Kaluza-Klein reduction on $T^d$, the $O(d,d+16,R)$
transformations map one configuration of background space-time fields to
another configuration.
Now naively, it may appear that the U-duality symmetry is broken
in the reduced action \reef{result}
by the appearance of the mass parameters $f_{abc}$.
However, this is only spontaneous symmetry breaking.
In the present case, we know that these couplings are
simply associated with the presence of additional nontrivial background
fields in the internal space. Hence our construction reveals that
U-duality symmetry consists of
the usual transformation rules for the fields in the ($10-d$)-dimensional
space-time supplemented by a compensating transformation
of the mass parameters, \ie the nontrivial fields on the internal space.
The action \reef{result} maintains U-duality invariance, however, the
transformations also act on the couplings $f_{abc}$
in the obvious way.
A novel feature in the case of the generalized reductions is then
that in transforming the internal fields, U-duality maps one reduced
theory to another with modified couplings.
This U-duality covariant formalism provides
a unified framework incorporating all of the previously ``distinct'' massive
supergravities, which various generalized reductions could have
produced for low energy heterotic string theory. It is natural to conjecture
that the $O(d,d+16,Z)$ subgroup, which is an exact symmetry of the full
heterotic string theory with a standard toroidal compactification\cite{exact},
will remain an exact symmetry of the full string theory for the
generalized axion reductions --- we mention some subtleties in section
\ref{conc}.

Some of our results above can be seen in the previous work
of ref.~\cite{berg}, where a
simpler Scherk-Schwarz reduction of the effective action of heterotic
string theory produced a gauged $N=4$ supergravity theory in four
dimensions. In the context of nine-dimensional massive Type II
supergravity, ref.~\cite{ortin} has constructed an effective action
which is manifestly invariant under the relevant U-duality group,
namely $SL(2,R)$. Hull\cite{cmhull} considered the effect of duality
transformations and their singular limits on gauged supergravities.
There the singular limits were used to construct new theories.
Further, Boucher\cite{boucher} also considered dimensional reductions
of $D=11$ supergravity similar to those constructed here.

The remainder of our paper is organized as follows: In section 2, we
review the standard Kaluza-Klein reduction of low energy
heterotic string theory on a $d$-torus. In particular, we describe
the global $SO(d,d+16)$ symmetries, which form the U-duality group
for the reduced theory. Section 3 presents a discussion
of generalized reductions which include constant fluxes of the
three-form or gauge field strengths on the torus.
The reduced action for
these compactifications is assembled in the form of eq.~\reef{result},
in which the $SO(d,d+16)$ invariance remains manifest.
Section 4 describes a generalized reduction ansatz which introduces
curvatures in the internal geometry, and produces masses
for the metric axions in the reduced theory. Section 5 then describes
the general massive reduction involving all three sources of the
masses. Again, the U-duality invariant form of the action \reef{result}
is recovered
with a modified set of structure constants $f_{abc}$. Section 6
provides a discussion of our results. In particular, we note that the
formalism introduced in section 4 and the resulting low energy action
applies for more general internal geometries than the $d$-torus.
This is followed by a number
of appendices, which contain details of the calculations made in
performing the generalized reductions. A final appendix presents a
new perspective on discussion of
the generalized axion reductions in ref.~\cite{kkm}.

\section{A Review of Kaluza-Klein Reduction}
\label{review}

We begin with a review of the standard Kaluza-Klein reduction
of low energy heterotic string theory on a $d$-torus. Our
notation will be such that $d+D=10$ and hence this
compactification yields an effective $D$-dimensional theory.
In ten dimensions, the low energy action is
\begin{equation}
S = \int d^{10}x \sqrt{-{\cal G}} e^{- \Phi} \Bigl\{ {\cal R}
 +  (\nabla \Phi)^2 - \frac{1}{12} {\cal H}_{\mu \nu \lambda}
{\cal H}^{\mu \nu \lambda} - \frac{1}{4} \Sigma_{I=1}^{16} {\cal
F}^{I}{}_{\mu\nu} {\cal F}^{I\,\mu\nu}\Bigr\}
\label{sact1}
\end{equation}
The ten-dimensional
fields, $\Phi$, ${\cal G}_{\mu\nu}$, ${\cal R}$, ${\cal
H}_{\mu\nu\lambda}$ and  ${\cal F}^I{}_{\mu\nu}$, denote the
dilaton,
(string-frame) metric, Ricci scalar, Kalb-Ramond three-form
field strength, and the Yang-Mills field strengths, respectively.
The $D$-dimensional counterparts of these fields will be denoted
with upper case latin letters, except the dilaton, which will be
$\phi$. Our convention for the
metric signature is ${\cal G} = (-,+,+,...,+)$, and that for the
curvature is $R^{\mu}{}_{\nu\lambda\sigma} =
\partial_\lambda \Gamma^{\mu}_{\nu\sigma} -
\partial_\sigma \Gamma^{\mu}_{\nu\lambda} + \ldots$.
We assume that the only nontrivial components of the
Yang-Mills potential reside in the Cartan subalgebra of
the gauge group\footnote{Wilson lines in a generic
toroidal compactification will break all but the corresponding
Abelian gauge symmetry.}, and hence ${\cal F}^I{}_{\mu\nu} =
\partial_\mu {\cal A}^I{}_\nu -\partial_\nu {\cal A}^I{}_\mu$.
The low energy action \reef{sact1}
has been truncated to terms with at most two
derivatives. Consistent with this truncation, the
three-form ${\cal H}$ is
defined by including only the Yang-Mills Chern-Simons term,
\begin{equation}
{\cal H} = d{\cal B} - \frac{1}{2} \Sigma_{I=1}^{16} {\cal A}^{I}
\wedge {\cal F}^{I}
\label{krdef}\end{equation}
In component notation, we have ${\cal
H}_{\mu\nu\rho}=\partial_{\mu} {\cal B}_{\nu\rho} - \frac{1}{2}
\Sigma_{I=1}^{16}{\cal A}^{I}_{\mu} {\cal F}^{I}{}_{\nu\rho}
+ ~cyclic~permutations$, by the antisymmetry of ${\cal B}_{\mu\nu}$
and ${\cal F}^I{}_{\mu\nu}$.
The normalization in eq.~\reef{krdef}
corresponds to choosing $\alpha'=1$.

We wish to consider the standard Kaluza-Klein dimensional
reduction of heterotic action (\ref{sact1}) on a $d$-torus,
to set the stage for Scherk-Schwarz reductions in the
following sections.
The starting point for any compactification is a
decomposition of the tensor degrees of freedom \cite{MS}
\ba
d{\cal S}^2 &=& {g}_{\mu\nu} (x,y) dx^{\mu} dx^{\nu} +
\hat {\cal G}_{MN}(x,y) (dy^M + \hat V^M{}_{\mu}(x,y) dx^{\mu})
(dy^N + \hat V^N{}_{\nu}(x,y) dx^{\nu})
\nonumber \\ {\cal B} &=& \frac{1}{2} \hat {\cal
B}_{\mu\nu}(x,y) dx^{\mu}dx^{\nu}+ \hat {\cal B}_{\mu M}(x,y)
dx^{\mu} dy^M +
\frac{1}{2} \hat {\cal B}_{MN}(x,y) dy^{M}dy^{N}
\nonumber \\
{\cal A}^I &=& \hat {\cal A}^I{}_{\mu}(x,y) dx^{\mu} +
\hat {\cal A}^I{}_M (x,y)dy^M
\label{dans}
\ea
where $x^\mu$ and $y^M$ denote the coordinates on the
uncompactified and the compact internal subspaces, respectively,
with $\mu=0,1,\ldots,D-1$ and $M=1,2,\ldots,d$. As usual,
there is a summation when such indices appear
repeated in a subscript-superscript pair.
Our convention will be that the gauge group indices,
$I,J=1,2,\ldots,16$, always appear as superscripts, and
in the following, a summation will also be implied
for a pair of repeated gauge superscripts. Note that
there are no assumptions about the structure of the space-time
involved in writing eq.~(\ref{dans}).
For a Kaluza-Klein reduction on a torus, the vector fields $\partial_M$
generate isometries of the system. Naively, this translates into the
statement that none of fields depend on the $y^M$'s.
Specifically, this means that
\ba
&&
\hat g_{\mu\nu}(x,y) = g_{\mu\nu}(x)\ \ \
~~~~~~~~ \ \ \ \hat {\cal B}_{\mu\nu}(x,y) = {\cal B}_{\mu\nu}(x)
\nonumber \\
&&\hat V^M{}_\mu(x,y) = V^M{}_\mu(x) ~~~~~~~~ \hat {\cal B}_{\mu
M}(x,y) = {\cal B}_{\mu M}(x) ~~~~~~~~ \hat {\cal A}^I{}_\mu(x,y) =
{\cal A}^I{}_\mu(x)
\nonumber \\
&&\hat {\cal G}_{MN}(x,y) = {\cal G}_{MN}(x) ~~~~~~~~ \hat {\cal
B}_{MN}(x,y) = {\cal B}_{MN}(x) ~~~~~~~~\hat {\cal A}^I{}_M(x,y) =
{\cal A}^I{}_M(x)\nonumber \\ &&
\Phi(x,y) = \phi(x) + \frac{1}{2} \ln |\det({\cal G}_{MN}(x))|
\label{kk}
\end{eqnarray}

The dimensional reduction produces a number of new
scalar and vector fields. The additional scalars are the internal components
of the metric, the two-form and the gauge fields, \ie the fields
in the third line in eq.~\reef{kk}. For the compactification on a
$d$-torus, the counting of these scalars is:
$d(d+1)/2$ from the metric, $d(d-1)/2$ from the
two-form, and $16d$ from the 16 gauge fields. Thus
there are a total of $d(d+16)$ moduli scalars in
the reduced theory.

The case of the vectors is more interesting. The off-diagonal
terms in the metric and the two-form, $V^{M}{}_{\mu}$ and ${\cal
B}_{\mu M}$, transform as vector fields with respect to the
space-time diffeomorphisms of the $D$-dimensional theory, and so
give rise to new $U(1)$ gauge fields. A closer scrutiny of
eq.~(\ref{dans}) shows that the split of the components of ${\cal
B}$ and ${\cal A}^I$ as given there is not ``gauge-invariant'' as
it stands. The forms $dy^M$ transform under residual
diffeomorphisms according to $dy^M \rightarrow dy'^M = dy^M + d
\omega^M(x)$. This translates into the Kaluza-Klein gauge
transformations $V^M{}_{\mu} \rightarrow V'^M{}_{\mu} = V^M{}_\mu
- \partial_\mu \omega^M$, which ensure the invariance of the
internal space $d$-bein, \ie $dy^M + V^M{}_{\mu} dx^{\mu}
\rightarrow dy^M + V^M{}_{\mu} dx^{\mu}$. However, in the
decomposition of eq.~\reef{dans}, the reduced Yang-Mills gauge
fields and the vectors ${\cal B}_{\mu M}$ transform nontrivially
under this symmetry. To demonstrate manifest gauge invariance,
one's best route is to consider the kinetic terms in the action on
the tangent space, and carry dimensional reduction there, using
the gauge-invariant vielbein. This is the standard procedure used
for the dimensional reduction of supergravity theories
\cite{crjul}. When the result of this calculation is pulled back
to the holonomic $D$-dimensional basis, it is guaranteed to be
manifestly gauge invariant. After straightforward but tedious
algebraic manipulations with the reduction formulas, we find that
the reduced degrees of freedom, with simple gauge transformation
properties, are given in terms of the original higher-dimensional
degrees of freedom as follows: \ba &&A^I{}_{\mu} = {\cal
A}^I{}_\mu - {\cal A}^I{}_M V^M{}_{\mu} \nonumber \\ &&B_{\mu M} =
{\cal B}_{\mu M} + {\cal B}_{MN} V^N{}_\mu + \frac12 {\cal
A}^I{}_M A^I_\mu \nonumber \\ && B_{\mu\nu} = {\cal B}_{\mu\nu} +
V^M{}_{[\mu} B_{\nu]M} - {\cal B}_{MN} V^M{}_\mu V^N{}_\nu -
{\cal A}^I{}_M V^M{}_{[\mu} A^I{}_{\nu]} \label{ansatz}\ea In
appendix \ref{cartan}, we give the generalization of the
calculation leading to eq.~(\ref{ansatz}) for the case when
various axionic masses are turned on. Eq.~(\ref{ansatz}) can be
easily deduced from there by setting all masses to zero. Hence we
will merely quote the result here. The vector fields $B_{\mu M}$
and $A^I{}_\mu$ together with $V^M{}_\mu$ comprise the full
multiplet of $2d+16$ Abelian $U(1)$ gauge fields, with simple,
decoupled, gauge transformation properties. Their field strengths
will be denoted: $V^M{}_{\mu\nu}= \partial_\mu V^M{}_\nu -
\partial_\nu V^M{}_\mu $, $H_{\mu\nu M} = \partial_\mu B_{\nu M} -
\partial_\nu B_{\mu M}$ and $F^I{}_{\mu\nu}= \partial_\mu
A^I{}_\nu - \partial_\nu A^I{}_\mu$.

The reduced action in $D$ dimensions may be decomposed as
follows \cite{MS}:
\be
S = S_{1} + S_{2} + S_{3}
\label{redact}
\ee
where the reduced metric-dilaton-two-form action is
\be
S_{1} = \int d^Dx \sqrt{-{g}} e^{- \phi} \Bigl\{{R}
 + ({\nabla} \phi)^2 - \frac{1}{12} {H}^2_{\mu\nu\lambda} \Bigr\}\ ,
\label{actmetdil}
\ee
the scalar moduli action is
\ba
S_{2} &=& \int d^Dx \sqrt{-{g}} e^{- \phi} \Bigl\{
 \frac14 (\nabla_\mu {\cal G}_{MN})
(\nabla^\mu {\cal G}^{MN}) - \frac12  {\cal G}^{MN}
(\nabla_\mu {\cal A}^I{}_M) (\nabla^\mu {\cal A}^I{}_N) \nonumber
\\ &&\qquad-\frac14 {\cal G}^{MP} {\cal G}^{NQ}
\Bigl(\nabla_\mu {\cal B}_{MN} +
{\cal A}^I{}_{[M} \nabla_\mu {\cal A}^I{}_{N]}\Bigr)
\Bigl(\nabla^\mu {\cal B}_{PQ} +
{\cal A}^J{}_{[P} \nabla^\mu {\cal A}^J{}_{Q]}\Bigr) \Bigr\}
\label{actmod}
\ea
and the gauge field action is
\be
S_{3} = -\frac14 \int d^Dx \sqrt{-{g}} e^{- \phi} \Bigl\{
 f^I_{\mu\nu} f^{I~\mu\nu} + {\cal G}^{MN}
h_{\mu\nu M} h^{\mu\nu}{}_N + {\cal G}_{MN} V^M{}_{\mu\nu}
V^{N\,\mu\nu}
\Bigr\}\ .
\label{redactgaug}
\ee
In the latter, we use the definitions
\be
f^I_{\mu\nu} = F^I{}_{\mu\nu} + {\cal A}^I{}_M V^M{}_{\mu\nu}
~~~~~~~~~~ h_{\mu\nu M} = H_{\mu\nu M}- {\cal A}^I{}_M
F^I{}_{\mu\nu}
- {\cal C}_{MN} V^N{}_{\mu\nu}
\ee
where ${\cal C}_{MN}={\cal B}_{MN}+{1\over2}
{\cal A}^I{}_M{\cal A}^I{}_N$.
The reduced three-form field strength in eq.~\reef{actmetdil} is
\be
H_{\mu\nu\rho} = \partial_\mu B_{\nu\rho}
- \frac12  A^I{}_\mu F^I{}_{\nu\rho}
- \frac12 V^M{}_\mu H_{\nu\rho M} - \frac12 B_{\mu M}
V^M{}_{\nu\rho} + ~cyclic~perm.
\label{not}
\ee
In addition to the original Yang-Mills Chern-Simons terms, the
three-form field strength now also contains the induced Chern-Simons
terms arising for the new gauge fields appearing in the reduction.
These additional terms essential in establishing the duality symmetries
(see below) as shown in \cite{MS}, and remain
important in that context when higher derivative corrections are
also included \cite{km}.

The reduced low energy theory has a global $O(d,d+16,R)$ symmetry \cite{MS,sen}.
This symmetry is a generalization of $T$-duality symmetry, which
interchanges string momentum and winding modes \cite{buscher}.
In order to make this duality symmetry manifest, we must introduce
some additional notation. Towards this end, let us define a new
set of (lower-case Latin) indices which take values $a,b=1,2,\ldots,2d+16$.
Then $O(d,d+16)$ transformations may be defined as any real matrices
$\Omega^a{}_b$ leaving invariant the $O(d,d+16)$ metric
\be
\Omega^a{}_c\Omega^b{}_dL^{cd}=L^{ab}\qquad
{\rm with}\
L^{ab}=~\pmatrix{~0~&{\bf 1} &~0~ \cr {\bf 1}&~0~&~0~\cr ~0~&~0~& {\bf
1'}}
\label{ldef}
\ee
where ${\bf 1}$ are $~ d \times d~$ unit matrices and ${\bf 1'}$ a $16 \times
16$ unit matrix. In
more conventional matrix notation, we may write
eq.~\reef{ldef} as $\Omega\,L\,\Omega^T=L$, where the superscript $T$
indicates matrix transposition. We will denote
the inverse of the metric \reef{ldef} as $L_{ab}$, even though it is
precisely the same matrix.

Now following \cite{MS,sen}, we can define the reduced
gauge multiplet according to
\begin{equation}
A^a_{\mu}~=~\pmatrix{V^M{}_{\mu} \cr
                          B_{\mu M}\cr
                          A^I{}_\mu \cr}
{}~~~~~~~~~~
F^a_{\mu\nu}~=~\pmatrix{   V^M{}_{\mu\nu} \cr
                                H_{\mu\nu M} \cr
                                F^I{}_{\mu\nu} \cr}
\label{vectm}\end{equation}
In keeping with the index notation, these gauge fields transform
in the fundamental representation of $O(d,d+16)$, \ie $A^a\rightarrow
{A'}^a=\Omega^a{}_b A^b$ and $F^a\rightarrow {F'}^a=\Omega^a{}_b F^b$.
{}From the notation
introduced here, we also note that it is useful to keep in mind that
an $O(d,d+16)$ superscript runs over a contravariant index on the
internal $d$-torus, a covariant index on the same space, and a gauge
index labeling the sixteen
Cartan generators, \ie $X^a=\lbrace X^M,X_M,X^I\rbrace$.
Lowering the $O(d,d+16)$ superscript with the inverse of the invariant metric
\reef{ldef} interchanges the first two sets of components,
\eg $X_a=L_{ab}X^b=\lbrace X_M, X^M, X^I\rbrace$.

The scalar moduli fields, which parameterize the internal components
of the metric, two-form and the Yang-Mills fields, may also be
assembled in a form covariant under the duality transformations.
Using the $d \times d$ matrices ${\cal G} =
({\cal G}_{MN})$, ${\cal B} = ({\cal B}_{MN})$, and mixed $d
\times 16$ matrix ${\cal A}
= ({\cal A}^I{}_M)$, we can introduce the
matrix\cite{MS,sen}
\begin{equation}
M^{ab}~=~\pmatrix{ {\cal G}^{-1}&-{\cal G}^{-1}{\cal C}&-{\cal G}^{-1}
{\cal A} \cr
-{\cal C}^{T}{\cal G}^{-1}&{\cal G} +
{\rm a} + {\cal C}^{T}{\cal G}^{-1}{\cal C} & {\cal A} + {\cal C}^T
{\cal G}^{-1} {\cal A}\cr
- {\cal A}^T {\cal G}^{-1} &
{\cal A}^T + {\cal A}^T {\cal G}^{-1}{\cal C} &{\bf 1'} + {\cal
A}^T {\cal G}^{-1} {\cal A}\cr}
\label{w4}\end{equation}
where  ${\rm a}_{MN} =
{\cal A}^I{}_M {\cal A}^I{}_N$ while ${\cal C}_{MN}$ and ${\bf 1'}$
are defined as above. By construction,
$M^{ab}$ is symmetric, and it has a fixed trace, $L_{ab}M^{ab}=16$.
Further, it is straightforward
to verify that $M^a{}_b=M^{ac}L_{cb}$ is an element of $O(d,d+16)$.
In fact\cite{coset}, one finds that this moduli matrix parameterizes
the coset $O(d,d+16)/O(d) \times O(d+16)$ --- as a quick check,
one may verify that the number of scalars, $d(d+16)$, coincides with
the number of parameters in the coset. Under the duality
transformations, these scalars transform as: $M^{ab}\rightarrow
M'^{ab}=\Omega^a{}_c\Omega^b{}_dM^{cd}$.

With the definitions above, we can rewrite the reduced
three-form field strength \reef{not} as
\begin{equation}
 {H}_{\mu\nu\lambda} = \partial_{\mu}
{B}_{\nu\lambda}
- \frac{1}{2}
{A}^a_{\mu}\, L_{ab}\, {F}^b_{\nu\lambda} + cyclic ~permutations
\label{axfsin}\end{equation}
where all of the gauge anomaly contributions now appear on the same
footing\cite{MS}.
The  action of the dimensionally
reduced theory \reef{redact} can be written in a form
for which the full classical $O(d, d+16)$ symmetry is manifest\cite{actor}:
\ba
S &=& \int d^{D} x \sqrt{-{g}} e^{-\phi} \left\{{R}
 + ({\nabla} \phi)^2 - \frac{1}{12} {H}^2_{\mu\nu\lambda}\right.
\nonumber\\
&&\left.\qquad\qquad\qquad + \frac{1}{8}
 L_{ab} \nabla_\mu M^{bc}L_{cd}\nabla^\mu M^{da}
- \frac{1}{4} {F}^a_{\mu\nu}
 L_{ab}M^{bc}L_{cd} { F}^{d\,\mu\nu}\right\}
\label{actodd}
\ea
Given the invariance of $L$ in eq.~\reef{ldef},
this action and hence the
equations of motion are obviously invariant under the $O(d,d+16)$
rotations: $ M \rightarrow
\Omega M \Omega^{T}$ and ${F} \rightarrow \Omega {F}$
(using matrix notation).
This group includes diffeomorphisms and gauge transformations
on the $d$-torus, which are inherited from the reduction of the metric,
torsion, and gauge fields. These elements of $O(d, d+16)$
essentially rearrange the components of these tensors.
The nontrivial part\cite{nont,modi} of the $O(d,d+16)$
group is the coset $O(d,d+16)/O(d) \times O(d+16)$.

Before proceeding, we comment on the
origin of the $U(1)^{2d+16}$ gauge symmetry of the reduced theory.
The factor $U(1)^{16}$ is obviously inherited from the sixteen Abelian
gauge fields already present in the ten-dimensional theory.
As commented above, a further $U(1)^d$ comes from residual
diffeomorphisms on the internal space $dy^M\rightarrow
dy'^M=dy^M+d\omega^M(x)$. One might think of this as a local
gauging of the translation symmetries
of the internal torus. The remaining $U(1)^d$ symmetry arises
as a residual one-form gauge invariance with ${\cal B}\rightarrow
{\cal B}'={\cal B} +d\omega$ where $\omega=\omega_M(x)dy^M$.
We will see below how this gauge symmetry group is modified by the
introduction of axionic masses.

\section{Reduction with Internal Fluxes}
\label{massone}

The standard Kaluza-Klein reduction, described above, can be generalized
by the introduction of a constant flux of the three-form or gauge field
strengths on a three- or two-cycle in the internal space. These
compactifications are then similar to the Type II string and M-theory
reductions considered in ref.~\cite{flux}.
Note that a constant internal flux requires
that the corresponding
potential necessarily depends on the internal coordinates. These
fluxes, or alternatively the slopes for the internal dependence for the
potentials, then appear as mass parameters in the reduced theory.
There is also another set of masses related to
certain components of the internal metric, but the discussion of
these contributions is
more involved and we will leave their discussion for the following section.

Using the same decomposition as in eq.~(\ref{dans}), the internal
fluxes are produced by including an explicit dependence on
the internal coordinates in the following fields:
\begin{eqnarray}
&&\hat {\cal A}^I{}_M (x, y) = {\cal A}^I{}_M(x) + m^{I}_{MN} y^N
\nonumber \\ &&\hat {\cal B}_{\mu M}(x, y) = {\cal B}_{\mu M}(x) +
\frac12  m^I_{MN} y^N ( A^I{}_{\mu} (x) + {\cal A}^I{}_P
V^P{}_{\mu} (x))
\nonumber \\
&& \hat{\cal B}_{MN} (x, y) = {\cal B}_{MN}(x) + \beta_{MNP} y^P +
{\cal A}^I{}_{[M}(x) m^I_{N]P}y^P
\label{gauaxio}
\end{eqnarray}
The other fields reduce exactly as in eq.~(\ref{kk}). In this
decomposition, the quantities without the caret are the local
fluctuations on the base space, \ie they are independent
of the $y^N$. The ansatz (\ref{gauaxio}) can
be easily deduced from considering the field strengths of
${\cal A}^I$ and ${\cal B}$ on the tangent space and
requiring that their fluxes on the compact space
are independent of the internal coordinates. Again, the
details represent the special case of the general situation
discussed in the appendices \ref{cartan} and \ref{three},
when we take the internal space to be a flat
$d$-torus without twists (\ie with vanishing curvatures),
so that the internal isometries commute. For
this reason, we will not delve here into a detailed derivation of
(\ref{gauaxio}) but merely quote it. The parameters $\beta_{MNP}$
comprise a totally antisymmetric constant tensor in $M,\,N,\,P$, and
stem from the three-form fluxes on the internal torus. The
constant matrices $m^I=(m^I_{MN})$ are counted by $I$ and
antisymmetric in $M,\,N$, and they correspond to the
internal fluxes in the Yang-Mills sector. Note that implicitly we are
following the (string theory) convention that the fields are dimensionless,
and hence the flux parameters, $\beta_{MNP}$ and $m^I_{MN}$, have
the dimension of length$^{-1}$. Hence these parameters will appear as
mass parameters in the reduced theory.

Not all $m^I_{MN}$'s are independent, however. As we mentioned
above, the reduction ansatz must ensure that the reduced field
strengths simultaneously satisfy both their equations of motion and
are independent of the internal coordinates $y^M$. If we apply the
latter demand to the tangent space components of the three-form field
strength, we find that $\partial {\cal H}_{MNP}/\partial y^{Q} =
0$. This means that the form $dH$ cannot have terms which
completely reside in the compact space, as long as it is a flat
torus without curvature, with the one-form basis $dy^M$ on its tangent
space. However, since $d{\cal H} = -\frac12 {\cal F}^I \wedge
{\cal F}^I$, this translates into the condition that the
Chern-Simons anomaly components must vanish in the internal space.
This condition yields
\be
 m^I_{[MN}m^I_{PQ]}=0
\label{mconstr}
\ee
Since $m^I_{MN}$ are antisymmetric
in the lower two indices, we can rewrite (\ref{mconstr})
as $m^I_{MN} m^I_{PQ} + m^I_{MP} m^I_{QN} + m^I_{MQ}
m^I_{NP} = 0$. Aside from these constraints, the mass parameters
$m^I_{MN}$ are arbitrary.

In passing, we note that the ansatz (\ref{gauaxio}) is linear in
the internal coordinates despite the explicit presence of the
corresponding axionic scalars without derivatives in the reduced
action. One might not have expected choosing this form to succeed,
given our previous results in ref.~\cite{kkm} on generalized axion
reductions. In fact, however, the analysis there and our present
results are not in conflict, as discussed in appendix \ref{more}.

Using eq.~(\ref{gauaxio}), we can proceed with the dimensional
reduction of the action (\ref{sact1}). We will keep the discussion here as
brief as possible, since this is just a special case of the more
general reduction on a twisted torus that we will consider later.
The best way of carrying out dimensional reduction is to follow the
approach based on
using gauge symmetry as the organizing
principle for identifying the reduced dynamics. To do so, we use
the tangent space basis defined by the zehnbein, with components
\begin{equation}
{\cal E}^A{}_M~=~\pmatrix{~e^\alpha{}_\mu~&~E^A{}_N V^N{}_\mu~ \cr
~0~&~E^A{}_M~} ~~~~~~~~~~ {\cal
E}_A{}^M~=~\pmatrix{~e_\alpha{}^\mu~&~- V^M{}_\mu e_\alpha{}^\mu~
\cr ~0~&~E_A{}^M~}
\label{viel}\end{equation}
where $g_{\mu\nu} = \eta_{\alpha\beta} e^{\alpha}{}_{\mu}
e^{\beta}{}_{\nu}$, ${\cal G}_{MN} = \delta_{AB} E^A{}_M E^B{}_N$
and $e^{\alpha}{}_{\mu} e_{\alpha}{}^{\nu}
= \delta_{\mu}{}^{\nu}$
and $E^A{}_N E_A{}^M = \delta_N{}^M$, and the flat
metric on the ten-dimensional tangent space is
$\hat \eta_{ab}~=~{\rm diag}(\eta_{\alpha\beta}, \delta_{AB})$.
To actually identify the reduced degrees of freedom, all one needs
to do is to take the reduction ansatz given by eqs.~(\ref{dans}) and
(\ref{gauaxio}), and compute the forms ${\cal R}_{\alpha\beta}$
(Riemann curvature forms), ${\cal F}^I$ and ${\cal H}$ in the
``intermediate" basis spanned by $dx^\mu,~ {\cal E}^A =
E^A{}_N(dy^N + V^N{}_\mu dx^\mu)$. A straightforward calculation
then shows that
\ba
{\cal F}^I &=& \frac12 (F^I{}_{\mu\nu} + {\cal A}^I{}_M V^M{}_{\mu\nu}
) dx^\mu \wedge dx^\nu \nonumber\\
&&\quad+ {\cal D}_\mu {\cal A}^I{}_M E_A{}^M dx^\mu
\wedge {\cal E}^A - m^I_{MN} E_A{}^M E_B{}^N {\cal E}^A \wedge
{\cal E}^B
\label{cymred}
\ea
and
\ba
{\cal H} &=& \Bigl( \frac12 \partial_{\mu} B_{\nu\lambda} -
\frac14
A^I{}_\mu F^I{}_{\nu\lambda} - \frac14
B_{\mu M} V^M{}_{\nu\lambda}
- \frac14 V^M{}_\mu H_{\nu\lambda M} \nonumber \\
&&~~~~~~~~~ + \frac14 \beta_{MNP} V^M{}_\mu V^N{}_\nu V^P{}_\lambda
-\frac12 m^I_{MN} A^I{}_\mu V^M{}_\nu V^N{}_\lambda \Bigr)
dx^\mu \wedge dx^\nu \wedge dx^\lambda
\nonumber \\
&&+ \frac12
\Bigl(H_{\mu\nu M} - {\cal A}^I{}_M F^I{}_{\mu\nu} - {\cal
C}_{MN} V^N{}_{\mu\nu} \Bigr) E_A{}^M dx^\mu \wedge dx^\nu \wedge
{\cal E}^A
\nonumber \\
&&+ \frac12 \Bigl({\cal D}_\mu \B_{MN} +
{\cal A}^I{}_{[M} {\cal D}_\mu {\cal A}^I{}_{N]} \Bigr)
E_A{}^M E_B{}^N dx^\mu \wedge {\cal E}^A \wedge {\cal E}^B
\nonumber \\
&&+ \frac16 \Bigl(3 \beta_{MNP} + 6{\cal A}_{[M} m^I_{NP]}
\Bigr) E_A{}^M E_B{}^N
E_C{}^P {\cal E}^A \wedge {\cal E}^B \wedge {\cal E}^C
\label{nsnsfredc}
\ea
where in addition to eqs.~(\ref{dans}), (\ref{kk}), (\ref{ansatz}) and
(\ref{gauaxio}), we use the following definitions:
\ba
F^I{}_{\mu\nu} &=& \partial_{\mu} A^I{}_\nu - \partial_\nu
A^I{}_\mu
- 2 m^I_{MN} V^M{}_\mu V^N{}_\nu \nonumber \\
H_{\mu\nu M} &=& \partial_{\mu} B_{\nu M} -
\partial_{\nu} B_{\mu M} + 3 \beta_{MNP} V^N{}_\mu V^P{}_\nu
+ 4 m^I_{MN} A^I{}_{[\mu} V^N{}_{\nu]}
\label{fieldstr}
\ea
for the reduced field strengths (while we still have
$V^M{}_{\mu\nu} = \partial_{\mu} V^M{}_\nu - \partial_\nu
V^M{}_\mu$) and
\ba
{\cal D}_\mu {\cal A}^I{}_M &=& \partial_\mu {\cal A}^I{}_M - 2
m^I_{MN} V^N{}_\mu
\nonumber \\
{\cal D}_\mu \B_{MN} &=&
\partial_\mu \B_{MN} + 2 m^I_{MN} A^I{}_\mu - 3 \beta_{MNP}
V^P{}_\mu - 2 {\cal A}^I{}_{[M} m^I_{N]P} V^P{}_\mu\ .
\label{covder}
\ea
The expressions for the Riemann curvature can be obtained by
straightforward methods. Since the metric and dilaton reduction
ansatz is the same as in the conventional Kaluza-Klein case, the
curvature and the dilaton make precisely the same contributions
as in the previous section.

We can now reduce the action (\ref{sact1}) using these results.
This will generalize the Kaluza-Klein reduction of the low energy
heterotic action considered in \cite{MS}. The net result of this
calculation is
\be
S = S_{1} + S_{2} + S_{3}
\label{reds}
\ee
where the individual contributions to the action are
\be
S_{1} = \int d^D x \sqrt{-g} e^{-\phi} \Bigl\{R + (\nabla \phi)^2
- \frac12 H^2_{\mu\nu\lambda} \Bigr\}
\label{dilmetac}
\ee
for the reduced metric-dilaton-two-form part,
\ba
S_2 &=& -\int d^D x \sqrt{-g} e^{-\phi} \Bigl\{W({\cal G},\A)
- \frac14 \nabla_\mu {\cal G}_{MN} \nabla^\mu {\cal G}^{MN}
+ \frac12 {\cal G}^{MN} {\cal D}_\mu {\cal A}^I{}_M {\cal
D}^\mu {\cal A}^I{}_{N}
\nonumber \\
&&+ \frac14 {\cal G}^{MN} {\cal G}^{PQ} ({\cal D}_\mu \B_{MP} +
{\cal A}^I{}_{[M} {\cal D}_\mu {\cal A}^I{}_{P]}) ({\cal
D}^\mu \B_{NQ} + {\cal A}^J{}_{[N} {\cal D}^\mu {\cal
A}^J{}_{Q]})
\Bigr\}
\label{modract}
\ea
for the moduli fields, and
\be
S_{3} = -\frac14 \int d^D x \sqrt{-g} e^{-\phi} \Bigl\{ {\cal
G}_{MN} V^{M}{}_{\mu\nu} V^{N\,\mu\nu} +
f^I{}_{\mu\nu} f^{I\,\mu\nu} +
{\cal G}^{MN} h_{\mu\nu M} h^{\mu\nu}{}_N \Bigr\}
\label{gfl}\ee
for the gauge field contributions. The auxiliary fields in
eq.~(\ref{gfl}) are defined according to
\begin{eqnarray}
f^I{}_{\mu\nu} &=& F^I{}_{\mu\nu} + {\cal A}^I{}_{M} V^M{}_{\mu\nu}
\nonumber \\ h_{\mu\nu M} &=& H_{\mu\nu M}
- {\cal A}^I{}_M F^I{}_{\mu\nu}
- {\cal C}_{MN} V^N{}_{\mu\nu}
\label{redfs} \end{eqnarray}
The reduced three-form field strength, with all Chern-Simons
contributions, is in component form
\begin{eqnarray}
H_{\mu\nu\lambda} &=& \partial_\mu B_{\nu\lambda}
- \frac12 A^I{}_\mu
F^I{}_{\nu\lambda}
- \frac 12 V^M{}_\mu H_{\nu\lambda M}
- \frac 12 B_{\mu M} V^M{}_{\nu\lambda}
\nonumber \\ &&
+\frac12 \beta_{MNP} V^M{}_\mu V^N{}_\nu V^P{}_\lambda
- m^I_{MN} A^I{}_\mu
V^M{}_\nu V^N{}_\lambda + ~cyclic ~ perm. ~
\label{krax}\end{eqnarray}
In the moduli action \reef{modract},
the function $W({\cal G}_{MN}, {\cal A}^I{}_M)$ denotes the
moduli potential, which arises because of the internal fluxes. They
will in general induce an effective scalar potential, via the terms
such as, \eg ${\cal F}^I{}_{MN} {\cal F}^{I\,MN} \sim \G^{MN} \G^{PQ}
m^I_{MP} m^I_{NQ}$.
For the reduction scheme given by (\ref{gauaxio}), we find that the
reduced moduli potential is
\begin{eqnarray}
W(\G,\A) &=& \frac{3}{4} {\cal G}^{MN} {\cal G}^{PQ} {\cal G}^{RS}
\Bigl(\beta_{MPR} +
2 {\cal A}^I{}_{[M} m^I_{PR]}\Bigr)
\Bigl(\beta_{NQS} +
2 {\cal A}^J{}_{[N} m^J_{QS]} \Bigr) \nonumber \\
&&+{\cal G}^{MN} {\cal G}^{PQ} m^I_{MP} m^I_{NQ}
\label{pot}\end{eqnarray}
Note that this moduli potential is independent of the two-form
axions $\B_{MN}$.

At this point we wish to rewrite the reduced action (\ref{reds}) in
a more tractable form, which would highlight the symmetries of the
reduced theory. In particular, we wish to determine what became of
the $O(d,d+16)$ symmetry which appeared in the standard Kaluza-Klein
reduction. Further, the definitions (\ref{fieldstr}) and (\ref{covder})
already suggest that the reduced gauge symmetries have become
nonabelian with the internal fluxes. Below, we will confirm this
with a detailed examination of
the gauge symmetries of the reduced theory. Let us first try to
intuitively understand how the nonabelian symmetry is produced. For
this purpose, we can focus on the Yang-Mills gauge
transformations. The gauge symmetries of the Kalb-Ramond gauge fields
(\ie the gauge fields arising from the reduction of the two-form)
behave exactly the same. The Yang-Mills gauge
transformations are the sixteen commuting symmetry transformations
of the gauge fields ${\cal A}^I$, under which ${\cal A}^I
\rightarrow {\cal A}'^I = {\cal A}^I + d \Lambda^I$. From the point
of view of the Scherk-Schwarz reduced theory, these gauge
transformations can be at most quadratic in $y^M$, in order not to
alter the general form of the reduction ansatz (\ref{gauaxio}).
Hence, the general allowed form of gauge transformations is
$\Lambda^I(x,y) =
\lambda^I(x) + \lambda^I{}_M(x) y^M + \lambda^I{}_{MN}(x) y^M y^N$.
Here, $\lambda^I{}_{MN}(x) = \lambda^I{}_{NM}(x)$ or else it would
disappear from the definition above. Further, if we turn off the
Kaluza-Klein gauge fields, $\lambda^I{}_M$ and $\lambda^I{}_{MN}$
must be constant. Using the reduction ansatz for ${\cal A}^I$, we
see that the gauge transformation properties of the fields ${\cal
A}^I$ decompose according to
\be
{\cal A}'^I{}_\mu = {\cal A}^I{}_\mu + \partial_\mu \lambda^I
~~~~~~~~ {\cal A}'^I{}_M = {\cal A}^I{}_M + \lambda^I{}_M ~~~~~~~~
m'^I_{MN} = m^I_{MN} + \lambda^I{}_{MN}
\label{gaugevec}
\ee
Since $\lambda^I{}_M$ and $\lambda^I{}_{MN}$ are constants, we see
that these transformations by $\Lambda^I$ has three separate effects:
First, the $\lambda^I(x)$ yield the
reduced version of the original $U(1)^{16}$ gauge symmetry appearing
in the action \reef{sact1}. Then,
we obtain the global axionic symmetries, which shift
the scalars $A^I{}_M$ by constants. Finally, we see that the
$\lambda^I{}_{MN}$
transformations simply ensure that only the antisymmetric part of
$m^I_{MN}$ are physical. In other words, even if we
had started with a general class of parameters $m^I_{MN}$, we can make
them antisymmetric by $\lambda^I{}_{MN}$-dependent gauge
transformations, leaving the rest unchanged.

The Kaluza-Klein vectors $V^M{}_\mu$ do not change under these
transformation. However, due to the Chern-Simons terms, the two-form potential
${\cal B}$ does transform, according to
\begin{equation}
\delta {\cal B} = \frac12 \Lambda^I {\cal F}^I + d \Lambda
\label{bagauge}
\end{equation}
where $\Lambda$ is a residual two-form gauge transformation playing
role of a custodian, which sweeps away total derivatives. It is
easy to verify that the explicit form of $\Lambda$ should be
$\Lambda
= \frac12  m^I_{MN} \lambda^I y^N dy^M - \frac12
\Lambda^I {\cal A}^I$. This guarantees that the reduction
ansatz for the two-form in eq.~(\ref{gauaxio}) is gauge invariant. This
induces nontrivial transformation properties of the fields which
emerge from the two-form ${\cal B}$ after the reduction. Their
explicit form will be given below. Here we first note that the
parameters $\beta_{MNP}$ change according to
\begin{equation}
\beta_{MNP} \rightarrow \beta'_{MNP} = \beta_{MNP}
- 2 \lambda^I_{[M} m^I_{N]P}
\label{bgtransf}
\end{equation}
{}From eq.~\reef{nsnsfredc}, we see that eq.~\reef{gaugevec} and
\reef{bgtransf} combine to ensure that the internal flux of $\cal H$
remains unchanged by the axionic shifts of $A^I{}_M$.

Turning the Kaluza-Klein gauge transformations on corresponds to
gauging the global axionic translations associated with the metric
axions. The resulting gauge group must be nonabelian because of the
anomaly in the two-form ${\cal B}$. The anomaly is manifest in the
presence of the vector supermultiplet Chern-Simons terms in
eq.~(\ref{krdef}). It gets enlarged further by reduction, because of
the gauge anomalous decomposition of the reduced two-form
$B_{\mu\nu}$.

Without further ado, we give here the final form of
the infinitesimal form of the full set of
reduced gauge transformations. They can be
obtained by straightforward, albeit lengthy computation, of which the
general case is given in appendix \ref{algebraaa}.
Hence the case of the flat torus, which is considered here,
can be recovered as a special case. We only
list the reduced vector field gauge transformations, and leave
aside the one-form gauge transformations of the reduced two-form
field. These reduced infinitesimal
gauge transformations fall into three categories,
listed here in the order of ascending complexity:

1) Kalb-Ramond gauge transformations:
\be
B'_{\mu M} = B_{\mu M} + \partial_\mu \lambda_M ~~~~~~~~
B'_{\mu\nu} = B_{\mu\nu} + \frac12 \lambda_M V^M{}_{\mu\nu}
\label{nsnsgauge}
\ee

2) Yang-Mills gauge transformations:
\begin{eqnarray}
&& {A'}^I{}_\mu = A^I{}_\mu + \partial_\mu \lambda^I
~~~~~~~~
B'_{\mu M} = B_{\mu M}- 2 \lambda^I m^I_{MN} V^N{}_\mu
\nonumber \\
&&B'_{MN}=B_{MN} - 2 \lambda^I m^I_{MN}
\label{ymcgauge}\\
&&B'_{\mu\nu} =B_{\mu\nu} + \frac12
\lambda^I F^I{}_{\mu\nu}
+ m^I_{MN} \lambda^I V^M{}_\mu V^N{}_\nu
\nonumber
\end{eqnarray}

3) Kaluza-Klein gauge transformations:
\begin{eqnarray}
V'^M{}_\mu &=& V^M{}_\mu + \partial_\mu
\omega^M ~~~~~~~~
~~~~~~~~ {A'}^I{}_\mu = A^I{}_\mu -2 m^I_{MN} \omega^N V^M{}_\mu +
O(\omega^2)
\nonumber \\
B'_{\mu M} &=& B_{\mu M} + 2
m^I_{MN} \omega^N A^I{}_\mu + 3\beta_{MNP} \omega^P V^N{}_\mu +
O(\omega^2)
\nonumber \\
{\cal A}'^I{}_M &=& {\cal A}^I{}_M + 2 m^I_{MN} \omega^N
\label{kkgauge}\\
B'_{MN} &=& B_{MN} + 3\beta_{MNP} \omega^P + 2
{\cal A}^I{}_{[M} m^I_{N]P} \omega^P + O(\omega^2)
\nonumber \\
B'_{\mu\nu} &=& B_{\mu\nu} + \frac12
\omega^M H_{\mu\nu M} - \frac32 \beta_{MNP} \omega^M V^N{}_\mu V^P{}_\nu
-2 \omega^M
m^I_{MN} A^I{}_{[\mu} V^N{}_{\nu]} + O(\omega^2)
\nonumber
\end{eqnarray}
Here those fields that are not listed above explicitly are invariant under the
corresponding transformations.

We can now determine the algebra of the gauge group defined by
eqs.~(\ref{nsnsgauge})-(\ref{kkgauge}). Understanding the full
structure of the gauge symmetry is facilitated by adopting the $O(d,d+16)$
notation introduced previously, which combines the three sets of
gauge fields in a single multiplet \reef{vectm}. We can then define
a combined gauge transformation parameter
\be
\hat{\omega}^a=(\omega^M,\lambda_M,\lambda^I)\ .
\label{parmg}
\ee
Now the proper way to examine the
gauge symmetries is to introduce the corresponding generators $T_a$, where the
index now takes values in the space of $\{~_M, ~^M, ~^I\}$, of
dimension $2d+ 16$. We  explicitly denote the separate generators as
 $T_a = (Z_M,
X^M,Y^I)$ --- note the order: first Kaluza-Klein, then Kalb-Ramond and
finally Yang-Mills. We expect that their algebra satisfies
\be
[T_a, T_b] = i f_{ab}{}^c T_c
\ee
where $f_{ab}{}^c$ are some structure constants to be determined. To
find them, we consider the successive application of gauge
transformations (\ref{nsnsgauge}--\ref{kkgauge}) of the form
$h^{-1} \cdot g^{-1}
\cdot h \cdot g$ where $h$ and $g$ are two of any of the three
above types of gauge transformations. We can use the
Baker-Hausdorff formula to determine the product. Infinitesimally,
for any two operators $A, B$ and a number $\alpha$ we have
$e^{\alpha A} B e^{-\alpha A} = B + \alpha [A,B] + O(\alpha^2)$
where $[~~,~~]$ denotes the commutator. If $g = \exp(i
\hat \omega_1^a T_a)$ and $h = \exp(i \hat \omega_2^a T_a)$, with
gauge parameters $\hat\omega_{1,2}^a$ as in eq.~\reef{parmg},
we have, to the lowest order in $\omega_1^a$,
$g^{-1} \cdot h \cdot g = h - i \hat \omega_1^a [T_a, h]$
and hence we find that, to the lowest order
\begin{equation}
h^{-1} \cdot g^{-1} \cdot h \cdot g = 1 + \hat \omega_1^a \hat
\omega_2^b [T_a, T_b] = 1 + i f_{ab}{}^{c} \hat \omega_1^a
\hat \omega_2^b T_c
\label{inftr3}
\end{equation}
Substituting the explicit form of the reduced gauge transformations
(\ref{nsnsgauge}--\ref{kkgauge}), we can deduce the structure
constants. Since they do not depend on a representation, we need to
project the operator expression (\ref{inftr3}) on a set of basis
vectors of a faithful irreducible representation of the gauge
group. In other words, we should evaluate $h^{-1} \cdot g^{-1}
\cdot h \cdot g | \Psi \rangle$ for a set of basis vectors $|\Psi\rangle$.
Vector fields must provide a faithful representation, and hence
using (\ref{nsnsgauge}--\ref{kkgauge}), we find the following
equations for the structure constants:
\begin{equation}
f^I{}_{MN} = f_{MN}{}^I=2 m^I_{MN} ~~~~~~~~~~ f_{MNP} = -3 \beta_{MNP}
\label{strc}
\end{equation}
These equations must hold in order to get the rules for the
combination of gauge transformation functions, $\lambda_M =
 \omega^N \lambda^I f^I{}_{NM}
= -2  m^I_{MN} \omega^N \lambda^I$ and
$\lambda_P = f_{MNP} \omega_1^M \omega_2^N = 3 \beta_{PNM}
\omega_1^M
\omega_2^N$,
as can be seen from the comparison of the compositions with the
original gauge transformations. Hence finally, the gauge algebra is
\begin{eqnarray}
&&[X^M, X^N] = [Y^I, Y^J] = [X^M, Y^I] = [X^M, Z_N] = 0 \nonumber
\\ &&[Y^I, Z_M] = 2i m^I_{MN} X^N ~~~~~~~~~~~~ [Z_M, Z_N] = -3i
\beta_{MNP} X^P + 2i  m^I_{MN} Y^I
\label{algebra}
\end{eqnarray}
Thus we see that the reduced gauge group is indeed nonabelian with the
internal flux parameters playing the role of the structure constants.

In order to rewrite the reduced theory in the form where the
nonabelian gauge symmetry is manifest, we first need to define a
metric on the Lie algebra of the gauge group. Normally, one would
use the standard Cartan metric on a Lie Algebra, given by $C_{ab} =
f_{ac}{}^{d} f_{bd}{}^{c}$. However, from eq.~(\ref{algebra})
in our example the gauge algebra is not semi-simple. This implies
that the Cartan metric is degenerate, and so without an inverse.
By comparing the gauge algebra with the algebra of the $O(d,
d+16)$ duality group which arises in the reduced action, one can
verify that the gauge algebra has become embedded in the duality
algebra. The reason is that the nonabelian gauge transformations,
as dictated by eq.~(\ref{algebra}), now mix the gauge generators, a
function previously reserved solely for the duality
transformations. However, even without this observation, and by
considering the case of the standard massless Kaluza-Klein
reduction from the previous section,
we see that the natural metric in the gauge algebra can be defined
by $\langle T_a, T_b\rangle \equiv L_{ab}$.
We will return to this interplay of the gauge and U-duality group
later.

Identifying our Lie-algebra-valued one-form gauge potential as
\begin{equation}
A =A^aT_a= V^M Z_M + B_M X^M +  A^I Y^I
\label{gfp}
\end{equation}
we find that the corresponding nonabelian field strength
\begin{equation}
F = d A + i A \wedge A = \frac12 F^a_{\mu\nu} T_a dx^\mu \wedge
dx^\nu
\label{gfla}
\end{equation}
has components which coincide with the expressions for field
strengths \reef{fieldstr} that come from dimensional reduction:
\be
F^a_{\mu\nu} = (V^M{}_{\mu\nu},~H_{\mu\nu M},~F^I{}_{\mu\nu})
\label{compns}
\ee
In component form, we have
\begin{eqnarray}
F_{\mu\nu}{}^M &=& \partial_\mu V^M{}_{\nu} - \partial_\nu
V^M{}_\mu ~~~~~~~~~~ F_{\mu\nu}{}^I = \partial_\mu A^I{}_{\nu} -
\partial_\nu A^I{}_\mu
- 2 m^I_{MN} V^M{}_\mu V^N{}_\nu
\nonumber \\
F_{\mu\nu M} &=&
\partial_\mu B_{\nu M} - \partial_\nu B_{\mu M}
+ 4  m^I_{MN} A^I{}_{[\mu} V^N{}_{\nu]} + 3 \beta_{MNP}
V^N{}_\mu V^P{}_\nu
\label{gfcomp}
\end{eqnarray}
The coupling constant may appear to be normalized to unity in eq.~\reef{gfla}.
More accurately, the coupling is absorbed into the structure constants
\reef{strc}, as the flux parameters are by the present analysis arbitrary
in magnitude.

We can also compute the Chern-Simons form for the gauge field \reef{gfp}.
The result is
\begin{eqnarray}
\omega_{CS} &=& \langle A \wedge F - \frac{i}3 A \wedge A \wedge A\rangle
\nonumber\\ &=& \frac13 dx^\mu \wedge dx^\nu \wedge dx^\lambda
\,\Bigl( \frac12  A^I{}_\mu F^I{}_{\nu\lambda} +
\frac12 V^M{}_\mu H_{\nu\lambda M} +
\frac12 B_{\mu M} V^M{}_{\nu\lambda} \nonumber \\
&&~~~ + \frac12 \beta_{MNP} V^M{}_\mu V^N{}_\nu V^P{}_\lambda -
 m^I_{MN} A^I{}_\mu V^M{}_\nu V^N{}_\lambda + {cyclic~perm.}
\Bigr)
\label{csfin}
\end{eqnarray}
which corresponds to
precisely the anomaly contribution in the reduced
three-form field strength, obtained by dimensional reduction, and
given in eq.~(\ref{krax}). Hence, we see that with the internal fluxes
all of the Chern-Simons contributions
terms in eq.~\reef{krax} can be organized into a single nonabelian
structure, and that, in form notation, the reduced three-form field
strength is then simply
\be
H = dB - \frac12 \omega_{CS}\ .
\ee
This is
exactly what is needed to maintain simultaneously U-duality and gauge
invariance. The transformation formulas for $B_{\mu\nu}$ always
contain the anomalous gauge transformation of $B$. This is
essential to establish the $O(d, d+16)$ duality invariance of
the reduced theory, as shown by Maharana and Schwarz \cite{MS}.
However,  in the case of the Scherk-Schwarz
dimensional reduction, there are additional terms arising in order
to account for additional nonlinearities present in the
Chern-Simons sector, and we have seen here that these are precisely
in the form of the nonabelian anomalous contributions.

Finally, we can put the moduli potential \reef{pot} in a symmetric form.
We lower the last index of the structure constants to define
the completely antisymmetric tensor in the Lie group
$f_{abc} = f_{[ab}{}^dL_{d|c]}$.
Using this tensor
and the  moduli matrix $M^{ab}$ introduced in eq.~\reef{w4},
we can write the potential for these scalars as:
\begin{equation}
W(M)=W({\cal G}, \B, {\cal A}) = \frac{1}{12} M^{ad}
M^{be} M^{cf} f_{abc} f_{def}
\label{potcov}
\end{equation}
Further, the constraint \reef{mconstr}
for the masses $ m^I_{[AB} m^I_{CD]} =0$ is encoded in
\be
L^{ad} f_{a[bc} f_{d|ef]} = \frac13 L^{ad} \Bigl(f_{abc} f_{def} +
f_{abe} f_{dfc} + f_{abf} f_{dce} \Bigr) = 0
\ee
That is, the flux constraint is precisely the only nontrivial component
of the Jacobi identity for the Lie algebra defined with the structure
constants $f_{ab}{}^c$ as in eq.~\reef{strc} and with the metric
$L_{ab}$.

Armed with these expressions, we can write down the reduced action
in the manifestly covariant form. Having squared away the anomaly,
we note that the gauge kinetic terms in the reduced action \reef{gfl} can be
rewritten as
\begin{equation}
{\cal F}^2 =  F^a_{\mu\nu} L_{ab} M^{bc} L_{cd}
F^{d\mu\nu}
\label{compkt}
\end{equation}
precisely as before, although the field strengths \reef{compns}
are now nonabelian. The nonabelian field strengths vary under gauge
transformations according to
\begin{equation}
F_{\mu\nu} \rightarrow F'_{\mu\nu} = U \cdot F_{\mu\nu} \cdot
U^{-1}
= F^a_{\mu\nu} U \cdot T_a \cdot U^{-1} = U^b{}_a F^a_{\mu\nu}\, T_b
\label{gtoff}
\end{equation}
Hence, the transformation law for the gauge field strength is
$F'^a_{\mu\nu} = U^a{}_b F^b_{\mu\nu}$. Gauge invariance of the kinetic
terms thus demands that the matrix $M$ must also transform nontrivially
under the nonabelian group, in contrast to the
simple Abelian situation encountered in standard Kaluza-Klein
reduction. Indeed, we find
\begin{equation}
M_{ab} \rightarrow M'_{ab} = U^c{}_a U^d{}_b M_{cd} ~~~~~~~ M^{ab}
\rightarrow M'^{ab} = U^a{}_c U^b{}_d M^{cd}
\label{mgauge}
\end{equation}
where $M_{ab}\equiv L_{ac}M^{cd}L_{db}$. The second transformation
is derived by noting that the gauge transformation $U^a{}_b$ is a
$O(d,d+16)$ matrix. Hence
since $F^a$ transform as an adjoint of the gauge group, $M^{ab}$
transforms as a (symmetrized) product of the two adjoints.
Therefore, the kinetic terms for the moduli fields must contain
gauge covariant derivatives
\begin{equation}
{\cal D}_\mu M^{ab} = \partial_\mu M^{ab} - f_{cd}{}^a A^c_{\mu}
M_{db} - f_{cd}{}^b A^c_{\mu} M^{ad}\ .
\label{covdermf}
\end{equation}
Combining the various expressions above, one finds that
these expressions are reproduced precisely by the derivatives \reef{covder}
appearing in the scalar kinetic terms in eq.~(\ref{modract}).

The  reduced action can therefore be put in a U-duality invariant
form:
\begin{eqnarray}
S &=& \int d^{D} x \sqrt{-{g}} e^{-\phi} \Bigl\{{R}
 + ({\nabla} \phi)^2 + \frac{1}{8}
L_{ab} {\cal D}_\mu M^{bc} L_{cd} {\cal D}^\mu M^{da}
 \nonumber \\
&&~~~~~~~~~~~~~~~~~~~~~ - \frac{1}{4} F^a_{\mu\nu}
 L_{ab} M^{bc} L_{cd} F^{d\mu\nu}
- \frac{1}{12} {H}^2_{\mu\nu\lambda} - W(M) \Bigr\}\ .
\label{actoddna}
\end{eqnarray}
Hence the generalized reduction has produced an effective
$D$-dimensional action of a form very similar to that \reef{actodd}
produced by the standard Kaluza-Klein reduction. The main differences
arise in the appearance of a nonabelian gauge symmetry and the
scalar potential $W(M)$. The parameters introduced by the
internal fluxes play a dual role as structure constants in the
nonabelian group, and mass couplings determining the potential
for the moduli.
With respect to the duality symmetry, it may appear from the form of the action
\reef{actoddna} that the $O(d,d+16)$ symmetry
is still present but in a slightly broken form, due
to the constants $f_{abc}$. An alternate point of view,
which we advocate, is that $O(d,d+16)$ is a symmetry
of the reduced theory, and that we are discovering here that
the usual transformations,
$M \rightarrow M'=\Omega M \Omega^{T}$ and $F
\rightarrow F' = \Omega F$, must be supplemented by a
transformation of the fluxes:
\be
f_{abc} \rightarrow f'_{abc} = \Omega_a{}^d \Omega_b{}^e
\Omega_c{}^f f_{def}
\label{strcontr}
\ee
where in the present notation,
$\Omega_a{}^b=L_{ac}\Omega^c{}_dL^{db}=(\Omega^{-1})^b{}_a$.
In the following sections, we will find that similar results
apply for the generalized reduction when mass parameters are also introduced
through the metric axions.

\section{Twisted Tori}
\label{twist}

We now turn to the question of inducing masses in general.
In particular, we wish to include mass parameters arising from
internal coordinate dependence of the metric.
Metric axion masses may be associated with
axionic symmetries for internal metric components, and so an easy way to
identify these is to first reduce the theory by the standard
Kaluza-Klein procedure and then look for axionic symmetries in the
reduced theory. In toroidal
standard reductions, a metric axion first appears
from the off-diagonal component of the internal metric in compactifying
on a two-torus. Hence a mass parameter can be generated by introducing
an internal coordinate dependence in this axion in a further $S^1$ reduction.
Geometrically, the axionic symmetry above corresponds to a constant
shift in the modular parameter of the two-torus. The subsequent reduction
then produces a nontrivial $SL(2,R)$ fibre bundle with $T^2$ fibre over $S^1$.
That is the $T^2$ geometry varies in traversing the $S^1$, and in one
cycle, its metric only returns to its original form up to an $SL(2,R)$
transformation. Thus the resulting geometry is similar to that appearing
around a stringy cosmic string \cite{comsic}, whose construction relies on
the same axionic symmetry of the two-torus reduction. This geometry
was also recently discussed\cite{hull,ortin} in relating M-theory to
massive IIa string theory.
We will refer to this three-torus with such nontrivial
curvature and its higher dimensional generalizations as ``twisted tori''.

A sequential dimensional descent extending the $T^3$ reduction described above
may not be the most useful tool to derive the most general
array of metric mass parameters. Further the geometric picture does not
give us a clear picture of the most general compactification.
In the following, we will consider a generalized ansatz
in which the internal components of the metric can have
complicated dependence on internal coordinates, restricted
only by the requirement that the internal manifold has an
isometry group acting transitively on it \cite{duffpope}.
Indeed, as we will see, the general procedure to turn on all available
metric mass terms is to perform dimensional reduction on all nonabelian
isometry groups consistent with the dimension of the compact space.
That is, while throughout this section, we have in mind the twisted
tori described above, the formalism developed here and the subsequent
results apply for more general compactifications. The essential
ingredient is an isometry group with a transitive action and with the
same dimension as the internal space. Hence
the topology of the internal manifold is not restricted to
be homeomorphic to fibered tori, and we will consider explicit
examples in section \ref{conc}.

Our construction is motivated by the general considerations of
\cite{schs}, and a concrete example provided
in ref.~\cite{kaloper}. The compactification considered there may be
interpreted as a twisted three-torus.
Explicitly, the basis one-forms on this torus were found to be
$\eta^1 = dx + Q (ydz - zdy), \eta^2 =dy, \eta^3 =dz$. This basis
contains a noncommutative structure, as can be seen from the fact that
$d\eta^1 = 2Q \eta^2 \wedge \eta^3$.
This in turn resulted in a nonabelian isometry group
of the internal space, with the nontrivial commutator between the
dual Killing vectors: $[Z_2, Z_3] = - 2Q Z_1$. This algebra is
known from the studies of homogeneous anisotropic spaces under the
name of Bianchi II algebra. The vector fields $Z_M$ are generators
of generalized translations, which leave the geometry invariant.
Hence the notion of homogeneity.

Our generalized ansatz builds on the above result with the introduction
of tangent space basis forms on the internal manifold of the form
\be
{\cal E}^A ={\cal E}^A{}_M(x)\,\eta^M(y,dy)
\label{nubase}
\ee
where the one-forms $\eta^M$ are allowed to depend on the internal coordinates,
in such a way that they yield all allowed
isometry groups with the dimension of the
Lie algebra equal to the given dimension of the internal space.
The Killing isometries generalize the internal space translations
present in the simplest reductions on flat tori.
The corresponding Killing vector fields $Z_M$,
which generate these isometries, are dual to the basis
one-forms $\eta^M$, \ie $Z_M(\eta^N) = \delta^M_N$.
To bring the full
force of symmetry to good use, we define the pullback functions
$N_N{}^M(y)$ from the holonomic basis $dy^M$ to the twisted
basis $\eta^M$:
\be
\eta^M = N_N{}^M(y) dy^N\ .
\label{etady}
\ee
Since the matrix $N_N{}^M(y)$ is a map between two bases, it must
be everywhere invertible on a smooth manifold.
Let its inverse be $N^M{}_N(y)$, so that the inverse map to
(\ref{etady}) is $dy^M = N^M{}_N(y) \eta^N$.
For the simple reduction in
the Bianchi II example, both $N_N{}^M(y)$ and
its inverse depended linearly on $y^M$.

The Killing vector fields can be expressed in terms of the holonomic
coordinate vector fields $\partial_M = \frac{\partial\ }{\partial
y^M}$ and the inverse pullback matrix $N^M{}_N(y)$ as
\be
Z_M = N^N{}_M(y) \partial_N
\label{vecfield}
\ee
They must satisfy a Lie algebra specified by a set of fixed structure
constants $2 \gamma^M_{NP}$
\be
[Z_M, Z_N] = 2\gamma^P_{MN} Z_P
\label{liekill}
\ee
where we have introduced the factor of $2$ for later convenience.
The dual map between $\eta^M$ and $Z_M$ then implies\footnote{This
can be seen as follows. From ${\cal L}_{Z_M}(Z_N(\eta^P)) = {\cal
L}_{Z_M} \delta^M_N = 0$ we have $Z_N({\cal L}_{Z_M} \eta^P)
= -[Z_M,Z_N](\eta^P) = - 2\gamma^Q_{MN} Z_Q(\eta^P) = - 2\gamma^P_{MN}$. Next,
${\cal L}_{Z_M} \eta^P = d(Z_M(\eta^P)) + Z_M (d \eta^P)$.
Therefore, if we expand the two-form $d \eta^P$ in terms of the
basis forms, $d\eta^P = \eta^P_{MN} \eta^M \wedge \eta^N$, we find
that $Z_N({\cal L}_{Z_M} \eta^P) = 2 \eta^P_{MN}$. Thus we arrive
at $\eta^P_{MN} = - \gamma^P_{MN}$, and $d \eta^P = -
\gamma^P_{MN} \eta^M \wedge \eta^N$, as claimed.}
\be
d \eta^M = - \gamma^M_{NP} \eta^N \wedge \eta^P
\label{formalg}
\ee
By the Bianchi identity for basis one-forms, $d^2 \eta^N = 0$, we
see that the structure constants must satisfy
\be
\gamma^M_{R[N}\, \gamma^R_{PQ]} = 0
\label{congamma2}
\ee
where the square brackets denote antisymmetrization of the enclosed
indices. This identity must hold for any set of values of $M,N,P,Q$.
Hence, the symmetry algebra is encoded in the commutation relations
of Killing isometries \reef{liekill} and derivatives of their dual
one-forms \reef{formalg}.
The structure constants $\gamma^M_{NP}$ of the
isometry group on the internal space
will play the role of axionic masses in the reduced theory.
This is evident from the simple Bianchi II example considered
in ref.~\cite{kaloper}.

A final constraint on $\gamma^M_{NP}$ can be obtained by considering
the volume form on the internal space,
$ V_{d} = \frac{1}{d!} \epsilon_{N_1 \ldots N_{d}} \eta^{N_1} \wedge ...
\wedge \eta^{N_{d}} = \det(N_N{}^M) d^{d} y$.
Invariance of this measure under the Killing isometries $Z_M$ implies
\cite{schs}
\be
\gamma^N_{NM} = 0\ .
\label{congamma3}
\ee
We can see this as follows:
We require that ${\cal L}_{Z_M} V_d=0$ where $\cal L$ denotes the Lie
derivative. For differential forms, one has ${\cal L}_\xi=di_\xi+i_\xi d$
where  $i_\xi$ denotes the interior product with a vector $\xi$.
Further in the case of interest $dV_d=0$ since it is the top form
on the internal space, and so
\be
{\cal L}_{Z_M} V_d=di_{Z_M} V_d
=(-)^{M+1} d\left(  \eta^{1} \wedge \ldots \widehat{\eta}{}^M\ldots
\wedge \eta^{d}\right)
\ee where the caret denotes that $\eta^M$ does not appear
in the wedge product. Given the exterior derivatives of the
basis forms in \reef{formalg}, it is not hard to see that
${\cal L}_{Z_M} V_d = \gamma^N_{NM}V_d$, and so invariance
requires eq.~\reef{congamma3}.

Let us now consider the infinitesimal form of the isometries,
which we determine from the Lie algebra. Since an infinitesimal transformation
of a tensor field $T$ generated by a Killing field $Z = \omega^P Z_P$ can be
written as $\delta T = {\cal L}_Z T$, we see that the basis $1$-forms
transform according to
\be
\eta^N \rightarrow (\eta')^N = {\cal S}^N{}_M \eta^M
\label{inftransf}
\ee
where the matrix ${\cal S}^M{}_N(\omega)$ belongs to the adjoint
of the group of diffeomorphisms on the internal space defined
by \reef{liekill}. Hence, this matrix can be
written in infinitesimal form as
\be
{\cal S}^M{}_N(\omega) = \delta^M_N - 2 \gamma^M_{NP}
\omega^P(x)
\label{odef}
\ee
It will be useful to also define the matrix ${\cal O}^M{}_N(\omega)$,
as formally a square root of ${\cal S}$. In infinitesimal form,
\be
{\cal O}^M{}_N(\omega) = \delta^M_N - \gamma^M_{NP} \omega^P(x)
\label{odefsqroot}
\ee
This matrix will encode the gauge transformations of the gauge fields
after dimensional reduction. To see this, we can take the parameters
$\omega^M(x)$ to depend on the base space coordinates, in which case the
transformations of the basis $1$-forms $\eta^M$ obey
\be
{\eta'}^M = {\cal S}^M{}_N \eta^N - {\cal O}^M{}_N d \omega^N
\label{actgroup}
\ee
where $d \omega^N = \partial_\mu \omega^N dx^\mu$ takes into
account the fact that $\omega^N$ are defined as local functions on
the reduced base space. The factor of $2$ in $S^M{}_N$ comes about
because the charge of $\eta^M$ is $1$, and the structure constants are
$2 \gamma^M_{NP}$.

For completeness we note that we can define the invariant forms
under the action of the symmetry group \reef{odef}. The
simplest way to construct them is to first look for invariant
vector fields, and then determine their duals. The invariant vector
fields are defined by ${\cal L}_{Z_M} {\cal Y} = 0$ for all $Z_M$.
Writing out this condition explicitly, we see that the most general
solution is
\be
{\cal Y} = {\cal X}^N N_N{}^M \partial_M = {\cal X}^M {\cal Y}_M
\label{invvect}
\ee
where ${\cal X}^M$ are arbitrary constants. Hence the maximal set
of vector fields invariant under (\ref{odef}) is the set of
${\cal Y}_M = N_M{}^N \partial_N$. Their dual one-forms
\be
\zeta^M = N^M{}_N dy^N
\label{invtforms}
\ee
are also invariant under the isometry group, ${\cal L}_{Z_M}
\zeta^N = 0$.
Note that in the terminology of
group manifolds the forms $\zeta^M$ correspond to the
right-invariant Maurer-Cartan forms on the group space,
whilst our basis $1$-forms $\eta^M$ correspond to the
left-invariant Maurer-Cartan forms. Having chosen $\eta^M$ as the
basis $1$-forms, we have elected to express
the internal metric in terms of the left-invariant forms.
One might be tempted to use the forms $\zeta^M$ for
the dimensional reduction of invariant objects, but
encoding the isometries with the use of the left-invariant
generators $Z_M$.
However, as we will discuss below, this is
not the optimal starting point for dimensional reduction in our
case, since it is incapable of recording axionic
shift symmetries, necessary to turn the axionic masses on. As it
turns out, this can be accomplished with the use of the basis
one-forms $\eta^M$ which are dual to the group generators.

At this point we need to define the proper ansatz for the
dimensional reduction with the group of isometries as
defined by eqs.~\reef{liekill} and \reef{formalg}.
This is a subtle issue due to the large number of
degrees of freedom and the interplay of different symmetries
present in the model. The ansatz for the reduction must correctly
convert the original symmetries of the compactification manifold
into the reduced gauge symmetries, while simultaneously ensuring
the equivalence of the dynamics. In other words, the ansatz must be
consistent with the ten-dimensional equations of motion in that the variations
of the reduced action produce the same set of equations.
We begin by considering first the reduction ansatz
for the metric.
The choice of $Z_M$'s as the symmetry generators
implies that the consistent ansatz for the nonabelian
Kaluza-Klein reduction where the symmetry group is defined by
eqs.~(\ref{liekill}) and \reef{formalg}
is found by expressing the metric on the internal manifold as
a bilinear combination of the basis one-forms $\eta^M$. When the
group is gauged, and the forms $\eta^M$ transform according to
(\ref{actgroup}), we must introduce the Kaluza-Klein gauge fields
$V^M{}_\mu$ to compensate the $d\omega^M$ terms in
(\ref{actgroup}). It is not hard to see that this implies the
following ansatz for the reduction of the metric:
\be
ds^2 = g_{\mu\nu}(x) dx^\mu dx^\nu + {\cal G}_{MN} (x) (\eta^M(y) +
V^M{}_\mu(x) dx^\mu) (\eta^N(y) + V^N{}_\nu(x) dx^\nu)
\label{nonabkkmet}
\ee
The matrix ${\cal G}_{MN}(x)$ contains the desired axionic degrees of freedom,
which can be seen as follows: First, the transformation rule for
$\eta^M$ in (\ref{actgroup}) requires that the vector field
$V^M{}_\mu$ transforms according to
\be
{V'}^M{}_\mu = {\cal S}^M{}_N V^N{}_\mu + {\cal O}^M{}_N
\partial_\mu \omega^N
\label{actgrvec}
\ee
and that the matrix ${\cal G}_{MN}$ transforms according to
\be
{\cal G'}_{MN} = {\cal S}_M{}^P {\cal S}_{N}{}^Q {\cal G}_{PQ}
\label{actgrmet}
\ee
where ${\cal S}_M{}^N$ is the inverse matrix of ${\cal S}^M{}_N$. If
we decompose the matrix ${\cal G}_{MN}$ in terms of the $d$-bein
${\cal E}^A{}_M$, defined by ${\cal G}_{MN} = \delta_{AB} {\cal
E}^A{}_M {\cal E}^B{}_N$, ${\cal E}^A{}_N {\cal E}_B{}^N = \delta^A_B$
and ${\cal E}^A{}_N {\cal E}_A{}^M = \delta^M_N$, we can see that the
$d$-bein transform according to
\be
{\cal E'}^A{}_N = {\cal S}_N{}^P {\cal E}^A{}_P
\label{actgrdbeinexact}
\ee
where ${\cal S}_N{}^P$ is the inverse of the matrix (\ref{odef}). In the
infinitesimal form, to linear order in $\omega^P$, this equation
becomes
\be
{\cal E'}^A{}_N = {\cal E}^A{}_N + 2 {\cal E}^A{}_P\, \gamma^P_{NM}
\,\omega^M
\label{actgrdbein}
\ee
It identifies the $d$-bein as an array of scalars belonging
to both the singlet and the adjoint (with charge $-1$)
representations of the isometry group.
As we alluded to at the beginning of this section,
this is precisely the transformation rule needed for
extracting the axionic degrees of freedom. They can be explicitly
found by carrying out the Gauss decomposition of the matrix ${\cal
E}^M{}_N$ and identifying the pivots with the dilatonic degrees of
freedom and the upper triangular matrix elements with the axions.
For example on $T^3$, one has
\ba
{\cal E}^A{}_M&=&\pmatrix{
e^{\phi_1}&e^{\phi_1} {\cal A}_0^{(12)}(y) &e^{\phi_1}
{\cal A}_0^{(13)}(y)\cr
 0&e^{\phi_2}&e^{\phi_2}{\cal A}_0^{(23)}(y)\cr
0& 0&e^{\phi_3} \cr}
\nonumber\\
&=& \pmatrix{e^{\phi_1}& 0 & 0 \cr
0&e^{\phi_2}&0\cr 0&0& e^{\phi_3} \cr}
\pmatrix{1&{\cal A}_0^{(12)}(y)&{\cal A}_0^{(13)}(y)\cr
0&1&{\cal A}_0^{(23)}(y)\cr 0&0&1\cr}
\label{zwei3}
\ea
with the dilatons $\phi_i$ and the axions ${\cal A}_0^{(ij)}$.
We will proceed with the dimensional reduction on the metric
(\ref{nonabkkmet}) in the next section. Here we merely pause to
note that the tangent space basis forms can be expressed in terms
of $e^\alpha = e^\alpha{}_\mu dx^\mu$ and
\be
{\cal E}^A = {\cal E}^A{}_N (\eta^N + V^N{}_\mu dx^\mu)
\label{tangtorbas}
\ee
which are invariant under the isometry group, as can be seen from
combining (\ref{actgroup}), (\ref{actgrvec}) and
(\ref{actgrdbein}).

We now sketch the Ans\"atze for the simultaneous reduction of the
vector fields ${\cal A}^I$ and the two-form potential ${\cal B}$ consistent
with the above metric Ans\"atze ---
see Appendices \ref{cartan} and \ref{three} for the complete details. It
is clear that the vector fields must be expressed as
\be
{\cal A}^I = {\cal A}^I{}_\mu(x) dx^\mu + {\cal A}^I{}_M(x) \eta^M
+ \sigma^I(y)
\label{vecnonabkk}
\ee
where the forms $\sigma^I$ are to be determined shortly. The reason
for this form of the ansatz is that it ensures the correct
transformation rules for ${\cal A}^I{}_M(x)$ after the reduction,
so that it retains its role as the axion of the reduced theory. If
we had attempted to reduce the gauge field ${\cal A}^I$ such that
this term were replaced by $\bar{\cal A}^I{}_M \zeta^M$, where
$\zeta^M$ were the invariant one-forms defined in
(\ref{invtforms}), the reduced quantity $\bar {\cal A}^I{}_M$ would
have been gauge singlets, and hence would not have transformed as
given in (\ref{kkgauge}) in the limit $\gamma^M_{LN} = 0$.
Eq.~\reef{vecnonabkk} is the only possibility for reduction which could
reduce to the correct limit as defined in (\ref{kkgauge}).
To determine $\sigma^I$, we turn to the gauge
field strength ${\cal F}^I = d {\cal A}^I$. Direct evaluation gives
\be
{\cal F}^I = \partial_{[\mu} {\cal A}^I{}_{\nu]} dx^\mu \wedge
dx^\nu + \partial_\mu {\cal A}^I{}_M \eta^M - {\cal A}^I{}_M
\gamma^M_{NP} \eta^N \wedge \eta^P + d\sigma^I
\label{agfnakk}
\ee
However by requiring the covariance of the reduced theory we must
have
\be
{\cal F}^I = \partial_{[\mu} {\cal A}^I{}_{\nu]} dx^\mu \wedge
dx^\nu + \partial_\mu {\cal A}^I{}_M \eta^M - (m^I_{NP} + {\cal
A}^I{}_M \gamma^M_{NP})\eta^N \wedge \eta^P
\label{agfnakka}
\ee
which formally coincides with (\ref{cymred}) in the limit
$\gamma^M_{NP}=0$. Note that this expression must be valid in
general, for any internal isometry group.
Here $m^I_{NP}$ are constants which are
antisymmetric in the lower two indices, and in the limit
$\gamma^M_{NP}=0$ they must become identical with the axionic
Yang-Mills masses discussed in the previous section. In
terms of the $m^I_{MN}$, we find that
\be
d \sigma^I = - m^I_{MN} \eta^M \wedge \eta^N
\label{integr}
\ee
The integrability condition for (\ref{integr}) gives
$m^I_{Q[M}\gamma^Q_{NP]} \eta^M \wedge \eta^N\wedge \eta^P = 0$, which
leads to
\be
m^I_{MQ} \gamma^Q_{NP} + m^I_{NQ}
\gamma^Q_{PM} + m^I_{PQ} \gamma^Q_{MN} = 0
\label{mgamma}
\ee
which, as we will see, will turn out to be exactly one of the
Jacobi identities for the structure constants of the reduced
theory. Comparing (\ref{vecnonabkk})
with the ansatz for $\hat {\cal A}^I(x,y)$ in (\ref{gauaxio}) shows that
they coincide when $\gamma^M_{NP}=0$, while the constraint
(\ref{mgamma}) disappears. Hence (\ref{vecnonabkk})
and the solution of (\ref{integr})
together comprise the reduction ansatz for the
Yang-Mills gauge fields.

The last remaining ingredient for the reduction ansatz is the
two-form potential ${\cal B}$. We can always write ${\cal B}$ in the
form given in (\ref{dans}). We have to assign the $y^M$ dependence
to the components of ${\cal B}$ in such a way as to ensure that the
three-form field strength ${\cal H}$ is independent of $y^M$ if
expanded in the $dx^\mu, \eta^M$, or equivalently, $e^\alpha, {\cal
E}^M$ basis. The former basis is much more convenient for
calculational simplicity, and we use it here. In this basis,
\ba
{\cal H} &=& \frac16 {\cal H}_{\mu\nu\lambda} dx^\mu \wedge dx^\nu
\wedge dx^\lambda +
\frac12 {\cal H}_{\mu\nu M}
dx^\mu \wedge dx^\nu \wedge \eta^M \nonumber \\ &&+ \frac12 {\cal
H}_{\mu MN} dx^\mu \wedge \eta^M \wedge \eta^N +
\frac16 {\cal H}_{MNP} \eta^M \wedge \eta^N \wedge \eta^P
\label{hred1}
\ea
On the other hand, ${\cal H} = d{\cal B} - \frac12
{\cal A}^I \wedge {\cal F}^I$. We can evaluate
the Chern-Simons contribution, using (\ref{vecnonabkk}) and
(\ref{agfnakka}), to find
\ba
{\cal A}^I \wedge {\cal F}^I &=& \frac12
{\cal A}^I{}_\mu {\cal F}^I{}_{\nu\lambda}
dx^\mu \wedge dx^\nu \wedge dx^\lambda \nonumber \\ &&+
\Bigl( \frac12 {\cal F}^I{}_{\mu\nu} ({\cal A}^I{}_M +
\sigma^I{}_M)) + {\cal A}^I{}_\mu \partial_\nu {\cal A}^I{}_M
\Bigr) dx^\mu \wedge dx^\nu \wedge \eta^M \nonumber \\ &&+
\Bigl( \partial_\mu {\cal A}^I{}_M ({\cal A}^I{}_N + \sigma^I{}_N)
- {\cal A}^I{}_\mu (m^I_{MN}
+ {\cal A}^I{}_P \gamma^P_{MN}) \Bigr) dx^\mu \wedge \eta^M \wedge
\eta^N \nonumber \\ &&-  ({\cal A}^I{}_M + \sigma^I{}_M)
(m^I_{NP} + {\cal A}^I{}_Q \gamma^Q_{NP})
\eta^M \wedge \eta^N \wedge \eta^P
\label{csexpansion}
\ea
Since this expression contains the terms which depend explicitly on
$y^M$'s, we must choose the ansatz for ${\cal B}$ such that this
dependence cancels, and the components of ${\cal H}$ in the
$dx^\mu, \eta^M$ basis depend explicitly only on $x^\mu$.
With this ansatz, the field strength becomes
\ba
{\cal H} &=& \Bigl( \frac12 \partial_{[\mu} {\cal B}_{\nu\lambda]}
- \frac14  {\cal A}^I{}_\mu {\cal F}^I{}_{\nu\lambda} \Bigr)
dx^\mu \wedge dx^\nu \wedge dx^\lambda
\nonumber \\
&&+
\Bigl(\partial_{[\mu}{\cal B}_{\nu]M} - \frac12  ( {\cal
F}^I{}_{\mu\nu} {\cal A}^I{}_M + {\cal A}^I{}_\mu \partial_\nu
{\cal A}^I{}_M) \Bigr) dx^\mu \wedge dx^\nu \wedge \eta^M
\nonumber\\
&&+\frac12 \Bigl(\partial_\mu {\cal B}_{MN} - {\cal B}_{\mu P}
\gamma^P_{MN} + \frac12 ({\cal A}^I{}_M \partial_\mu {\cal
A}^I{}_N + {\cal A}^I{}_\mu (2 m^I_{MN} + {\cal A}^I{}_P
\gamma^P_{MN})) \Bigr) dx^\mu \wedge \eta^M \wedge \eta^N
\nonumber\\
&& + \frac12  {\cal A}^I{}_M (2 m^I_{NP} + {\cal
A}^I{}_Q\gamma^Q_{NP})
\eta^M \wedge \eta^N \wedge \eta^P + \frac12 \beta_{MNP}
\eta^M \wedge \eta^N \wedge \eta^P
\label{hred2}
\ea
The constants $\beta_{MNP}$ are the axionic masses for the two-form, which
become identical to their corresponding parameters discussed in the
previous section when $\gamma^M_{NP} = 0$. Here we also have an
additional constraint. The integrability condition for the field
strength, which follows from its definition, is $d{\cal H} = -
\frac12  {\cal F}^I \wedge {\cal F}^I$. Substituting in
this equation (\ref{hred2}) and (\ref{agfnakka}), we find the
following expression relating $\beta_{MNP}$, $\gamma^M_{NP}$, and
$m^I_{MN}$:
\be
3 \beta_{L[MN} \gamma^L_{PQ]} =  m^I_{[MN} m^I_{PQ]}
\label{mbetagamma}
\ee
This relation represents the last of the Jacobi identities for the
structure constants of the gauge group of the reduced theory
$\gamma^M_{NP}$, $m^I_{MN}$ and $\beta_{MNP}$. Obviously, when
$\gamma^M_{NP} = 0$ this equation reduces to (\ref{mconstr}), as it
should. Note however that the general form of (\ref{hred2})
must remain completely unchanged regardless of the
explicit ansatz for reduction,
which contains information about the symmetry algebra.
This is a consequence of symmetry. The only effect
of a more complicated realization of internal symmetry is to increase
the number of independent structure constants,
but without changing the form of (\ref{hred2}).

Hence the consistent reduction Ans\"atze for our most general massive
supergravity which can be derived from any Scherk-Schwarz
reduction is given by equations (\ref{nonabkkmet}) and
(\ref{vecnonabkk}), plus the appropriate reduction
ansatz for the ${\cal B}$ field yielding \reef{hred2}.
Note that we have not needed to present the explicit form of
the internal coordinate dependence appearing in the basis forms $\eta^M$.
This dependence is implicitly fixed by the symmetry algebra given in
\reef{liekill} and \reef{formalg}. In the simplest examples, this dependence
may be simply polynomial, but more complicated transcendental functions
can also arise.
We will complete the reduction of the action to the form which is
manifestly invariant under both gauge and U-duality transformations
in the next section. As we
have discussed in the previous two sections, the gauge symmetries
of the reduced theory come in four flavors: the Yang-Mills
gauge symmetries which already present in ten-dimensions; the Kaluza-Klein
gauge symmetries which arise from diffeomorphisms on the internal space;
the Kalb-Ramond gauge symmetries, which are similarly related to
gauge transformations of the two-form on the internal space; and the
remaining two-form $U(1)$ gauge symmetry, which however
remains decoupled from the one-form gauge symmetries by Lorentz
invariance.

\section{General Massive Reductions} \label{general}

At this point, we are ready to carry out the reduction of the action
(\ref{sact1}) from $10$ to $D$ dimensions,
using the Ans\"atze (\ref{nonabkkmet}), (\ref{vecnonabkk})
and \reef{hred2},
with the set of mass parameters $\gamma^M_{NP}$, $m^I_{MN}$ and
$\beta_{MNP}$. We will not give the detailed computation here,
which can be found in the appendix, but merely quote the results.
First, it is convenient to split the action into three sectors: the
metric-dilaton, Yang-Mills, and Kalb-Ramond three-from, and we discuss
each of them separately.

We begin with the metric-dilaton sector, which is given by
\be
S_{g\phi} = \int d^{10}x \sqrt{-{\cal G}} e^{-\Phi}
\Bigl\{ {\cal R}({\cal G}) + (\nabla \Phi)^2 \Bigr\}
\label{metdil}
\ee
Using the ansatz (\ref{nonabkkmet}) and the splitting of the
zehnbein in terms of the $D$-bein $e^\alpha$ and $d$-bein ${\cal
E}^A$ defined in (\ref{tangtorbas}), we can expand the ten-dimensional Ricci
scalar and dilaton in terms of fields in the $D$-dimensional space-time.
The action (\ref{metdil}) becomes
\ba
S_{g\phi} &=& \int d^D x \sqrt{-g} e^{-\phi} \Bigl\{R + (\nabla
\phi)^2 + \frac14 {\cal D}_\mu {\cal G}_{MN} {\cal D}^\mu {\cal
G}^{MN} - \frac14 {\cal G}_{MN} V^M{}_{\mu\nu} V^{N\mu\nu}
\nonumber \\
&& ~~~~~~~~~~~~
- {\cal G}_{MN}
{\cal G}^{PQ} {\cal G}^{RS} \gamma^M_{PR} \gamma^N_{QS}
-2 {\cal G}^{MN} \gamma^P_{MQ} \gamma^Q_{NP} \Bigr\}
\label{redmetdil}
\ea
where we have used the following definitions:
\ba
\phi &=& \Phi - \ln(\sqrt{\det({\cal G})}) \nonumber \\
{\cal D}_\mu {\cal G}_{MN} &=& \partial_\mu {\cal G}_{MN}
- 2 {\cal G}_{MP} \gamma^P_{NQ} V^Q{}_\mu - 2 {\cal G}_{NP} \gamma^P_{MQ}
V^Q{}_\mu
\nonumber \\
V^M{}_{\mu\nu} &=& \partial_\mu V^M{}_\nu - \partial_\nu V^M{}_\mu
- 2 \gamma^M_{NP} V^N{}_\mu V^P{}_\nu
\label{reddefs1}
\ea
and where the covariant derivative of the moduli ${\cal G}_{MN}$
emerges because of the axionic degrees of freedom
contained in the matrix ${\cal G}_{MN}$:
we have ${\cal G}_{MN} = \delta_{AB} {\cal E}^A{}_M {\cal
E}^B{}_N$ and the transformation rule (\ref{actgrdbein}), which
induces (\ref{actgrmet}). Hence the local derivatives of ${\cal
G}_{MN}$ must be defined covariantly, since it contains a symmetric
bilinear of adjoint fields with charge $1$ with respect to the
nonabelian Kaluza-Klein group. It can be easily verified that the
determinant of ${\cal G}_{MN}$ does not contain any axionic fields,
however, and so is a gauge singlet. That it why we can still shift
the ten-dimensional dilaton in the usual way in (\ref{reddefs1}) to get the
$D$-dimensional dilaton.

We can now reduce the Yang-Mills sector. The ten-dimensional action is
\be
S_{CYM} = - \frac14  \int d^{10}x \sqrt{-{\cal G}} e^{-\Phi}
{\cal F}^I{}_{\mu\nu} {\cal F}^{I~\mu\nu}
\label{carym}
\ee
Using (\ref{nonabkkmet}) and (\ref{vecnonabkk})
as well as the definition $A^I{}_\mu = {\cal A}^I{}_\mu - {\cal
A}^I{}_M V^M{}_\mu$, which is the same as in (\ref{ansatz}), we
arrive at
\ba
S_{CYM} &=& -  \int d^Dx \sqrt{-g} e^{-\phi} \Bigl\{
\frac14 (F^I{}_{\mu\nu} + {\cal A}^I{}_M V^M{}_{\mu\nu})
(F^{I~\mu\nu} + {\cal A}^I{}_M V^{A~\mu\nu}) \nonumber \\ &&
~~~~~~~~~~~~~~~~ + \frac12 {\cal G}^{MN} {\cal D}_\mu {\cal
A}^I{}_M {\cal D}^\mu {\cal A}^I{}_N
\nonumber \\
&& ~~~~~~~~~~~~~~~~ + {\cal G}^{MP} {\cal G}^{NQ} (m^I_{MN} + {\cal
A}^I{}_R \gamma^R_{MN}) (m^I_{PQ} + {\cal A}^I{}_S \gamma^S_{PQ})
\Bigr\}
\label{redcym}
\ea
where we use
\ba
{\cal D}_\mu {\cal A}^I{}_M &=& \partial_\mu {\cal A}^I{}_M - 2
(m^I_{MN} + {\cal A}^I{}_P \gamma^P_{MN}) V^N{}_\mu \nonumber \\
F^I{}_{\mu\nu} &=& \partial_\mu A^I{}_\nu - \partial_\nu A^I{}_\mu
- 2 m^I_{MN} V^M{}_\mu V^N{}_\nu
\label{yukc}
\ea
which again follow straightforwardly by dimensional reduction. More
details can be found in appendix \ref{cartan}

The last contribution to the action comes from the three-form kinetic
terms in ten dimensions
\be
S_{NS} = - \frac{1}{12} \int d^{10}x \sqrt{-{\cal G}} e^{-\Phi}
{\cal H}_{\mu\nu\lambda} {\cal H}^{\mu\nu\lambda}
\label{nsact}
\ee
The reduction of this action, using once again the Ans\"atze
(\ref{nonabkkmet}), (\ref{vecnonabkk}), \reef{hred2}
and the field redefinition (\ref{ansatz}),
produces the following action in $D$ dimensions:
\ba
S_{NS} &=& - \int d^D x \sqrt{-g} e^{-\phi} \Bigl\{ \frac{1}{12}
H_{\mu\nu\lambda} H^{\mu\nu\lambda} \nonumber \\ &&+ \frac14 {\cal
G}^{MN} (H_{\mu\nu M} -  {\cal A}^I{}_M F^I{}_{\mu\nu}
- {\cal C}_{MP} V^P{}_{\mu\nu})
(H^{\mu\nu}{}_N -  {\cal A}^I{}_N F^{I~\mu\nu}
- {\cal C}_{NQ} V^{Q\mu\nu})  \nonumber \\
&&+ \frac14 {\cal G}^{MP} {\cal G}^{NQ} ({\cal D}_\mu \B_{MN} +
 {\cal A}^I{}_{[M} {\cal D}_\mu {\cal A}^I{}_{N]}) ({\cal
D}^\mu \B_{PQ} +  {\cal A}^J{}_{[P} {\cal D}^\mu {\cal
A}^J{}_{Q]})
\nonumber \\
&&+ \frac34 {\cal G}^{MQ} {\cal G}^{NR} {\cal G}^{PS} (\beta_{MNP}
+ 2  {\cal A}^I{}_{[M} m^I_{NP]}
- 2 {\cal C}_{T[M} \gamma^G_{NP]}) \nonumber \\
&&~~~~~~~~~~~~~~~~~~\quad \times (\beta_{QRS} + 2  {\cal
A}^J{}_{[Q} m^J_{RS]}- 2 {\cal C}_{U[Q} \gamma^U_{RS]}) \Bigr\}
\label{nsredact}
\ea
The new definitions here are
\ba
{\cal D}_\mu \B_{MN} &=& \partial_\mu \B_{MN} + 2  m^I_{MN}
A^I{}_\mu + 2 \gamma^P_{MN} B_{\mu P} \nonumber \\ &&
- \beta_{MNP} V^P{}_\mu + 4 \B_{Q[M} \gamma^Q_{N]P} V^P{}_\mu
-2  {\cal A}^I{}_{[M} m^I_{N]P} V^P{}_\mu \nonumber \\
H_{\mu\nu M} &=& \partial_\mu B_{\nu M} - \partial_\nu B_{\mu M} +
3\beta_{MNP} V^N{}_\mu V^P{}_\nu \nonumber \\ &&+ 4 \gamma^P_{MN}
B_{[\mu P} V^N{}_{\nu]} + 4  m^I_{MN} A^I{}_{[\mu}
V^N{}_{\nu]}
\label{covdefs}
\ea
and the reduced three-form field strength is
\ba
H_{\mu\nu\lambda} &=& \partial_\mu B_{\nu\lambda}
- \frac12  A^I{}_\mu
F^I{}_{\nu\lambda}
- \frac 12 V^M{}_\mu H_{\nu\lambda M}
- \frac 12 B_{\mu M} V^M{}_{\nu\lambda}
+\frac12 \beta_{MNP} V^M{}_\mu V^N{}_\nu V^P{}_\lambda
\nonumber \\
&&
-  m^I_{MN} A^I{}_\mu
V^M{}_\nu V^N{}_\lambda - \gamma^M_{NP} B_{\mu M} V^N{}_\nu
V^P{}_\lambda + ~cyclic ~ perm. ~
\label{rednsnsthree}
\ea
where $B_{\mu M}$ and $B_{\mu\nu}$ are defined in (\ref{ansatz}),
and still are the correct quantities to express the reduced action,
in a manifestly gauge and U-duality symmetric way.

To reassemble the reduced terms (\ref{redmetdil}), (\ref{redcym})
and (\ref{nsredact}) into a manifestly symmetric action in $D$
dimensions, we first need to establish the correct gauge algebra of
the reduced theory, and identify the gauge invariant couplings of
fields. Proceeding as before, we first give the infinitesimal
reduced gauge transformations. Details of their derivation can be
found in appendix \ref{algebraaa}. As before, the small gauge transformations
fall into three categories, listed here in the order of ascending
complexity. The fields not listed below explicitly are invariant
under the corresponding transformations. The reduced gauge transformations
are now:

1) Kalb-Ramond gauge transformations:
\ba
&&\B'_{MN} = \B_{MN} - 2 \lambda_P \gamma^P_{MN} \nonumber \\
&&B'_{\mu M} = B_{\mu M} + \partial_\mu \lambda_M - 2
\lambda_P\gamma^P_{MN} V^N{}_\mu \nonumber \\
&&B'_{\mu\nu} = B_{\mu\nu} + \frac12 \lambda_M V^M{}_{\mu\nu} +
\gamma^M_{NP} \lambda_M V^N{}_\mu V^P{}_\nu
\label{nsnsgauge1}
\ea

2) Yang-Mills gauge transformations:
\begin{eqnarray}
&&\B'_{MN} = \B_{MN} - 2  \lambda^I m^I_{MN} \nonumber \\ &&
{A'}^I{}_\mu = A^I{}_\mu + \partial_\mu \lambda^I \nonumber \\ &&
B'_{\mu M} = B_{\mu M}
- 2  \lambda^I m^I_{MN} V^N{}_\mu
\nonumber \\
&& B'_{\mu\nu} = B_{\mu\nu} + \frac12
 \lambda^I F^I{}_{\mu\nu}
+  m^I_{MN} \lambda^I V^M{}_\mu V^N{}_\nu
\label{ymcgauge1}
\end{eqnarray}
(\ie unchanged from the case of flat torus discussed in section
\ref{massone}), and

3) Kaluza-Klein gauge transformations:
\begin{eqnarray}
&&{\cal A}'^I{}_M = {\cal A}^I{}_M +2 \gamma^N_{MP} \omega^P {\cal
A}^I{}_N + 2 m^I_{MN} \omega^N \nonumber \\ &&\B'_{MN} = \B_{MN} +
3\beta_{MNP} \omega^P + 2
{\cal A}^I{}_{[M} m^I_{N]P} \omega^P + O(\omega^2)
\nonumber \\
&& {\cal G'}_{MN} = {\cal G}_{MN} + 2 \gamma^P_{MQ} \omega^Q {\cal G}_{PN}
+ 2 \gamma^P_{NQ} \omega^Q {\cal G}_{MP} + O(\omega^2)
\nonumber \\
&& V'^M{}_\mu = V^M{}_\mu -2 \gamma^M_{NP} \omega^P V^N{}_\mu +
\partial_\mu
\omega^M + O(\omega^2) \nonumber \\
&& {A'}^I{}_\mu = A^I{}_\mu -2 m^I_{MN} \omega^N V^M{}_\mu +
O(\omega^2) \nonumber \\ &&B'_{\mu M} = B_{\mu M} + 2 \gamma^N_{MP}
\omega^P B_{\mu N} + 2  m^I_{MN} \omega^N A^I{}_\mu +
3\beta_{MNP} \omega^P V^N{}_\mu + O(\omega^2)
\nonumber \\
&&B'_{\mu\nu} = B_{\mu\nu} + \frac12
\omega^M H_{\mu\nu M} - \frac32 \beta_{MNP} \omega^M V^N{}_\mu V^P{}_\nu
\nonumber \\
&& ~~~~~~~~~~~~~
- 2 \gamma^P_{MN} \omega^M B_{[\mu P} V^N{}_{\nu]} -2  \omega^M
m^I_{MN} A^I{}_{[\mu} V^N{}_{\nu]} + O(\omega^2)
\label{kkgauge1}
\end{eqnarray}
Note that in the last set of gauge transformations,
we have nontrivial transformation rules for the moduli
${\cal G}_{MN}$. This arises from the nontrivial couplings of
the metric axions, which were absent in the section \ref{massone} where
the metric mass parameters $\gamma^{M}_{NP}$ were set to zero.

Now as in eq.~\reef{parmg}, we define a combined gauge parameter
\be
\hat{\omega}^a=(\omega^M,\lambda_M,\lambda^I)
\label{parmg1}
\ee
and generators: $T_a = (Z_M,
X^M,Y^I)$. The algebra of the latter
\be
[T_a, T_b] = i f_{ab}{}^c T_c
\ee
defines the new set of structure constants, $f^{ab}{}_c$.
 To compute these, we again consider the products of
transformations (\ref{nsnsgauge1}--\ref{kkgauge1}) of the form
$h^{-1} \cdot g^{-1} \cdot h \cdot g$ where $h$ and $g$ are two of
gauge transformations with $g = \exp(i
\hat \omega_1^a T_a)$ and $h = \exp(i \hat \omega_2^a T_a)$.
Hence
substituting the explicit form of the gauge transformations
(\ref{nsnsgauge1}--\ref{kkgauge1}), we can deduce the structure
constants. We evaluate $h^{-1} \cdot g^{-1}
\cdot h \cdot g | \Psi \rangle$ for the set of basis states defined
by the vector fields. The structure constants are, as expected,
\begin{equation}
f^M{}_{NP} =f_{NP}{}^M= 2 \gamma^M_{NP}
~~~~~~~~~~ f^I{}_{MN} =f_{MN}{}^I= 2 m^I_{MN}
~~~~~~~~~~ f_{MNP} = -3 \beta_{MNP}
\label{strc1}
\end{equation}
The resulting gauge algebra, which encompasses the case given in
(\ref{algebra}), is
\ba
&&[X^M, X^N] = [Y^I, Y^J] = [X^M, Y^I] = 0 \nonumber \\ &&[X^M,
Z_N] = 2 i \gamma^M_{NP} X^P ~~~~~~~~~~ [Y^I, Z_M] = 2i m^I_{MN}
X^N \nonumber \\ &&[Z_M, Z_N] = -3i \beta_{MNP} X^P + 2i
m^I_{MN} Y^I + 2i \gamma^P_{MN} Z_P
\label{algebra1}
\ea
While the standard Cartan metric on this Lie algebra
(\ref{algebra1}) is still degenerate, since the gauge algebra is
not semi-simple, we can nevertheless define the metric on the gauge
algebra by $\langle T_a, T_b\rangle = L_{ab}$, exactly as before.

Formally keeping the definition of the Lie-algebra-valued gauge
field one-form potential given in eq.~(\ref{gfp}), we find that the
Lie-algebra-valued gauge field strength
\begin{equation}
F = d A + i A \wedge A = \frac12 F^a_{\mu\nu} T_a dx^\mu \wedge
dx^\nu
\label{gfla1}
\end{equation}
again has components which coincide with the expressions for field
strengths that come from dimensional reduction:
$F^a_{\mu\nu}=(V^M{}_{\mu\nu},H_{\mu\nu M},F^I_{\mu\nu})$.
Explicitly, the components of the gauge field strength are
\begin{eqnarray}
F_{\mu\nu}{}^M &=& \partial_\mu V^M{}_{\nu} - \partial_\nu
V^M{}_\mu - 2 \gamma^M_{NP} V^N{}_\mu V^P{}_\nu
\nonumber\\
F_{\mu\nu M} &=&
\partial_\mu B_{\nu M} - \partial_\nu B_{\mu M}
+ 3\beta_{MNP} V^N{}_\mu V^P{}_\nu + 4  m^I_{MN}
A^I{}_{[\mu} V^N{}_{\nu]}
+ 4 \gamma^P_{MN} B_{[\mu P} V^N{}_{\nu]}
\nonumber\\
F_{\mu\nu}{}^I &=& \partial_\mu A^I{}_{\nu} -
\partial_\nu A^I{}_\mu
- 2 m^I_{MN} V^M{}_\mu V^N{}_\nu
\label{gfcomp1}
\end{eqnarray}

The nonabelian Chern-Simons form can be computed as usual, to yield
\begin{eqnarray}
\omega_{CS} &=& \langle A \wedge F - \frac{i}3 A \wedge A \wedge A\rangle
\label{csfin1}\\
&=&\frac13 dx^\mu \wedge dx^\nu \wedge dx^\lambda
\Bigl( \frac12  A^I{}_\mu F^I{}_{\nu\lambda} +
\frac12 V^M{}_\mu H_{\nu\lambda M} +
\frac12 B_{\mu M} V^M{}_{\nu\lambda} \nonumber \\
&+&\frac12 \beta_{MNP} V^M{}_\mu V^N{}_\nu V^P{}_\lambda -
 m^I_{MN} A^I{}_\mu V^M{}_\nu V^N{}_\lambda
- \gamma^M_{NP} B_{\mu M} V^N{}_\nu V^P{}_\nu + c.p.
\hfill\Bigr)
\nonumber
\end{eqnarray}
This is exactly the anomaly contribution to the reduced three-form
field strength. Again the Chern-Simons terms get twisted together
into a single nonabelian structure, such that, in form notation,
the reduced three-form field strength is simply
\be
H = dB - \frac12 \omega_{CS}
\ee

We can also put the moduli potential in
a duality symmetric form. Lowering the last of  index on the
structure constants,
$f_{abc} = f_{[ab}{}^dL_{d|c]}$, we can write the scalar
potential entirely in terms of the moduli matrix $M^{ab}$:
\begin{equation}
W(M)=
\frac{1}{12} M^{ad}
M^{be} M^{cf} f_{abc} f_{def}
- \frac{1}{4} M^{ad} L^{be}L^{cf} f_{abc}f_{def}\ .
\label{potcov1}
\end{equation}
Note the additional term linear in $M$ which was absent in the
formula (\ref{potcov}) obtained with nontrivial fluxes in the
matter sector. This is, of course, consistent with our previous
results. Using the structure constants found in eq.~\reef{strc},
\ie $\gamma^M_{NP}=0$, this linear term automatically vanishes.
This new term's only nontrivial contribution here is the last
interaction appearing in eq.~\reef{redmetdil}, which is linear
in $\G$ and quadratic in the $\gamma^M_{NP}$.

The mass parameters
$\gamma^M_{NP}$, $m^I_{MN}$ and $\beta_{MNP}$ must
satisfy the constraints
\ba
&&\gamma^M_{N[P} \gamma^N_{QR]} = 0 ~~~~~~~~~~ \gamma^M_{MN} = 0
\nonumber \\ && m^I_{Q[M} \gamma^Q_{NP]} = 0 \nonumber \\ && 3
\beta_{R[MN} \gamma^R_{PQ]} =  m^I_{[MN} m^I_{PQ]} \ .
\label{allconstr}
\ea
These relations will ensure that the structure constants
satisfy the Jacobi identity:
\be
L^{ad} f_{a[bc} f_{d|ef]} = \frac13 L^{ad} \Bigl(f_{abc} f_{def} +
f_{abe} f_{dfc} + f_{abf} f_{dce} \Bigr) = 0
\ee

It now remains to note that we can also collect the gauge kinetic terms
and the moduli terms in the manifestly gauge- and duality-invariant
fashion. The gauge fields transform according to
\begin{equation}
F'^a_{\mu\nu} = U^a{}_b F_{\mu\nu}^{b}
\label{gtoff1}
\end{equation}
where $U^a{}_b$ is an $O(d,d+16)$ matrix. Thus
the moduli fields transform according to
\begin{equation}
M'^{ab} = U^a{}_c U^b{}_d M^{cd}
\label{mgauge1}
\end{equation}
under the same gauge transformations, and
we see that the covariant derivative of the scalar moduli can be
written as
\begin{equation}
{\cal D}_\mu M^{ab} = \partial_\mu M^{ab} - f_{cd}{}^a A^c_{\mu }
M^{db} - f_{cd}{}^b A^c_{\mu} M^{ad}\ .
\label{covdermf1}
\end{equation}
These are indeed identical with the expressions which were obtained
by reduction. With
this it can be easily shown that the gauge kinetic terms can be
collected into the covariant expression
\begin{equation}
{\cal F}^2 = F^a_{\mu\nu} L_{ab} M^{bc} L_{cd}
F^{d\mu\nu}
\label{compkt1}
\end{equation}

Therefore, the reduced action can be again rewritten as
\begin{eqnarray}
S &=& \int d^{D} x \sqrt{-{g}} e^{-\phi} \Bigl\{{R}
 + ({\nabla} \phi)^2 + \frac{1}{8}
L_{ab} {\cal D}_\mu M^{bc} L_{cd} {\cal D}^\mu M^{da}
 \nonumber \\
&&~~~~~~~~~~~~~~~~~~~~~ - \frac{1}{4} F^a_{\mu\nu}
 L_{ab} M^{bc} L_{cd} F^{d\mu\nu}
- \frac{1}{12} {H}^2_{\mu\nu\lambda} - W(M) \Bigr\}
\label{actoddna1}
\end{eqnarray}
which is formally the same as (\ref{actoddna}), albeit the
structure of the gauge group is more complicated.
Hence as before, the general massive reductions produce reduced theories with
a remarkably symmetric form, where a part of the $O(d, d+16)$
duality must be gauged in order to accommodate the couplings induced
by the mass terms.

\section{Discussion}
\label{conc}

Above we have shown that, under generalized Scherk-Schwarz reductions
on a twisted $d$-dimensional torus,
low energy heterotic string theory is manifestly
invariant under $O(d,d+16)$. The essential step in the construction of the
reduced action
is to realize that the mass parameters can be assembled into a
completely antisymmetric three-index tensor $f_{abc}$ with
$O(d,d+16)$ indices. These parameters serve two roles in the reduced
theory: as the mass parameters defining the scalar potential
\reef{potcov1}, and also
as the structure constants in the {\it non-abelian}
gauge group. Thus the reduced theory is now a massive gauged
supergravity. By inspecting the reduced action (\ref{actoddna1}), we
discover that usual U-duality transformations,
$M \rightarrow M'=\Omega M \Omega^{T}$, $F\rightarrow F' = \Omega F$,
are now accompanied by
\be
f_{abc} \rightarrow f'_{abc} = \Omega_a{}^d \Omega_b{}^e
\Omega_c{}^e f_{def}
\label{strcontrb}
\ee
Given a certain set of mass parameters, U-duality transformations
will in general map these to a new collection of modified couplings.
In so doing,  U-duality maps one {\it reduced}
theory to another. Thus we can think that while U-duality remains
a symmetry of the full theory, it is spontaneously broken in any
given compactification. Our U-duality covariant formalism provides
a unified framework in which to investigate the
``disparate'' massive supergravities, appearing in generalized reductions of
heterotic string theory.

It is interesting to consider the symmetries identified here
in the framework of ref.~\cite{pseudo}. There, proper symmetries are
identified as those that act on fields, while pseudo-symmetries act
on both fields and coupling constants. In the present case, from
the point of view of the ten-dimensional heterotic string theory
(or its low energy supergravity limit), the $O(d,d+16)$ transformations
are proper symmetries. The key point is that the mass parameters
$f_{abc}$ in eq.~\reef{strcontrb} were identified with nontrivial
background fields in the internal space. However having compactified,
one could truncate the reduced theory to
the $D$-dimensional gauged supergravity for which eq.~\reef{actoddna1}
is the bosonic part of the action. From the point of view of this
truncated theory, the $O(d,d+16)$ duality group contains both
proper symmetries and pseudo-symmetries since the $f_{abc}$ are
now simply coupling constants. The proper symetries would be the
subgroup of transformations leaving a given set of couplings invariant
under the above action \reef{strcontrb}, while the pseudo-symmetries
act nontrivially. It is in this sense that we mean that in general
U-duality maps one {\it reduced} theory to another.

A novel feature in the present case is that the
background fields that produce this spontaneous breaking are topological.
That is the corresponding fluxes or geometric curvatures
on the internal
space are not associated with dynamical fields in the theory. One aspect
of the mass parameters, which we have not considered above, is the fact
that they must be quantized. The usual Dirac quantization conditions
arise for the Yang-Mills fluxes from the consideration of charged string
states moving in such a background. Similarly, the three-form fluxes
are shown to be quantized by considering strings with nontrivial winding
numbers \cite{three}. Finally for compactification on a twisted torus,
properly matching the geometry of the
$d$-torus will require that the metric twists lie in $SL(d,Z)$. Given
the quantized nature of the mass parameters, it would seem that the
U-duality symmetry is broken down to $O(d,d+16,Z)$. It is natural
to conjecture then that these discrete symmetries are in fact an
exact symmetry of the full string theory, just as in the case of the
standard toroidal compactifications \cite{exact}.

Here, we are lead to a slight puzzle. Quantization conditions seem to
break the $O(d,d+16,R)$ symmetry of the reduced action \reef{actoddna1}
to $O(d,d+16,Z)$. Another aspect of the reduced theory was that a part
of the global $O(d,d+16,R)$ symmetry becomes a local gauge symmetry.
Thus it would seem that this continuous subgroup of the U-duality
group must also be exact since it corresponds to a constant gauge
transformations. The puzzle is to understand the interplay of these
two apparent exact symmetries. We will argue that in fact these symmetries
are distinct symmetries, despite their apparent common origin
in $O(d,d+16,R)$.

This distinction is most easily understood by examining a concrete example.
The simplest case to consider is a generalized reduction on
$T^3$ with a metric twist. As described in section \ref{twist},
one can think of the resulting geometry as an $SL(2,Z)$
fibre bundle with a $T^2$ fibre over an $S^1$ base.
That is as one moves around the $S^1$, the $T^2$ geometry is
varying by a real shift of the modular parameter, $\tau\rightarrow
\tau+a$.
Having circumnavigated the circle once, in the simplest
case, the total shift is $a=1$, allowing the $T^2$ fibres to
be identified by an $SL(2,Z)$ transformation.
Now we note that in the reduced theory
an arbitrary shift of the metric axion, \ie the $T^2$ modulus $\tau$,
is a symmetry. In the ten-dimensional space,
while this shift changes the $T^2$ geometry
locally over a fixed point on the $S^1$, the geometry is
in fact invariant if the axion shift is accompanied
by a translation in the $S^1$ direction. That is all
possible $T^2$ geometries can be found in the fibre over some point in
the $S^1$, hence a shift of the $T^2$ modulus along with a
corresponding translation along the $S^1$ must be a symmetry.
Another point of view would be that in a standard
reduction, a constant translation along the $S^1$ corresponds
to a constant $U(1)$ symmetry transformation, which acts trivially
in the reduced theory. In the generalized reduction considered here, these
translations alone are no longer a symmetry because of the twisted
geometry. Instead, to produce a symmetry the translation is accompanied
by a shift of the metric axion. In any event, the global
gauge transformation which produces $\tau\rightarrow\tau+1$ is accompanied
by a translation once around the $S^1$. The U-duality symmetry
producing the same shift of $\tau$ makes no translation
in the $S^1$ direction. Hence these two $O(d,d+16)$ transformations
are actually distinct ten-dimensional symmetries.

The nontrivial interplay of the gauge and U-duality symmetries seems
to modify the form of the moduli space of these theories. For a standard
toroidal compactification of the heterotic string, the moduli space\cite{modi}
is $SO(d,d+16,Z)\backslash SO(d,d+16,R)/(SO(d,R)\times SO(d+16,R))$. For the
generalized axion compactifications considered here, it appears that the
moduli space is reduced to $(H\times SO(d,d+16,Z))\backslash
SO(d,d+16,R)/(SO(d,R)\times SO(d+16,R))$ where $H$ is the group of
global gauge transformations. One might have included the
global gauge transforms for the standard compactification,
however, their action is trivial in that case and so they have no effect.
Note that the additional reduction of modding out by $H$ affects even
the local structure of the moduli space.

Here we use the term ``moduli space" loosely, as typically the scalar
potential \reef{potcov1} pushes the scalar moduli to evolve under its
influence. For example, one might choose to set $M^{ab}=\delta^{ab}$ as
initial data, but this configuration typically would not extremize
the potential and thus the scalars would begin to vary.
One might observe that naively a simple solution for moduli that always
extremizes the potential is $M^{ab}\propto L^{ab}$. Unfortunately,
this is not a physical solution as it corresponds to a vanishing internal
metric. Instead
the natural vacua of such theories are therefore similar to the linear
dilaton vacua \cite{rlindil}. The natural supersymmetric vacua
are typically domain walls in the reduced theory \cite{cow,others}.
One is also led to consider cosmological evolutions in such theories
\cite{cosmo}.

Although it should be obvious, we also mention that the appearance
of a nonabelian gauge symmetry in the generalized reductions is
very different than the symmetry enhancement that arises at special
points in the moduli space of the standard toroidal reduction.
In the latter case, new massless degrees of freedom associated with
string winding modes appear at these special points to enlarge the
gauge group. In the generalized reductions, there are no extra
massless fields appearing in the reduced theory. Actually quite the
opposite effect should take place. If one examines the detailed form
of the covariant derivatives of the moduli in eqs.~\reef{yukc} and
\reef{covdefs}, one recognizes that these axion scalars are actually
Stuckelberg fields. The same is true in eq.~\reef{reddefs1} for the metric
axions if one considers fluctuations around a fixed background metric.
The Stuckelberg nature of the axions is also evident from the gauge
transformations in eqs.~(\ref{nsnsgauge1}--\ref{kkgauge1}). Of course,
this should have been anticipated given that it was the axionic shift
symmetry which was gauged in the generalized reduction. In any event,
the associated gauge fields will in fact become massive, and the
original axions are traded for massive longitudinal modes. So in
contrast with the usual enhanced symmetry points where
the number of massless modes increases, the generalized
reductions actually produce a reduction in the number of (strictly) massless
degrees of freedom as compared to the standard compactification.

It should be mentioned, however, that in ref.~\cite{GPR}
a unified framework was constructed to incorporate all of the
enhanced symmetry points of the standard toroidal reduction
to four dimensions.
The form of the ``completely duality-invariant low-energy effective
action'' constructed there is in fact very similar to our action
in eq.~\reef{actoddna1}. The similarity of these two actions for four
dimensions is essentially dictated by the fact that they are both versions
of  gauged $N=4$ supergravity in which the form of the action is completely
fixed given the gauge group \cite{more}. Undoubtedly, it is also true
that supersymmetry essentially fixes the elegant form of our bosonic
action \reef{actoddna1} in higher dimensions, as well.
In particular, the simple form of the moduli potential \reef{potcov1}
is probably a result of the large number of supersymmetries in the theory.

Given the above discussion, a natural question is under what
conditions do we expect the action \reef{actoddna1} to give a
reasonable description of the low energy degrees of freedom. In
our derivation, we are ignoring\footnote{More accurately, we
should think that these states are integrated out in constructing
the effective theory.} massive string states with masses
$m_s\simeq 1/\sqrt{\alpha'}=1/\ell_s$, Kaluza-Klein momentum modes
with $m_{KK} \simeq 1/L$ and string winding modes with
$m_{w}\simeq L/\ell_s^2$, where $L$ is the typical size of the
compact dimensions. Hence we should require that any of the masses
appearing in our generalized Scherk-Schwarz reductions are smaller
than these. Otherwise, the corresponding massive states should
also be integrated out, or alternatively we could consider the
full string theory including all of the massive states. Given the
flux and curvature quantization conditions discussed above, one
must be careful in considering this question.

Considering each type of the mass contribution individually, we
begin with the three-form fluxes. In this case \cite{three}, the
quantization condition takes the form: $\int_{3-cycle} H \simeq n\,\ell_s^2$
where $n$ is an integer. Therefore one finds that the corresponding
mass parameters must be $\beta_{MNP}\simeq \ell_s^2/L^3$. Thus if one
is considering a regime where $\ell_s/L<<1$, \ie the internal space is
large compared to the string scale, then these masses are much smaller than
those of the massive states listed above. Similarly, considering the
gauge field fluxes, we have $\int_{2-cycle} F^I \simeq n\,\ell_s$
and $m^I_{MN}\simeq \ell_s/L^2$. Thus in the regime $\ell_s/L<<1$,
these gauge flux masses are smaller than those of the states which have
been ignored, however, they are systematically larger than those generated
by the three-form fluxes. The latter though assumes that all of the compact
dimensions are essentially the same size.

Finally we consider the geometric masses for compactification on a
twisted torus. In this case, it is easiest to gain insight by considering
the case of $T^3$ for which some of the explicit formulae were presented
at the beginning of section \ref{twist}. It is straightforward to see
that in general $\gamma^M_{NP}\simeq L_M/(L_N L_P)$ where $L_M$ is the
compactification size associated with $y^M$. Therefore if all of the compact
dimensions are the same size, these masses will be of the same order as
that of the Kaluza-Klein modes, and so one is not really justified in
using the action \reef{actoddna1} as it stands. By arranging a less democratic
compactification, however, in which $L_M/L_N<<1$ for certain directions, one
can safely introduce certain $\gamma^M_{NP}$ while keeping only the
degrees of freedom appearing in \reef{actoddna1}.

The preceding analysis assumed that the sizes of the compact directions
were fixed. However, as discussed above, this will likely not be the
case due to the scalar potential \reef{potcov1}. Hence
in examining solutions of the equations of motion, one should
be aware that the scalars may evolve to (or from) a regime where
the low energy theory described by \reef{actoddna1} is no longer valid.

We have already noted that in section \ref{twist} the explicit form of
the internal coordinate dependence appearing in the basis forms $\eta^M$
is not explicitly needed. Rather, this dependence is implicitly fixed by the
internal symmetry algebra given in \reef{liekill} and \reef{formalg}.
The key for inducing the metric mass parameters was the
presence of a nonabelian isometry group on the internal manifold.
Thus our formalism includes more general
nonabelian Kaluza-Klein reductions \cite{schs,kkna,duffpope}
than the case of twisted tori. In general a
reduction must satisfy the consistency condition that the procedures of
dimensional reduction and variation of the action commute.
Further one  usually wants to construct a reduced theory with a finite number
of field degrees of freedom, which are generally in one-to-one correspondence
with the degrees of freedom of the abelian reductions on
internal tori. As it has been discussed
in particular in \cite{duffpope}, these two conditions are in fact
related. A sufficient condition for finite-dimensional truncations
is that the pullbacks of generators of the isometry algebra
on the tangent space span a complete, linearly independent,
orthogonal basis.
The end result is then that the permissible internal manifolds
should be homogeneous spaces with isometry algebra of dimension
equal to the dimension of the manifold on which it acts.

In lower dimensions, the relevant isometry algebras can be completely
classified \cite{ryanshepley}. In two dimensions, the only admissible Lie
algebras with two generators are the abelian one
and an algebra defined by $[Z_1, Z_2] = Q Z_2$, however the
only structure constant of the latter is $\gamma^2_{12} = Q$,
which violates eq.~\reef{congamma3} when $Q\ne 0$.
In three-dimensions, the algebras of interest are given
by five of the nine Bianchi model geometries \cite{ryanshepley}. They are
Bianchi I (abelian), the two algebras
corresponding to the twisted tori:
Bianchi II (discussed in \cite{kaloper}
and in section (\ref{twist}) and Bianchi VIII (a
twisted torus with a cubic ansatz in internal coordinates)
and two non-toroidal algebras:
a special Bianchi VI with the $h=-1$
(which must be chosen to satisfy eq.~\reef{congamma3}), and
Bianchi IX (corresponding to a group manifold of
$SO(3) \approx SU(2)/Z_2$ or its universal covering space
$SU(2) \approx S^3$, and with the $SU(2)$ algebra).
The latter two algebras give rise
to the reduced theories of the same form as
\reef{actoddna1}. The 3-sphere with antipodal identification
$SU(2)/Z_2$ has nontrivial $1$-cycles, and Wilson loops
could still be present, breaking the vector supermultiplet
gauge group to $U(1)^{16}$. Hence it
appears to be covered by our consideration, at least
in the supergravity limit. In fact reductions of
the $N=1 D=10$ supergravity on $S^3 \times S^3$
have been considered \cite{chams}, and in this case
because of the absence of the vector supermultiplet,
the global topology on the sphere did not play a role.
Nevertheless the complete understanding
of the case of $S^3$, and higher-dimensional simply
connected manifolds in heterotic string theory
would require a more detailed scrutiny of the
gauge sector. In addition, since the internal subalgebra is
semisimple, the sizes of the internal radii would have
to be equal, giving all the geometric masses $\gamma^M_{NP}$
of the order of $1/L$. Thus as indicated in the discussion
above, we can not use the action \reef{actoddna1} as it stands
to consistently study the dynamics of the low energy theory.
One must either extend it to include other massive
Kaluza-Klein states,
which have been integrated out in our construction, or
exclude the modes appearing in our action with masses of the
order of $1/L$. In any event,
this case therefore requires further scrutiny.

In closing, we underline that the Scherk-Schwarz reduction \cite{schs}
may be viewed as an application
of the nonabelian Kaluza-Klein
reduction \cite{kkna,duffpope} on a group which
admits finite-dimensional truncations
to supergravity theories. As discussed
by Scherk and Schwarz in their
work \cite{schs}, their main motivation
was seeking ways to break supersymmetry.
Actually, this approach to supersymmetry
breaking is really similar to supersymmetry
breaking in various $p$-brane configurations, with the only
additional requirement that the
$p$-brane solutions admit a simply transitive
isometry group in 
spatial directions.
Such solutions can be interpreted
as spontaneous compactification,
such as the Freund-Rubin
solutions \cite{fr} (which happen not to break any
supersymmetry, since they correspond to
spontaneous compactifications
to $AdS_4 \times S^7$ spacetime,
but their extensions \cite{englert} may do so).
Note that the original terminology of Scherk
and Schwarz \cite{schs} distinguished
between the so-called internal and
external axionic symmetries. The former
were associated with the axions that
emerge from the metric of the internal
manifold, while the latter come from
the form-field sector. However, in light
of our interpretation of the $O(d,d+16)$
dualities as the maps between
the internal and 
external masses,
it is clear that this distinction between
the internal and external symmetries is somewhat artificial
in string theory.

Moreover, the physically meaningful
properties of the reduced theory, given by the masses and the
structure constants, depend on the directions and types of fields
which are excited on the internal space.
In general, the internal fields can be turned
on by using the
tensor representations of isometries.
In this context, a very natural question to ask is
what is the most general set of admissible structure
constants $f_{abc}$ which arise from the Scherk-Schwarz reduction?
It is obvious from our discussion that the answer is equivalent to
classifying all inequivalent isometry groups which
may act on a compact manifold of dimension $d$. In other words,
this means that to determine all massive
supergravities which come from the
Scherk-Schwarz dimensional reduction, one
should find all inequivalent internal axionic groups, and using
them deduce the external groups, which satisfy the Jacobi
consistency conditions. The mass terms are related by
duality symmetry. If the reductions are performed over group spaces
without nontrivial $1$-cycles, or if the
Wilson loops in the Yang-Mills sector vanish in their own right,
the original Yang-Mills gauge symmetry remains unbroken.
Hence, the full nonabelian structure of the vector supermultiplet
fields should be considered. It may be interesting
to consider the relationship of duality and Lie algebras of
the Yang-Mills sector and the
internal manifold in this case. One may further ask if there
may be massive supergravities which cannot be obtained by Scherk-Schwarz
reductions, but could be dual to them,
in a manner which we have discussed here. Other
interesting questions which may be asked in this
context are what happens in type I and II superstring theories.
These tasks are beyond the scope of the present article, but
clearly merit further investigation. It therefore appears
appropriate to end our discussion
here on this note, and leave these questions for future work.

\vspace{1cm}
{\bf Acknowledgements}

We would like to thank Ramzi Khuri for useful discussions.
NK was supported by NSF Grant PHY-9219345.
RCM was supported by NSERC of Canada and Fonds FCAR du
Qu\'ebec.
RCM would also like to thank the Institute for
Theoretical Physics at the University of California, Santa
Barbara, for hospitality while part of this work was being done.
Research at the ITP was supported by NSF Grant PHY94-07194.

\appendix

\section{Reduction of Curvature and Dilaton}
\label{curv}

Here we give the main formulas for the dimensional reduction of the
metric-dilaton sector of the action (\ref{sact1}). The
metric-dilaton part is given in Eq. (\ref{metdil}), which we repeat
here:
\be
S_{g\phi} = \int d^{10}x \sqrt{-{\cal G}} e^{-\Phi}
\Bigl\{ {\cal R}({\cal G}) + (\nabla \Phi)^2 \Bigr\}
\label{ametdil}
\ee
The reduction ansatz for the metric is as in eq.~(\ref{nonabkkmet})
\be
ds^2 = g_{\mu\nu}(x) dx^\mu dx^\nu + {\cal G}_{MN} (x) (\eta^M(y) +
V^M{}_\mu(x) dx^\mu) (\eta^N(y) + V^N{}_\nu(x) dx^\nu)
\label{anonabkkmet}
\ee
wherefore we see that we can define the torsion-free tangent space
basis according to $\hat e^a = \{e^\alpha, {\cal E}^A\}$, where
\ba
e^{\alpha} &=& e^{\alpha}{}_\mu dx^\mu \nonumber \\ {\cal E}^A &=&
{\cal E}^A{}_N (\eta^N + V^N{}_\mu dx^\mu)
\label{atangbasis}
\ea
Using (\ref{atangbasis}), the torsion-free condition $d \hat e^a +
\hat \omega^a{}_b \wedge \hat e^b = 0$, and the metric
compatibility condition $\hat \omega_{ab} = - \hat \omega_{ba}$,
where the indices are raised and lowered with the flat-space metric
$\hat \eta_{ab} = (\eta_{\alpha\beta}, \delta_{AB})$, we can
determine the spin connexion. Recalling that the non-zero structure
constants imply $d \eta^M = - \gamma^M_{NP} \eta^N \wedge \eta^P$
(\ref{formalg}), after straightforward but tedious algebra we find
\ba
\hat \omega^\alpha{}_\beta &=& \omega^{\alpha}{}_\beta -
\frac12 {\cal E}_{MN} V^N{}_{\mu\nu} e^{\alpha \mu}
e_\beta{}^\mu {\cal E}^M \nonumber \\
\hat \omega^A{}_\alpha &=& \frac12
{\cal E}^A{}_N V^N{}_{\mu\nu} e_\alpha{}^\mu
dx^\nu + \frac12 ({\cal E}^{AM} {\cal D}_\mu {\cal
E}_{BM} + {\cal E}_B{}^M {\cal D}_\mu {\cal E}^A{}_M)
e_\alpha{}^\mu {\cal E}^B \nonumber \\
\hat \omega^A{}_B &=& (\gamma^M_{NP} {\cal E}_{CM} {\cal E}^{AN}
{\cal E}_B{}^P
+ \gamma^M_{NP} {\cal E}_{BM} {\cal E}^{AN} {\cal E}_C{}^P
- \gamma^M_{NP} {\cal E}^A{}_{M} {\cal E}_B{}^{N} {\cal E}_C{}^P)
{\cal E}^C
\nonumber \\
&&-\frac12 ({\cal E}_B{}^M {\cal D}_\mu {\cal E}^A{}_M - {\cal
E}^{AM} {\cal D}_\mu {\cal E}_{BM}) dx^\mu
\label{aspinconn}
\ea
Here $\omega^{\alpha}{}_\beta $ is the spin connexion of the
reduced metric $g_{\mu\nu}$. The covariant derivative ${\cal
D}_\mu$ is defined according to
\be
{\cal D}_\mu {\cal E}^A{}_N = \partial_\mu {\cal E}^A{}_N - 2 {\cal
E}^A{}_M
\gamma^M_{NP} V^P{}_\mu
\label{acovder}
\ee
and we see that the first index on ${\cal E}^A{}_N$ is counting the
number of independent objects ${\cal E}$, while the second counts
the components of each object and hence carries gauge charge with
respect to $V^M{}_\mu$. Comparing this with the transformation rule
for ${\cal E}^A{}_N$ given in (\ref{actgrdbein}) and the
transformation rule for the gauge field $V^M{}_\mu$
(\ref{actgrvec}), we find that
\be
{\cal D'}_\mu {\cal E'}^A{}_N = {\cal S}_N{}^M {\cal D}_\mu {\cal
E}^A{}_M
\label{acovdertrans}
\ee
Hence the ansatz for the reduction of the metric
(\ref{anonabkkmet}) ensures that the reduced quantities are
automatically gauge-covariant, as intended. Next, we can use the
form-notation for the Einstein-Hilbert Lagrangian to simplify the
algebra of the reduction of the action. We have in general
\be
{\cal L}_{EH} = \frac{1}{(D+d-2)!} \epsilon_{a_1 ... a_{D+d}} {\cal
R}^{a_1 a_2}
\wedge \hat e^{a_3} \wedge ... \wedge \hat e^{a_{D+d}}
= \sqrt{-\hat {\cal G}} {\cal R}
d^{D}x d^dy
\label{aehact}
\ee
The curvature forms are defined in terms of the spin connexion as
${\cal R}_{a_1 a_b} = d\hat \omega_{a_1 a_2} + \hat \omega_{a_1 b}
\wedge \hat \omega^b{}_{a_2}$. If we now
look at the specific contractions in (\ref{aehact}), we see that we
can write
\be
{\cal L}_{EH} = {\cal J}_1 + 2 {\cal J}_2 + {\cal J}_3
\label{ajs}
\ee
where
\ba
{\cal J}_1 &=& \frac{1}{(D+d-2)!}
\epsilon_{\alpha \beta a_1 ... a_{D+d-2}} {\cal R}^{\alpha\beta} \wedge
\hat e^{a_1} \wedge ... \wedge \hat e^{a_{D+d-2}} \nonumber \\
{\cal J}_2 &=& \frac{1}{(D+d-2)!}
\epsilon_{\alpha A a_1 ... a_{D+d-2}} {\cal R}^{\alpha A} \wedge
\hat e^{a_1} \wedge ... \wedge \hat e^{a_{D+d-2}} \nonumber \\
{\cal J}_3 &=& \frac{1}{(D+d-2)!}
\epsilon_{AB a_1 ... a_{D+d-2}} {\cal R}^{AB} \wedge
\hat e^{a_1} \wedge ... \wedge \hat e^{a_{D+d-2}}
\label{ajsexp}
\ea
Now, these formulas give useful calculational shortcuts. By the
structure of the indices in the $\epsilon$-tensor of the first
expression, one can readily see that the factor
$\hat e^{a_1} \wedge ... \wedge\hat e^{a_{D+d-2}}$
must contain all $d$-beins ${\cal E}$. As a result, only the
those terms in ${\cal R}^{\alpha\beta}$ which are independent of
${\cal E}^A$ contribute to ${\cal J}_1$. Since
\be
{\cal R}^{\alpha\beta} = R^{\alpha\beta} - \frac14 {\cal G}_{MN}
V^M{}_{\mu\nu} V^N{}_{\lambda\sigma} e^{\alpha\mu}
(e^{\beta\nu} dx^\lambda + e^{\beta\lambda} dx^\nu
) \wedge dx^\sigma + O({\cal E}^M)
\label{acurva}
\ee
where $V^M{}_{\mu\nu}= \partial_\mu V^M{}_\nu - \partial_\nu
V^M{}_\mu
- 2 \gamma^M_{NP} V^N{}_\mu V^P{}_\nu$ is precisely the correct
nonabelian field strength as defined in (\ref{reddefs1}).
Factoring out the invariant measure on the internal space,
we have $\hat V_{10} = e^{1} \wedge ... \wedge
e^{D+N} =
\sqrt{-g} \sqrt{\cal G} d^D x d^{d}y$. With this, we finally find
\be
{\cal J}_1 = \sqrt{-g} \sqrt{\cal G} d^Dx d^{d}y (R -
\frac34 {\cal G}_{MN} V^M{}_{\mu\nu} V^{N\mu\nu})
\label{ajone}
\ee
Similar calculations lead to the expressions for ${\cal J}_2$ and
${\cal J}_3$. Noting that it is convenient to define
\be
{\cal D}_\mu {\cal G}_{MN} = \partial_\mu {\cal G}_{MN}
- 2{\cal G}_{MP} \gamma^P_{NQ} V^Q{}_\mu
- 2{\cal G}_{PN} \gamma^P_{MQ} V^Q{}_\mu
\label{acovdermet}
\ee
which comes directly from (\ref{acovdertrans}) and the fact that
since ${\cal E}^A{}_N$ are $d$ objects which transform in the
adjoint representation of the reduced gauge group, the matrix
${\cal G}_{MN}$ transforms in the symmetric direct product of two
adjoints. With this, we can show that
\ba
{\cal J}_3 &=& \sqrt{-g}\sqrt{\cal G} d^Dx d^{d}y (
\frac14 {\cal G}^{MN} {\cal G}^{PQ} {\cal D}_\mu {\cal G}_{MP}
{\cal D}^\mu
{\cal G}_{NQ} - \frac14 {\cal G}^{MN} {\cal D}_\mu {\cal G}_{MN}
{\cal G}^{PQ} {\cal D}_\mu {\cal G}_{PQ} \nonumber \\
&&~~~~~~~~~~~~~~~~~~
- {\cal G}_{MN} {\cal G}^{PQ} {\cal G}^{RS} \gamma^M_{PR} \gamma^N_{QS}
- 2 {\cal G}^{MN} \gamma^P_{MQ} \gamma^Q_{NP} )
\label{ajthree}
\ea
and after some more involved algebra, and the definition of the
covariant derivative of a gauge-charged reduced base space tensor
according to
\ba
{\cal D}_\mu \Phi^{M_1...M_k~\nu_1 ... \nu_p} &=& {\partial}_\mu
\Phi^{M_1...M_k~\nu_1 ... \nu_p} + \Gamma^{\nu_1}_{\mu\rho}
\Phi^{M_1...M_k~\rho ... \nu_p} +...+ \Gamma^{\nu_p}_{\mu\rho}
\Phi^{M_1...M_k~\nu_1 ... \rho}
\nonumber \\
&&+ 2q \gamma^{M_1}_{NP} V^P{}_\mu \Phi^{N...M_k~\nu_1 ... \nu_p} +
... + 2q \gamma^{M_k}_{NP} V^P{}_\mu \Phi^{M_1 ... N~\nu_1 ...
\nu_p}
\label{agencovder}
\ea
where $q$ is the unit of charge of $\Phi^{M_1 ... M_k~\nu_1 ...
\nu_p}$, we find that
\be
{\cal J}_2 = \sqrt{-g} \sqrt{\cal G} d^Dx d^{d}y \Bigl(
\frac14 {\cal G}_{MN} V^M{}_{\mu\nu} V^{N\mu\nu}
+ \frac14 {\cal D}_\mu {\cal G}_{MN} {\cal D}^\mu {\cal G}^{MN}
- \frac12 {\cal D}_\mu ({\cal G}^{MN} {\cal D}^\mu {\cal G}_{MN}) \Bigr)
\label{ajtwo}
\ee
Hence by combining (\ref{ajone}), (\ref{ajthree}) and
(\ref{ajtwo}), we can write down the metric-dilaton action
(\ref{ametdil}) as
\ba
S_{g\phi} &=& \int d^Dx \sqrt{-g} \sqrt{\cal G} e^{-\Phi}
\Bigl(R - \frac14 {\cal G}_{MN} V^M{}_{\mu\nu} V^{N\mu\nu}
- \frac14 {\cal G}^{MN} {\cal D}_\mu {\cal G}_{MN}
{\cal G}^{PQ} {\cal D}^\mu {\cal G}_{PQ} \nonumber \\
&&~~~~~~~~~~~~~~~~~~ + \frac14 {\cal D}^\mu {\cal G}^{MN} {\cal
D}_\mu {\cal G}_{MN}
- \frac12 {\cal D}_\mu ({\cal G}^{MN} {\cal D}_\mu {\cal G}_{MN})
\nonumber \\
&&~~~~~~~~~~~~~~~~~~
- {\cal G}_{MN} {\cal G}^{PQ} {\cal G}^{RS} \gamma^M_{PR} \gamma^N_{QS}
- 2 {\cal G}^{MN} \gamma^P_{MQ} \gamma^Q_{NP} + (\nabla \Phi)^2 \Bigr)
\label{ametdilred}
\ea
where we have canceled the constant factor $\int d^{d} y$.
Defining the reduced dilaton field as in (\ref{kk}), $\exp(-\phi) =
\sqrt{\cal G} \exp(-\Phi)$, we can integrate terms in
(\ref{ametdilred}) by parts, and dropping the boundary terms (which
is in fact required by duality as discussed in \cite{wilczlars}),
we find the final answer
\ba
S_{g\phi} &=& \int d^D x \sqrt{-g} e^{-\phi} \Bigl\{R + (\nabla
\phi)^2 + \frac14 {\cal D}_\mu {\cal G}_{MN} {\cal D}^\mu {\cal
G}^{MN} - \frac14 {\cal G}_{MN} V^M{}_{\mu\nu} V^{N\mu\nu}
\nonumber \\
&& ~~~~~~~~~~~~
- {\cal G}_{MN}
{\cal G}^{PQ} {\cal G}^{RS} \gamma^M_{PR} \gamma^N_{QS}
- 2 {\cal G}^{MN} \gamma^P_{MQ} \gamma^Q_{NP} \Bigr\}
\label{aredmetdil}
\ea
This completes the reduction of the metric-dilaton action.

\section{Reduction of Yang-Mills Fields}
\label{cartan}

Here we outline the dimensional reduction of the Yang-Mills
gauge fields. We recall the ten-dimensional action in eq.~(\ref{carym})
\be
S_{CYM} = - \frac14  \int d^{10}x \sqrt{-{\cal G}} e^{-\Phi}
{\cal F}^I{}_{\mu\nu} {\cal F}^{I\mu\nu}
\label{acarym}
\ee
To proceed, we need to determine the
Ans\"atze for the reduction of the
vector fields ${\cal A}^I$. As we discussed in section 3,
eqs.~(\ref{dans}) and (\ref{gauaxio}) comprise
the ansatz for the case of flat internal torus.
The ensuing mass terms $m^I_{MN}$ already affect the
gauge algebra, which becomes manifest when one considers
the effect of Kaluza-Klein gauge transformations on the
Yang-Mills fields, and the axions which arise
from reducing these fields. For the scalar axion field,
we had from (\ref{kkgauge})
\be
{\cal A}'^I{}_M = {\cal A}^I{}_M + 2 m^I_{MN} \omega^N
\label{akkgauge}
\ee
The inhomogeneous term selects ${\cal A}^I{}_M$ as one of the
axions of the theory. This must remain true even
when the internal space
is not flat, after
appropriate generalizations of the gauge transformations
are made, to account for $\gamma^M_{NP}$. This leads to
the following ansatz for the Yang-Mills
vector fields:
\be
{\cal A}^I = {\cal A}^I{}_\mu(x) dx^\mu + {\cal A}^I{}_M(x) \eta^M
+ \sigma^I(y)
\label{avecnonabkk}
\ee
where the forms $\sigma^I$ are to be defined below. Let us
discuss this ansatz. First, it ensures the correct
transformation rules for ${\cal A}^I{}_M(x)$ after the reduction,
so that it retains its role as the axion of the reduced theory. An
alternative possibility, to which one might be tempted to resort, would have
been to reduce the gauge field ${\cal A}^I$ such that
the cross-term were $\bar{\cal A}^I{}_M \zeta^M$, where
$\zeta^M$ were the invariant one-forms defined in
(\ref{invtforms}). However, by the invariance of $\zeta^M$
under Kaluza-Klein gauge transformations, this would have
implied that the reduced quantity $\bar {\cal A}^I{}_M$ would
have been gauge singlets, and hence would not have transformed as
given in (\ref{akkgauge}) in the limit $\gamma^M_{NP} = 0$.
On the other hand, (\ref{avecnonabkk}) correctly produces
(\ref{akkgauge}) when $\gamma^M_{NP} \rightarrow 0$.
Thus, the only possibility for reduction which could reduce to the
correct limit $\gamma^M_{NP}$ as defined in (\ref{akkgauge}) is
(\ref{avecnonabkk}).

The fields $\sigma^I$ generalize the
terms $m^I_{MN}y^N$ in (\ref{gauaxio}) to the case
of a general internal space.
They encode the information about the gauge field fluxes on
internal two-cycles.
This means, that if we consider
the components of the gauge field strengths ${\cal F}^I$ in the
internal space, we should expect to find
${\cal F}^I{}_{MN} \sim m^I_{MN}$ in the basis spanned by
$\eta^M$'s. Note that these fields can only be defined locally,
and not globally. This is because we see that the mass terms
must emerge as $m^I_{MN} \eta^M \wedge \eta^N \sim d \sigma^I$.
Further, the mass terms must be harmonic forms by the projections
of the gauge field equations of motion in the internal space:
$d {\cal F}^I = 0$, $d{~^*} {\cal F}^I \sim 0$ (where for
simplicity we ignore the dilaton-dependent terms in the
latter equation). Clearly, if mass terms were exact forms,
as suggested by the derivation from $\sim d \sigma^I$,
they would solve the equations of motion, but would have vanishing
internal fluxes: $\int_{2-cycle} {\cal F}^I = 0$. Hence
$\sigma^I$ must be defined only locally, with nontrivial
transition functions between different charts on the internal
space. This of course merely generalizes the considerations
involved in the construction of the Dirac monopole.
However, this implies that the field ${\cal A}^I$ is also defined
only locally, but in such a way that
the perturbations around some background value are globally
well-defined. Since the reduced fields are nothing else but such
perturbations, they are well-defined in the reduced theory.

To determine the explicit form of $\sigma^I$ when $\gamma^M_{NP} \ne 0$,
we need to require
that the gauge field strength evaluated from (\ref{avecnonabkk})
has nonzero internal flux contributions.
Since ${\cal F}^I = d {\cal A}^I$, we find
\be
{\cal F}^I = \partial_{[\mu} {\cal A}^I{}_{\nu]} dx^\mu \wedge
dx^\nu + \partial_\mu {\cal A}^I{}_M dx^\mu\wedge\eta^M - {\cal A}^I{}_M
\gamma^M_{NP} \eta^N \wedge \eta^P + d\sigma^I
\label{aagfnakk}
\ee
This must coincide with
\be
{\cal F}^I = \partial_{[\mu} {\cal A}^I{}_{\nu]} dx^\mu \wedge
dx^\nu + \partial_\mu {\cal A}^I{}_Mdx^\mu\wedge\eta^M - (m^I_{BC} + {\cal
A}^I{}_M \gamma^M_{NP})\eta^N \wedge \eta^P
\label{aagfnakka}
\ee
leading to
\be
d \sigma^I = - m^I_{MN} \eta^M \wedge \eta^N
\label{aintegr}
\ee
This coincides with (\ref{cymred}) in the limit
$\gamma^M_{NP}=0$. The quantities $m^I_{NP}$ are constants which are
antisymmetric in the lower two indices, and in the limit
$\gamma^M_{NP}=0$ they become identical with the axionic
Yang-Mills masses discussed in Sec. (3).
The integrability condition for (\ref{aintegr}), which is
found by looking at
\be
0 = d^2 \sigma^I = 2m^I_{MN} \gamma^N_{PQ} \eta^M \wedge \eta^Q
\wedge \eta^P
\label{aintcind}
\ee
where we have used (\ref{formalg}), is more important, and is
completely general. It comes from demanding that the field strength
is independent of $y^N$. This leads to
\be
m^I_{MN} \gamma^N_{PQ} + m^I_{PN}
\gamma^N_{QM} + m^I_{QN} \gamma^N_{MP} = 0
\label{amgamma}
\ee
which will be interpreted as nothing else but one of the
Jacobi identities for the structure constants of the reduced
theory. Hence the information about the reduced nonabelian
gauge symmetry is in fact encoded in the reduction ansatz,
as a consistency condition between the various modes that
are excited on the internal space.
Comparing (\ref{avecnonabkk})
with the ansatz for $\hat {\cal A}^I(x,y)$ in (\ref{gauaxio}) shows that
they coincide when $\gamma^M_{NP}=0$, while the constraint
(\ref{amgamma}) disappears. Hence indeed (\ref{avecnonabkk}) and
the solution of (\ref{aintegr})
together comprise the reduction ansatz for the
Yang-Mills gauge fields.

With this ansatz, we can now reduce the action (\ref{acarym}).
We can reexpress the field strength (\ref{aagfnakka}) in
terms of the tangent space basis (\ref{atangbasis}) as
\ba
{\cal F}^I &=& \frac12 (F^I{}_{\mu\nu} + {\cal A}^I{}_M V^M{}_{\mu\nu})
 dx^\mu \wedge dx^\nu + {\cal D}_\mu {\cal A}^I{}_M {\cal E}_A{}^M dx^\mu
 \wedge{\cal E}^A
\nonumber \\
&& - (m^I_{MN} + {\cal A}^I{}_P \gamma^P_{MN})
{\cal E}_A{}^M {\cal E}_B{}^N {\cal E}^A \wedge {\cal E}^B
\label{aagfnatb}
\ea
where we use
\ba
{\cal D}_\mu {\cal A}^I{}_M &=& \partial_\mu {\cal A}^I{}_M - 2
(m^I_{MN} + {\cal A}^I{}_P \gamma^P_{MN}) V^N{}_\mu \nonumber \\
F^I{}_{\mu\nu} &=& \partial_\mu A^I{}_\nu - \partial_\nu A^I{}_\mu
- 2 m^I_{MN} V^M{}_\mu V^N{}_\nu
\label{acovdersred}
\ea
as well as the definition of the nonabelian Kaluza-Klein gauge
field $V^M{}_{\mu\nu}$ given in (\ref{reddefs1}). The
reduced gauge fields $A^I{}_\mu$ are defined by
\be
A^I{}_\mu = {\cal A}^I{}_\mu - {\cal A}^I{}_M V^M{}_\mu
\ee
precisely as in (\ref{ansatz}), which diagonalizes the gauge field
sector of the reduced theory. It then takes a very simple
calculation to show that (\ref{aagfnatb}) gives the following
expression for the reduced Yang-Mills action:
\ba
S_{CYM} &=& -  \int d^Dx \sqrt{-g} e^{-\phi} \Bigl\{
\frac14 (F^I{}_{\mu\nu} + {\cal A}^I{}_M V^M{}_{\mu\nu})
(F^{I~\mu\nu} + {\cal A}^I{}_M V^{M\mu\nu}) \nonumber \\ &&
~~~~~~~~~~~~~~~~ + \frac12 {\cal G}^{MN} {\cal D}_\mu {\cal
A}^I{}_M {\cal D}^\mu {\cal A}^I{}_N
\nonumber \\
&& ~~~~~~~~~~~~~~~~ + {\cal G}^{MP} {\cal G}^{NQ} (m^I_{MN} + {\cal
A}^I{}_R \gamma^R_{MN}) (m^I_{PQ} + {\cal A}^I{}_S \gamma^S_{PQ})
\Bigr\}
\label{aredcym}
\ea
That completes this step of the reduction procedure.

\section{Reduction of Two-Form}
\label{three}

The reduction of the Kalb-Ramond two-form ${\cal B}$ contributions
is done in a fashion similar to the reduction of the Yang-Mills
gauge fields, except for the complications that are caused by
the more involved tensor structure of ${\cal H}$ field, and the
anomalous gauge transformation properties of ${\cal B}$, which
require the inclusion of the Yang-Mills Chern-Simons terms
in the definition of ${\cal H}$.
The Kalb-Ramond action is
\be
S_{NS} = - \frac{1}{12} \int d^{10}x \sqrt{-{\cal G}} e^{-\Phi}
{\cal H}_{\mu\nu\lambda} {\cal H}^{\mu\nu\lambda}
\label{ansact}
\ee
We can always write ${\cal B}$ in the
form given in (\ref{dans}), but need to determine the proper
ansatz for the $y^M$ dependence such that the components of
${\cal B}$ produce internal fluxes of the three-form field
strength.
Further, we must ensure that when the limit $\gamma^M_{NP}=0$
is taken, the ansatz correctly produces the
gauge transformation properties given in (\ref{ymcgauge}) and
(\ref{kkgauge}). We can start in much the same way as we
did previously,
when we considered the Yang-Mills fluxes. First, we define the
two-form potential
\be
\hat {\cal B} = {\cal B} + \frac12 \alpha + \frac12 \beta + \gamma
\label{abfred} \ee where the two-form ${\cal B}$ is independent of
$y^M$, and hence can be viewed simply as a globally-defined
perturbation around the Dirac-type configuration of $\hat {\cal
B}$ producing an internal flux.
This is encoded in the forms $\alpha$,
$\beta$ and $\gamma$, which all depend on the internal
coordinates. To define these forms, we recall that there are
Yang-Mills Chern-Simons terms in the definition of the field
strength ${\cal H}$: ${\cal H} = d\hat {\cal B} - \frac12
{\cal A}^I \wedge {\cal F}^I$ (\ref{krdef}). In terms of the
potential and field strength for the Yang-Mills fields
(\ref{avecnonabkk}) and (\ref{aagfnakka}) from the previous
section, we find
\ba  {\cal A}^I \wedge {\cal F}^I &=&
\frac12  {\cal A}^I{}_\mu {\cal F}^I{}_{\nu\lambda} dx^\mu
\wedge dx^\nu \wedge dx^\lambda \nonumber \\ &&+
\Bigl\{\frac12 {\cal F}^I{}_{\mu\nu} ({\cal A}^I{}_M +
\sigma^I{}_M) + {\cal A}^I{}_\mu \partial_\nu {\cal A}^I{}_M
\Bigr\} dx^\mu \wedge dx^\nu \wedge \eta^M \nonumber  \\
&&+ \Bigl\{\partial_\mu {\cal A}^I{}_M ({\cal A}^I{}_N +
\sigma^I{}_N) - {\cal A}^I{}_\mu ( m^I_{MN} + {\cal A}^I{}_P
\gamma^P_{MN}) \Bigr\} dx^\mu \wedge \eta^M \wedge \eta^N
\nonumber  \\ && -  ({\cal A}^I{}_M + \sigma^I{}_M)
(m^I_{NP} + {\cal A}^I{}_Q \gamma^Q_{NP}) \eta^M \wedge \eta^N
\wedge \eta^P \label{acs} \ea
In this expression, there is several
terms which depend explicitly on the internal space coordinates
$y^M$. However to be able to dimensionally reduce the theory, we
must require that the three-form field strength must be
$y^M$-independent. Hence the $y^M$-dependent terms in
$d{\cal B}$ and ${\cal A} {\cal F}$ terms must cancel against each
other, leaving only $y^M$-independent contributions in the
components of ${\cal H}$ expressed in the covariant basis $\eta^M$
(or the tangent basis (\ref{atangbasis}), which is the equivalent
and perhaps more usual statement).

For this purpose, we define the forms
\be
\alpha =  {\cal A}^I{}_M \eta^M \wedge \sigma^I
~~~~~~~~~~~~ \beta =  {\cal A}^I{}_\mu dx^\mu \sigma^I
\label{aalphbeta}
\ee
With this, we can after some algebra write down the
expression for the three-form field strength ${\cal H}$ as
\ba
{\cal H} &=& d {\cal B}  - \frac14
 {\cal A}^I{}_\mu {\cal F}^I{}_{\mu\nu} dx^\mu
\wedge dx^\nu \wedge dx^\lambda \nonumber \\
&&-\frac12  \Bigl\{\frac12 {\cal F}^I{}_{\mu\nu}
{\cal A}^I{}_M  + {\cal A}^I{}_\mu \partial_\nu
{\cal A}^I{}_M \Bigr\} dx^\mu \wedge dx^\nu \wedge \eta^M
\nonumber  \\
&&+\frac12  \Bigl\{\partial_\mu {\cal A}^I{}_M
{\cal A}^I{}_N - {\cal A}^I{}_\mu (
2m^I_{MN} + {\cal A}^I{}_P \gamma^P_{MN})
\Bigr\} dx^\mu \wedge \eta^M \wedge \eta^N
\nonumber  \\
&&+ \frac12  {\cal A}^I{}_M
(2m^I_{NP} + {\cal A}^I{}_Q \gamma^Q_{NP})
\eta^M \wedge \eta^N \wedge \eta^P \nonumber \\
&&+ d \gamma +  \frac12   \sigma^I{}_M m^I_{NP}
\eta^M \wedge \eta^N \wedge \eta^P
\label{ahfield}
\ea
and in this equation, only the last two terms contain
explicit $y^M$ dependence. This dependence however is now
quite nontrivial. Namely, we can consider the exterior
derivative of these two terms. We find
\ba
 d^2 \gamma +  \frac12  d\Bigl\{ \sigma^I{}_M m^I_{NP}
\eta^M \wedge \eta^N \wedge \eta^P \Bigr\} &=&
- \frac12  d\sigma^I \wedge d\sigma^I \nonumber \\
&=& -\frac12  m^I_{[MN} m^I_{PQ]} \eta^M \wedge \eta^N \wedge
\eta^P \wedge \eta^Q
\label{diffcondgamma}
\ea
On the other hand, we the three-form field strength
${\cal H}$ must have components in the internal space
which are independent of the internal space coordinates.
Given the expression (\ref{ahfield}), that means
that we must have
\be
d \gamma +  \frac12   \sigma^I{}_M m^I_{NP}
\eta^M \wedge \eta^N \wedge \eta^P = \frac12 \beta_{MNP}
\eta^M \wedge \eta^N \wedge \eta^P
\label{gammadiffeq}
\ee
where $\beta_{MNP}$ are constants as a consequence of equations
of motion. Combining (\ref{diffcondgamma}) and (\ref{gammadiffeq}),
we get the following constraint on the $\beta_{MNP}$: since
$d \beta_{MNP} (\eta^M \wedge \eta^N \wedge \eta^P) =
3 \beta_{MNP} \gamma^M_{QR} \eta^Q \wedge \eta^R \wedge \eta^N \wedge \eta^P$,
comparing the sides of the equation we get
\be
3 \beta_{R[MN} \gamma^R_{PQ]} =  m^I_{[MN} m^I_{PQ]}
\label{ajachf}
\ee
This is the last of the independent classes of Jacobi
identities of the reduced theory.

Hence regardless of the specifics of the ansatz, that depends on the
symmetry of the internal manifold, the general expression for ${\cal H}$
is inevitably
\ba
{\cal H} &=& \Bigl( \frac12 \partial_{[\mu} {\cal B}_{\nu\lambda]}
- \frac14  {\cal A}^I{}_\mu {\cal F}^I{}_{\nu\lambda} \Bigr)
dx^\mu \wedge dx^\nu \wedge dx^\lambda
\nonumber \\
&&+
\Bigl(\partial_{[\mu}{\cal B}_{\nu]M} - \frac12  ( {\cal
F}^I{}_{\mu\nu} {\cal A}^I{}_M + {\cal A}^I{}_\mu \partial_\nu
{\cal A}^I{}_M) \Bigr) dx^\mu \wedge dx^\nu \wedge \eta^M
\nonumber\\
&&+\frac12 \Bigl(\partial_\mu {\cal B}_{MN} - {\cal B}_{\mu P}
\gamma^P_{MN} + \frac12 ({\cal A}^I{}_M \partial_\mu {\cal
A}^I{}_N + {\cal A}^I{}_\mu (2 m^I_{MN} + {\cal A}^I{}_P
\gamma^P_{MN})) \Bigr) dx^\mu \wedge \eta^M \wedge \eta^N
\nonumber\\
&& + \frac12  {\cal A}^I{}_M (2 m^I_{NP} + {\cal
A}^I{}_Q \gamma^Q_{NP}) \eta^M \wedge \eta^N \wedge \eta^P +
\frac12 \beta_{MNP} \eta^M \wedge \eta^N \wedge \eta^P
\label{ahred2} \ea
Then to complete the reduction of the action
(\ref{ansact}), we need to rewrite (\ref{ahred2}) in the
orthogonal basis (\ref{atangbasis}). Using these definitions, we
find after another lengthy calculation that
\ba
{\cal H} &=&
\frac16 H_{\mu\nu\lambda} dx^\mu \wedge dx^\nu \wedge dx^\lambda
\label{moreh}\\
&&+ \frac12 \Bigl(H_{\mu\nu M} -  {\cal A}^I_M
F^I{}_{\mu\nu} - {\cal C}_{MN} V^N{}_{\mu\nu} \Bigr){\cal E}_A{}^M
dx^\mu  \wedge dx^\nu \wedge {\cal E}^A
\nonumber \\
&&+ \frac12 \Bigl({\cal D}_\mu \B_{MN} +
{\cal A}^I{}_{[M} {\cal D}_\mu {\cal A}^I{}_{N]} \Bigr)
{\cal E}_A{}^M {\cal E}_B{}^N
dx^\mu  \wedge {\cal E}^A \wedge {\cal E}^B
\nonumber \\
&&\frac12 \Bigl(\beta_{MNP} + 2  {\cal
A}^I{}_M m^I_{NP}  + 2 {\cal C}_{MQ} \gamma^Q_{NP}
\Bigr) {\cal E}_A{}^M {\cal E}_B{}^N {\cal E}_C{}^P
{\cal E}^A \wedge {\cal E}^B \wedge {\cal E}^C
\nonumber
\ea
Here we are using the new definitions
\ba
{\cal D}_\mu \B_{MN}&=& \partial_\mu \B_{MN} + 2  m^I_{MN} A^I{}_\mu + 2
\gamma^P_{MN} B_{\mu P}
\nonumber \\
&& - \beta_{MNP} V^P{}_\mu +
4 \B_{Q[M} \gamma^Q_{N]P} V^P{}_\mu -2  {\cal A}^I{}_{[M}
m^I_{N]P} V^P{}_\mu
\nonumber \\
H_{\mu\nu M} &=& \partial_\mu
B_{\nu M} - \partial_\nu B_{\mu M} + 3\beta_{MNP} V^N{}_\mu
V^P{}_\nu
\nonumber \\
&&+ 4 \gamma^P_{MN} B_{[\mu P} V^N{}_{\nu]}
+ 4  m^I_{MN} A^I{}_{[\mu} V^N{}_{\nu]}
\label{acovdefs}
\ea
and
\ba
H_{\mu\nu\lambda} &=& \partial_\mu B_{\nu\lambda} -
\frac12  A^I{}_\mu F^I{}_{\nu\lambda} - \frac 12 V^M{}_\mu
H_{\nu\lambda M} - \frac 12 B_{\mu M} V^M{}_{\nu\lambda} +\frac12
\beta_{MNP} V^M{}_\mu V^N{}_\nu V^P{}_\lambda
\nonumber \\
&& - m^I_{MN} A^I{}_\mu V^M{}_\nu V^N{}_\lambda -
\gamma^M_{NP} B_{\mu M} V^N{}_\nu V^P{}_\lambda + ~cyclic ~ perm. ~
\label{arednsnsthree}
\ea
in addition to (\ref{nonabkkmet}),
(\ref{vecnonabkk}),
and (\ref{ansatz}). We have redefined $B_{\mu M}$ and $B_{\mu\nu}$
according to (\ref{ansatz}), in order to express the reduced
action, in a manifestly gauge- and duality-symmetric way. With
this, we finally find the reduced Kalb-Ramond action in $D$ dimensions:
\ba
S_{KR} &=& - \int d^D x \sqrt{-g} e^{-\phi} \Bigl\{
\frac{1}{12} H_{\mu\nu\lambda} H^{\mu\nu\lambda} \nonumber \\ &&+
\frac14 {\cal G}^{MN} (B_{\mu\nu M} -  {\cal A}^I{}_M
F^I{}_{\mu\nu} - {\cal C}_{MP} V^P{}_{\mu\nu}) (B^{\mu\nu}{}_N -
 {\cal A}^I{}_N F^{I~\mu\nu} - {\cal C}_{NQ} V^{Q\mu\nu})
\nonumber \\
&&+ \frac14 {\cal G}^{MP} {\cal G}^{NQ} ({\cal D}_\mu
B_{MN} +  {\cal A}^I{}_{[M} {\cal D}_\mu {\cal
A}^I{}_{N]}) ({\cal D}^\mu B_{PQ} +  {\cal A}^J{}_{[P}
{\cal D}^\mu {\cal A}^J{}_{Q]})
\nonumber \\
&&+ \frac34 {\cal
G}^{MQ} {\cal G}^{NR} {\cal G}^{PS} (\beta_{MNP} + 2
{\cal A}^I{}_{[M} m^I_{NP]} - 2 {\cal C}_{T[M} \gamma^G_{NP]})
\nonumber \\ &&~~~~~~~~~~~~~~~~~~ \quad\times (\beta_{QRS} + 2
 {\cal A}^J{}_{[Q} m^J_{RS]} - 2 {\cal C}_{U[Q}
\gamma^H_{RS]}) \Bigr\}
\label{ansredact}
\ea
This is the last
step in the reduction of the effective action.

\section{Gauge Algebra}
\label{algebraaa}

Here we give a detailed derivation of the gauge algebra
(\ref{algebra1}). Our approach is to first compute the gauge-dependent
variations of the reduced gauge fields under the gauge transformations,
and then to reassemble them into the algebra using the homomorphism
between the neighbourhood of identity of the gauge group and the gauge
algebra. We will work in the ascending order of complexity. Thus, we
start with the reduced form of the gauge transformations associated
with the Kalb-Ramond two-form field.

Recall that the two-form field introduces a one-form gauge symmetry
in the original ten-dimensional theory, ${\cal B} \rightarrow {\cal B}' =
{\cal B} + d \Lambda$, where $\Lambda = \lambda_M \eta^M + \lambda_\mu dx^\mu$
is a one-form. Hence, upon dimensional
reduction to $D$ dimensions, we find $d$ new $U(1)$ gauge symmetries,
counted by the components of $\Lambda$ in the internal space. Of course, there
still remains the reduced one-form of the original gauge transformation
$\Lambda = \lambda_\mu dx^\mu$, which produces variations of
the reduced field $B = \frac12 B_{\mu\nu} dx^\mu \wedge dx^\nu$. These
transformations must allow the reduced fields of the theory, which emerge from
the two-form, to retain their $y^M$-independent form, in order that
the reduction is consistent. Hence all such reduced gauge transformations
must be independent of $y^M$ as well. The physical meaning of this statement
can be seen as follows: From the point of view of the reduced theory, the
modes which are excited in  the internal dimensions (\ie they depend
on $y^M$) are all very heavy, and hence cannot be excited by the low
energy physics, which is described by the effective action in $D$ dimensions.
So from this and the reduction ansatz for the two-form,
we see that the only fields which transform nontrivially under the gauge
transformations generated by $\lambda_M$ are the reduced components
${\cal B}_{\mu\nu}$, ${\cal B}_{\mu M}$ and ${\cal B}_{MN}$. An explicit
computation shows that (where we set the reduced two-form gauge transformation
to zero, $\lambda_\mu = 0$)
\ba
{\cal B}'_{MN} &=& {\cal B}_{MN} - 2 \lambda_P \gamma^P_{MN} \nonumber \\
{\cal B}'_{\mu M} &=& {\cal B}_{\mu M} + \partial_\mu \lambda_M \nonumber \\
{\cal B}'_{\mu\nu} &=& {\cal B}_{\mu\nu}
\ea
Using field redefinitions (\ref{ansatz}) which diagonalize the reduced gauge
transformations, we can rewrite the previous equation as
\ba
&&\B'_{MN} = \B_{MN} - 2 \lambda_P \gamma^P_{MN} \nonumber \\
&&B'_{\mu A} = B_{\mu A} + \partial_\mu \lambda_M - 2
\lambda_P \gamma^P_{MN} V^N{}_\mu \nonumber \\
&&B'_{\mu\nu} = B_{\mu\nu} + \frac12 \lambda_M V^M{}_{\mu\nu} +
\gamma^M_{np} \lambda_M V^N{}_\mu V^P{}_\nu
\label{ansnsgauge1}
\ea
It is clear that these are the
only nontrivial gauge transformations rules in the
case of gauge symmetries arising from the two-form
sector. Note that the $B_{\mu\nu}$ transforms nontrivially,
which is a signature of the anomaly that emerges from the reduction.

Next, we turn to the reduction of the Yang-Mills subalgebra of the
original ten-dimensional gauge group. In the ten-dimensional case, these
symmetries correspond to ${\cal A}^I \rightarrow
{\cal A'}^I = {\cal A}^I + d \Lambda^I$. If we wish to preserve the
reduction ansatz, and the fact that the reduced field strengths are
independent of $y^M$, again we must require that the residual
gauge symmetries after dimensional reduction are independent
of $y^M$. In the Yang-Mills sector after the reduction, and diagonalization
of the reduced degrees of freedom using (\ref{ansatz}), it is easy to see that
\be
{\cal A'}^I{}_M = {\cal A}^I{}_M ~~~~~~~~~~ {\cal A'}^I{}_\mu =
{\cal A}^I{}_\mu + \partial_\mu \lambda^I
\ee
These transformations obviously leave the Kaluza-Klein sector unchanged.
However, they  induce nontrivial transformation properties in the two-form
sector. To start with, due to the anomaly, the two-form field in
ten-dimensions transforms
according to $\hat {\cal B}' = \hat {\cal B} + \frac12
\lambda^I {\cal F}^I + d \Lambda$. If we consider the
decomposition of $\hat {\cal B}$ in terms of the reduced degrees
of freedom, we see that in addition to the
${\cal B}$ terms, also the terms proportional to ${\cal A}^I{}_\mu$
are gauge-noninvariant. Hence we must have the following transformation
rule for the combination of these fields:
\ba
&& \frac12 {\cal B'}_{\mu\nu}(x) dx^\mu \wedge dx^\nu +
{\cal B'}_{\mu M}(x) dx^\mu \wedge \eta^M + \frac12 {\cal B'}_{MN}(x)
\eta^M \wedge \eta^N
+  {\cal A'}^I{}_\mu dx^\mu \sigma^I = \nonumber \\
&& ~~~~~~~~~~~~ \frac12 {\cal B'}_{\mu\nu}(x) dx^\mu \wedge dx^\nu +
{\cal B'}_{\mu M}(x) dx^\mu \wedge \eta^M + \frac12 {\cal B'}_{MN}(x)
\eta^M \wedge \eta^N \nonumber \\
&& ~~~~~~~~~~~~
+  {\cal A'}^I{}_\mu dx^\mu \sigma^I + \frac12  \lambda^I
{\cal F}^I + d \Lambda
\ea
Next, using the fact that
\be
 {\cal A'}^I{}_\mu dx^\mu \wedge \sigma^I =
 {\cal A}^I{}_\mu dx^\mu \wedge \sigma^I
+ d( \lambda^I \sigma^I)
+  \lambda^I m^I_{MN} \eta^M \wedge \eta^N
\ee
where we have integrated the gauge-dependent piece by parts, and have
used
$d \sigma^I = - m^I_{MN} \eta^M \wedge \eta^N$, we get that
\ba
&& \frac12 {\cal B'}_{\mu\nu}(x) dx^\mu \wedge dx^\nu +
{\cal B'}_{\mu M}(x) dx^\mu \wedge \eta^M + \frac12 {\cal B'}_{MN}(x)
\eta^M \wedge \eta^N = \nonumber \\
&& ~~~~~~~~~~~~ \frac12 {\cal B}_{\mu\nu}(x) dx^\mu \wedge dx^\nu +
{\cal B}_{\mu M}(x) dx^\mu \wedge \eta^M + \frac12 {\cal B}_{MN}(x)
\eta^M \wedge \eta^N \nonumber \\
&& ~~~~~~~~~~~~
+ \frac12  \lambda^I
({\cal F}^I - m^I_{MN} \eta^M \wedge \eta^N)
+ d (\Lambda - \frac12  \lambda^I \sigma^I) = \nonumber \\
&& ~~~~~~~~~~~~ \frac12 {\cal B}_{\mu\nu}(x) dx^\mu \wedge dx^\nu +
{\cal B}_{\mu M}(x) dx^\mu \wedge \eta^M + \frac12 {\cal B}_{MN}(x)
\eta^M \wedge \eta^N \nonumber \\
&& ~~~~~~~~~~~~
+ \frac14  \lambda^I {\cal F}^I{}_{\mu\nu} dx^\mu \wedge dx^\nu +
\frac12  \lambda^I \partial {\cal A}^I{}_M dx^\mu \wedge \eta^M
\nonumber \\
&& ~~~~~~~~~~~~
-\frac12  \lambda^I(2 m^I_{MN} + {\cal A}^I{}_P \gamma^P_{MN})
\eta^M \wedge \eta^N
+ d (\Lambda - \frac12  \lambda^I \sigma^I)
\ea
Now, using this equation and the formulae for diagonalizing the reduced
degrees of freedom given in (\ref{ansatz}), after some tedious but
straightforward algebra we finally arrive at
\begin{eqnarray}
&&\B'_{MN} = \B_{MN} - 2  \lambda^I m^I_{MN} \nonumber \\ &&
{A'}^I{}_\mu = A^I{}_\mu + \partial_\mu \lambda^I \nonumber \\ &&
B'_{\mu M} = B_{\mu M}
- 2  \lambda^I m^I_{MN} V^N{}_\mu
\nonumber \\
&& B'_{\mu\nu} = B_{\mu\nu} + \frac12
 \lambda^I F^I{}_{\mu\nu}
+  m^I_{MN} \lambda^I V^M{}_\mu V^N{}_\nu
\label{aymcgauge1}
\end{eqnarray}
which as we see are independent of $\gamma^M_{NP}$, and hence these
identical to the simple case of a flat internal torus.

Finally, we consider the reduction of the Kaluza-Klein gauge
transformations. They have the most complicated structure, since
they affect the matter degrees of freedom from all three sectors.
We start
with the metric moduli ${\cal G}_{MN}$ and the gauge fields
$V^M{}_\mu$. In fact, we have already determined how these fields
transform; the results are given in (\ref{actgrvec}) and
(\ref{actgrmet})
\be
{\cal G'}_{MN} = {\cal S}_M{}^P {\cal S}_{N}{}^Q {\cal G}_{PQ}
~~~~~~~~~~
{V'}^M{}_\mu = {\cal S}^M{}_N V^N{}_\mu + {\cal O}^M{}_N
\partial_\mu \omega^N
\ee
Using the infinitesimal forms of ${\cal O}^M{}_N$ and ${\cal S}^M{}_N$
given in (\ref{odef}), it is straightforward to compute the infinitesimal
form of the gauge transformations of ${\cal G}_{MN}$ and $V^M{}_\mu$. They are
\ba
&& {\cal G'}_{MN} = {\cal G}_{MN} + 2 \gamma^P_{MQ} \omega^Q {\cal G}_{PN}
+ 2 \gamma^P_{NQ} \omega^Q {\cal G}_{MP} + O(\omega^2)
\nonumber \\
&& V'^M{}_\mu = V^M{}_\mu -2 \gamma^M_{NP} \omega^P V^N{}_\mu +
\partial_\mu
\omega^M + O(\omega^2)
\ea
{}From the transformation rule for the moduli matrix ${\cal G}_{MN}$, we
see that it is a direct product of scalars belonging
to the singlet and two adjoints (with charge $-1$)
of the isometry group. This is precisely the transformation rule
of the axions hidden in ${\cal G}_{MN}$. They can be explicitly
found by Gauss decomposition of the matrix ${\cal E}^A{}_M$
(with the property that ${\cal G}_{MN} = \delta_{AB}
{\cal E}^A{}_M {\cal E}^B{}_N$)
and identifying the pivots with the dilaton-like fields
and the upper triangular matrix elements with the axions
--- see, \eg eq.~\reef{zwei3}.

Next we consider the effect of Kaluza-Klein gauge transformations
on Yang-Mills gauge fields. From (\ref{vecnonabkk}), it is
clear that the reduced fields transform nontrivially under the
Kaluza-Klein transformations.
A lengthy but straightforward calculation shows that due to the form of the
compensator forms $\sigma^I$, we find
\be
{\sigma'}^I = \sigma^I + 2 m^I_{MN} \omega^M \eta^N + d\, \Xi
\label{asigmagkk}
\ee
Hence if we accompany the Kaluza-Klein gauge transformation
by a Yang-Mills custodian transformation
\be
{\cal A'}^I = {\cal A}^I + d\,\Xi
\label{acusto}
\ee
the components of ${\cal A}$ transform infinitesimally according to
\ba
{\cal A'}^I{}_M &=& {\cal A}^I{}_M + 2 m^I_{MN} \omega^N +
2 \gamma^N_{MP} {\cal A}^I{}_N \omega^P + O(\omega^2) \nonumber \\
{\cal A'}^I{}_\mu &=& {\cal A}^I{}_\mu + {\cal A}^I{}_M \partial_\mu
\omega^M + O(\omega^2)
\label{aacompskk}
\ea
so that the diagonalized reduced Yang-Mills degrees of freedom
transform according to
\ba
{\cal A}'^I{}_M &=& {\cal A}^I{}_M +2 \gamma^N_{MP} \omega^P {\cal
A}^I{}_N + 2 m^I_{MN} \omega^N + O(\omega^2)\nonumber \\
{A'}^I{}_\mu &=& A^I{}_\mu -2 m^I_{MN} \omega^N V^M{}_\mu +
O(\omega^2)
\label{acymkkg}
\ea

Finally we come to consider the transformation properties of the
two-form degrees of freedom. First, we recall the notation
of (\ref{abfred}),
$\hat {\cal B} = {\cal B} + \frac12 \alpha + \frac12 \beta + \gamma$,
and the definition (\ref{aalphbeta}).
Note that $\alpha + \beta =  {\cal A}^I \wedge \sigma^I$. Now
we use an algebraic trick to compute the gauge dependence of the
compensator forms $\gamma$. Instead of working with $\gamma$ itself,
we consider its exterior derivative.
Using the fact that
\be
d\gamma = - \frac12  m^I_{MN} \sigma^I \wedge \eta^M \wedge \eta^N
+ \frac12 \beta_{MNP} \eta^M \wedge \eta^N \wedge \eta^P
\ee
if we denote the gauge variation $\delta \gamma = \gamma' - \gamma$
we find after a straightforward calculation
\be
d \delta \gamma =  m^I_{MN} \sigma^I \wedge d(\omega^M \eta^N)
+ \frac12  \delta \sigma^I \wedge \sigma^I +
\frac12 \beta_{MNP} \delta(\eta^M \wedge \eta^N \wedge \eta^P)
\ee
Next, using the transformation properties of $\sigma^I$, given in
(\ref{asigmagkk}), we can deduce that the explicit form of the
exterior derivative of the gauge dependence of $\gamma$ is
\ba
d \delta \gamma &=& - d \vartheta + 2  m^I_{MN}m^I_{PQ}
\omega^N \eta^M \wedge \eta^P \wedge \eta^Q
\nonumber\\
&&\qquad\qquad -
3 \beta_{MNP} \gamma^M_{QR} \omega^R \eta^Q \wedge \eta^N \wedge \eta^P
- \frac32 \beta_{MNP} d \omega^M \wedge \eta^N \wedge \eta^P
\label{agammaderiv}
\ea
where
\be
\vartheta =  \sigma^I \wedge \Bigl(m^I_{MN} \omega^M \eta^N
- \frac12 d\,\Xi
\Bigr)
\ee
Now we can integrate eq.~(\ref{agammaderiv}). Using the Jacobi
identity (\ref{ajachf}), we find after some straightforward algebra that
\be
\gamma' = \gamma - \vartheta - \frac32 \beta_{MNP} \omega^M \eta^N
\wedge \eta^P + d \tilde \Lambda \label{agammakk} \ee where
$\tilde \Lambda$ is an arbitrary globally defined one-form. Then,
using (\ref{asigmagkk}) and (\ref{aacompskk}), we can compute the
gauge variation of $\alpha + \beta$. We find
\ba
\alpha'+ \beta'
&=& \alpha + \beta + 2  m^I_{MN} \omega^M \sigma^I
\wedge \eta^N  \\
&& +  ({\cal A}^I{}_\mu dx^\mu + {\cal
A}^I{}_M \eta^M) \wedge \Bigl(2 m^I_{MN} \omega^M \eta^N + d\,\Xi
\Bigr) \nonumber
\label{aalphabetakk}
\ea
We can now compute the transformation properties of the reduced
components. Recalling that the Yang-Mills fields should be
gauge transformed by the custodian transformation (\ref{acusto}),
which induces $\delta {\cal B} = \frac12  \lambda^I {\cal F}^I$,
after some more algebra we find that
\ba
{\cal B'} &=& {\cal B} + \delta {\cal B} - \delta \gamma -
\frac12 \delta (\alpha + \beta) \nonumber \\
&=& d\Bigl(\Lambda - \tilde \Lambda + \frac12 {\cal A}^I \wedge \Xi
\Bigr) \nonumber \\
&&+\frac32 \beta_{MNP} \omega^M \eta^N \wedge \eta^P
-  \Bigl({\cal A}^I{}_\mu dx^\mu + {\cal A}^I{}_M \eta^M \Bigr)
m^I_{NP} \omega^N \eta^P
\ea
We should mention again here that the role of the custodian transformations,
as well as the anomaly, is essential to ensure the cancellation of the
$y^M$-dependent terms in the gauge transformation laws. This is precisely
what will eventually produce the nonabelian symmetry structure among the
reduced vector fields.
Choosing the ${\cal B}$ field gauge transformation according to
$\Lambda = \tilde \Lambda - \frac12 {\cal A}^I \wedge \Xi
$, we finally find that the infinitesimal
transformations of ${\cal B}$ are
\be
{\cal B}' = {\cal B} +\frac32 \beta_{MNP} \omega^M \eta^N \wedge \eta^P
-  \Bigl({\cal A}^I{}_\mu dx^\mu + {\cal A}^I{}_M \eta^M \Bigr)
m^I_{NP} \omega^N \eta^P + O(\omega^2)
\ee
Using the expressions for the diagonalized reduced degrees of freedom
(\ref{ansatz}), after a straightforward calculation we finally find
\begin{eqnarray}
&&\B'_{MN} = \B_{MN} +
3\beta_{MNP} \omega^P + 2
 a^I{}_{[M} m^I_{N]P} \omega^P + O(\omega^2)
\nonumber \\
&&B'_{\mu M} = B_{\mu M} + 2 \gamma^N_{MP}
\omega^P B_{\mu N} + 2  m^I_{MN} \omega^N A^I{}_\mu +
3\beta_{MNP} \omega^P V^N{}_\mu + O(\omega^2)
\nonumber \\
&&B'_{\mu\nu} = B_{\mu\nu} + \frac12
\omega^M H_{\mu\nu M} - \frac32 \beta_{MNP} \omega^M V^N{}_\mu V^P{}_\nu
\nonumber \\
&& ~~~~~~~~~~~~~
- 2 \gamma^P_{MN} \omega^M B_{[\mu |P} V^N{}_{\nu]} -2  \omega^M
m^I_{MN} A^I{}_{[\mu} V^N{}_{\nu]} + O(\omega^2)
\label{akknsnsgauge}
\end{eqnarray}
So the nontrivial infinitesimal Kaluza-Klein transformations for all the reduced
degrees of freedom are
\begin{eqnarray}
&&{\cal A}'^I{}_M = {\cal A}^I{}_M +2 \gamma^N_{MP} \omega^P {\cal
A}^I{}_N + 2 m^I_{MN} \omega^N
\nonumber \\
&&\B'_{MN} = \B_{MN} +
3\beta_{MNP} \omega^P + 2
 a^I{}_{[M} m^I_{N]P} \omega^P + O(\omega^2)
\nonumber \\
&& {\cal G'}_{MN} = {\cal G}_{MN} + 2 \gamma^P_{MQ} \omega^Q {\cal G}_{PN}
+ 2 \gamma^P_{NQ} \omega^Q {\cal G}_{MP} + O(\omega^2)
\nonumber \\
&& V'^M{}_\mu = V^M{}_\mu -2 \gamma^M_{NP} \omega^P V^N{}_\mu +
\partial_\mu
\omega^M + O(\omega^2) \nonumber \\
&& {A'}^I{}_\mu = A^I{}_\mu -2 m^I_{MN} \omega^N V^M{}_\mu +
O(\omega^2)
\nonumber \\
&&B'_{\mu M} = B_{\mu M} + 2 \gamma^N_{MP}
\omega^P B_{\mu N} + 2  m^I_{MN} \omega^N A^I{}_\mu +
3\beta_{MNP} \omega^P V^N{}_\mu + O(\omega^2)
\nonumber \\
&&B'_{\mu\nu} = B_{\mu\nu} + \frac12
\omega^M H_{\mu\nu M} - \frac32 \beta_{MNP} \omega^M V^N{}_\mu V^P{}_\nu
\nonumber \\
&& ~~~~~~~~~~~~~
- 2 \gamma^P_{MN} \omega^M B_{[\mu |P} V^N{}_{\nu]} -2  \omega^M
m^I_{MN} A^I{}_{[\mu} V^N{}_{\nu]} + O(\omega^2)
\label{aakkgaugefin}
\end{eqnarray}

We are now finally ready to reassemble the results of this appendix into
the algebra of the reduced gauge theory. The calculation is fairly
straightforward, and hence we merely describe the method
here and list the results. First, we recapitulate the explicit form
of the gauge group in the neighbourhood of identity. We have

1) reduced two-form gauge transformations:
\ba
&&\B'_{MN} = \B_{MN} - 2 \lambda_P \gamma^P_{MN} \nonumber \\
&&B'_{\mu M} = B_{\mu M} + \partial_\mu \lambda_M -
2\lambda_P \gamma^P_{MN} V^N{}_\mu \nonumber \\
&&B'_{\mu\nu} = B_{\mu\nu} + \frac12 \lambda_M V^M{}_{\mu\nu} +
\gamma^M_{NP} \lambda_M V^N{}_\mu V^P{}_\nu
\label{aansnsgauge1}
\ea

2) reduced Yang-Mills gauge transformations:
\begin{eqnarray}
&&\B'_{MN} = \B_{MN} - 2  \lambda^I m^I_{MN} \nonumber \\ &&
{A'}^I{}_\mu = A^I{}_\mu + \partial_\mu \lambda^I \nonumber \\ &&
B'_{\mu M} = B_{\mu M}
- 2  \lambda^I m^I_{MN} V^N{}_\mu
\nonumber \\
&& B'_{\mu\nu} = B_{\mu\nu} + \frac12
 \lambda^I F^I{}_{\mu\nu}
+  m^I_{MN} \lambda^I V^M{}_\mu V^N{}_\nu
\label{aaymcgauge1}
\end{eqnarray}

3) reduced Kaluza-Klein gauge transformations:
\begin{eqnarray}
&&{\cal A}'^I{}_M = {\cal A}^I{}_M +2 \gamma^N_{MP} \omega^P {\cal
A}^I{}_N + 2 m^I_{MN} \omega^N \nonumber \\
&&\B'_{MN} = \B_{MN} +
3\beta_{MNP} \omega^P + 2
 a^I{}_{[M} m^I_{N]P} \omega^P + O(\omega^2)
\nonumber \\
&& {\cal G'}_{MN} = {\cal G}_{MN} + 2 \gamma^P_{MQ} \omega^Q {\cal G}_{PN}
+ 2 \gamma^P_{NQ} \omega^Q {\cal G}_{MP} + O(\omega^2)
\nonumber \\
&& V'^M{}_\mu = V^M{}_\mu -2 \gamma^M_{NP} \omega^P V^N{}_\mu +
\partial_\mu
\omega^M + O(\omega^2) \nonumber \\
&& {A'}^I{}_\mu = A^I{}_\mu -2 m^I_{MN} \omega^N V^M{}_\mu +
O(\omega^2) \nonumber \\
&&B'_{\mu M} = B_{\mu M} + 2 \gamma^N_{MP}
\omega^P B_{\mu N} + 2  m^I_{MN} \omega^N A^I{}_\mu +
3\beta_{MNP} \omega^P V^N{}_\mu + O(\omega^2)
\nonumber \\
&&B'_{\mu\nu} = B_{\mu\nu} + \frac12
\omega^M H_{\mu\nu M} - \frac32 \beta_{MNP} \omega^M V^N{}_\mu V^P{}_\nu
\nonumber \\
&& ~~~~~~~~~~~~~
- 2 \gamma^P_{MN} \omega^M B_{[\mu| P} V^N{}_{\nu]} -2  \omega^M
m^I_{MN} A^I{}_{[\mu} V^N{}_{\nu]} + O(\omega^2)
\label{aakkgauge1}
\end{eqnarray}

We can now calculate the algebra of gauge generators. As in section
\ref{massone}, the group generators are denoted $T_a$, where the
indices take values in the space of $\{~_M, ~^M, ~^I\}$, with
dimension $2d + 16$, and in the explicit form arise from the three sectors
described above: $T_a = (Z_M,X^M,Y^I)$. The gauge algebra must close,
and hence it satisfies
\be
[T_a, T_b] = i f_{ab}{}^c T_c
\ee
where $f_{ab}{}^c$ are the structure constants which we need to
determine. We can compute them by considering the products of
transformations (\ref{aansnsgauge1})-(\ref{aakkgauge1}), which are
in general of the form
$h^{-1} \cdot g^{-1} \cdot h \cdot g$, with $h$ and $g$ any two infinitesimal
gauge transformations.
Infinitesimally, for any two operators $A, B$ and a number $\alpha <<1$, we
have $e^{\alpha A} B e^{-\alpha A} = B + \alpha [A,B] + O(\alpha^2)$.
So using $g = \exp(i\hat \omega_1^a T_a)$ and $h =
\exp(i \hat \omega_2^a T_a)$, where
$\hat \omega^a = (\omega^M, \lambda_M, \lambda^I)$ are the gauge
transformation parameters defined above, we have, to linear order
in $\hat\omega^a_1$,
\begin{equation}
g^{-1} \cdot h \cdot g = h - i \hat \omega_1^a [T_a, h]
\label{inftr2}
\end{equation}
and hence we find, to the lowest nontrivial order (quadratic)
\be
h^{-1} \cdot g^{-1} \cdot h \cdot g = 1 + \hat \omega_1^a \hat \omega_2^b
[T_a, T_b] = 1 + i f_{ab}{}^{c} \hat \omega_1^a \hat \omega_2^b T_c\ .
\label{ainftr3}
\ee
Hence substituting the explicit form of the gauge transformations
(\ref{aansnsgauge1})-(\ref{aakkgauge1}), we can calculate the structure
constants.
To do it, we must project the algebra composition laws on the set
of basis states of our irreducible Lorentz representations, \ie we will
compute $h^{-1} \cdot g^{-1} \cdot h \cdot g | \Psi \rangle$. At first glance,
it may be tempting to apply gauge transformations to scalar fields,
because these are simpler. However, while the scalars transform
nontrivially, in general they do not span a faithful representation
of the gauge group. Thus we evaluate $h^{-1} \cdot g^{-1}
\cdot h \cdot g | \Psi \rangle$ on the set of basis states defined
by the vector fields, because they do span a faithful representation
of the gauge group. Clearly, we should consider terms
of the form $[X,X]$, $[Y,Y]$, $[Z,Z]$, $[X,Y]$, $[X,Z]$ and
$[Y,Z]$. The explicit computation, which can be carried out
straightforwardly using the formulas above, produces the structure constants.
They are
\begin{equation}
f^M{}_{NP} =f_{NP}{}^M= 2 \gamma^M_{NP}
~~~~~~~~~~ f^I{}_{MN} =f_{MN}{}^I= 2 m^I_{MN}
~~~~~~~~~~ f_{MNP} = -3 \beta_{MNP}
\label{aastrc1}
\end{equation}
With this, we can finally write down the explicit form of the
reduced gauge algebra:
\ba
&&[X^M, X^N] = [Y^I, Y^J] = [X^M, Y^I] = 0 \nonumber \\ &&[X^M,
Z_N] = 2 i \gamma^M_{NP} X^P ~~~~~~~~~~ [Y^I, Z_M] = 2i m^I_{MN}
X^N \nonumber \\ &&[Z_M, Z_N] = -3i \beta_{MNP} X^P + 2i
m^I_{MN} Y^I + 2i \gamma^P_{MN} Z_P
\label{aalgebra1}
\ea

\section{Generalized Axion Reductions of Ref.~\cite{kkm}}
\label{more}

Above, we noted that the ansatz (\ref{gauaxio}) is linear in the
internal coordinates despite the explicit presence of the axionic
potentials in the action, and not just their derivatives. Naively,
this may seem to contradict our discussion of ref.~\cite{kkm}.
There, we considered a similar situation, where the explicit
appearance of undifferentiated axion potentials
required that to generate all the axion masses one had to
introduce an ansatz where the  dependence on  the internal space
coordinates was
of order higher than linear. We explicitly considered a case in which
the reduction ansatz was quadratic in internal coordinates. However, we
also pointed out that these quadratic terms could be removed by a
suitable field redefinition. To clarify the situation,
let us re-examine that case here.
There were three scalar axions ${\cal A}_1, {\cal A}_2$
and ${\cal A}_3$ appearing as off-diagonal metric
components\footnote{The fields in ref.~\cite{kkm} and above
in eq.~\reef{zwei3} correspond to those here as follows:
${\cal A}_1={\cal A}_0^{(12)}$, ${\cal A}_2={\cal A}_0^{(23)}$
and ${\cal A}_3={\cal A}_0^{(13)}$.}
in a reduction on $T^3$. The corresponding ``field strengths,''
which naturally appeared in the dimensional reduction, were
\be
{\cal F}_1 = d {\cal A}_1 ~~~~~~ {\cal F}_2 = d {\cal A}_2 ~~~~~~
{\cal F}_3 = d {\cal A}_3 - {\cal A}_2 d {\cal A}_1
\label{fs}
\ee
As we have showed in \cite{kkm}, the ansatz which
simultaneously induces all three possible mass terms is
\ba
{\cal A}_1(x,y) &=& {\cal A}_1(x) + m_1 y \nonumber\\ {\cal
A}_2(x,y) &=& {\cal A}_2(x) + m_2 y \nonumber\\ {\cal A}_3(x,y) &=&
{\cal A}_3(x) + m_3 y + m_2 y {\cal A}_1(x) + \frac12 m_1 m_2 y^2
\label{quad}
\ea
A direct substitution of (\ref{quad}) into (\ref{fs}) shows that
the field strengths do not depend on the coordinate $y$.

However, it is easy to see that replacing the axion ${\cal A}_3$ by
\be
{\cal A}_3 = \bar {\cal A}_3 + \frac12 {\cal A}_1 {\cal A}_2
\ee
redefines its field strength to
\be
{\cal F}_3 = d\bar {\cal A}_3 + \frac12 {\cal A}_1 d {\cal A}_2
-\frac12 {\cal A}_2 d {\cal A}_1
\label{symf}
\ee
and also completely
removes the terms of order of $y^2$ in the reduction ansatz.
Indeed, the last formula in (\ref{quad}) is replaced by
\be
\bar {\cal A}_3(x,y) = {\cal A}_3(x) + m_3 y +
\frac12 m_2 y {\cal A}_1(x)
- \frac12 m_1 y {\cal A}_2(x)
\label{syma}
\ee
Note that the (anti)symmetric form of the ``Chern-Simons'' contribution
to ${\cal F}_3$ in \reef{symf} closely resembles that in the three-form
field strength \reef{axfsin}, in the case of interest in the present
paper. This seems to be the essential ingredient in the linearization
of the reduction ansatz. Note that
the democratic form of the revised ansatz \reef{syma} is also
similar to that for $\hat{\cal B}_{MN}$ in eq.~\reef{gauaxio}.

The ``meaning'' of the disappearance of the $O(y^2)$
terms can be deduced as follows: The quadratic ansatz (\ref{quad})
can be rewritten in the matrix form \cite{kkm} as\footnote{These
matrices correspond to the dreibein of the internal torus \cite{kkm},
where we have set to zero the dilatonic fields that determine the
scale of the internal dimensions. Alternatively,
the latter can be factored out as in eq.~\reef{zwei3}.
The fields of eq.~\reef{gauaxio} however all come from
the $p$-forms of the theory. Nevertheless the reduction
algorithms are closely related.}
\ba
\pmatrix{1&{\cal A}_1(x,y)&{\cal A}_3(x,y) \cr
0&1&{\cal A}_2(x,y) \cr 0&0&1\cr} &=&
\pmatrix{1&{\cal A}_1(x)&{\cal A}_3(x) \cr
0&1&{\cal A}_2(x) \cr 0&0&1\cr}
\pmatrix{1&m_1 y&m_3 y + \frac{m_1 m_2}{2} y^2 \cr
0&1&m_2 y\cr 0&0&1\cr} \nonumber \\ &=&
\pmatrix{1&{\cal A}_1(x)&{\cal A}_3(x) \cr
0&1&{\cal A}_2(x) \cr 0&0&1\cr}
\exp\left[y\pmatrix{0&m_1&m_3\cr
0&0&m_2\cr 0&0&0\cr}\right]
\label{matr}
\ea
Noting that for ${\cal A}_3 = \bar {\cal A}_3 + \frac12 {\cal A}_1
{\cal A}_2$, we can relate the field redefinitions in the matrix
form as well,
\be
\exp\pmatrix{0&{\cal A}_1&\bar {\cal A}_3\cr
0&0&{\cal A}_2\cr 0&0&0\cr} =
\pmatrix{1&{\cal A}_1&\bar {\cal A}_3 +
\frac{{\cal A}_1 {\cal A}_2}{2}\cr
0&1&{\cal A}_2\cr 0&0&1\cr}
\ee
we can rewrite the reduction ansatz as
\be
\exp\pmatrix{0&{\cal A}_1(x,y)&\bar {\cal A}_3(x,y) \cr
0&0&{\cal A}_2(x,y) \cr 0&0&0\cr} =
\exp\pmatrix{0&{\cal A}_1(x)&\bar{\cal A}_3(x) \cr
0&0&{\cal A}_2(x) \cr 0&0&0\cr}
\exp\left[y\pmatrix{0&m_1&m_3\cr
0&0&m_2\cr 0&0&0\cr}\right]
\label{expon}
\ee
Now, we recall a version of the Baker-Campbell-Hausdorff formula,
which allows us to evaluate products of exponentials of matrices.
Namely, for as long as $[A,[A,B]] = [B,[A,B]] = 0$ for any two
matrices $A,B$ we have $\exp(A) \exp(yB) = \exp(A+yB + \frac12
y[A,B])$. In our case,
\ba
&&A=\pmatrix{0&{\cal A}_1(x)&\bar {\cal A}_3(x) \cr 0&0&{\cal
A}_2(x) \cr 0&0&0\cr} ~~~~~~ B=\pmatrix{0&m_1&m_3\cr 0&0&m_2\cr
0&0&0\cr} \nonumber \\ &&~~~~~~~~~ [A,B] = \pmatrix{0&0&m_2 {\cal
A}_1(x) - m_1 {\cal A}_2(x)\cr 0&0&0\cr 0&0&0\cr}
\ea
and hence using the Baker-Campbell-Hausdorff formula we find, upon
the substitution into (\ref{expon}) and taking the logarithm of
both sides, we find
\be
\pmatrix{0&{\cal A}_1(x,y)&\bar {\cal A}_3(x,y) \cr
0&0&{\cal A}_2(x,y) \cr 0&0&0\cr} =
\pmatrix{0&{\cal A}_1(x) + m_1 y&\bar{\cal A}_3(x)
+\frac12 y(m_2 {\cal A}_1(x) - m_1 {\cal A}_2(x))\cr 0&0&{\cal
A}_2(x) + m_2 y\cr 0&0&0\cr}
\ee
which is precisely the linear reduction ansatz. Hence, we see that
the $O(y^2)$ disappeared essentially by making the
field redefinition which amounts to taking a functional logarithm of the
axionic degrees of freedom. Nevertheless, we note that in general
this procedure of linearization may turn out to be quite awkward.


\begin{thebibliography}{99}


\bibitem{cow} E. Bergshoeff, M. de Roo, M.B. Green, G.
Papadopoulos and  P.K. Townsend, Nucl. Phys. {\bf B470} (1996) 113
[hep-th/9601150];
P.M. Cowdall, H. Lu, C.N. Pope, K.S. Stelle and P.K.
Townsend, Nucl. Phys. {\bf B486} (1997) 49 [hep-th/9608173].

\bibitem{others} P.M. Cowdall, Class. Quant. Grav. {\bf 15} (1998)
2937 [hep-th/9710214]; ``On Gauged Maximal
Supergravity in Six-Dimensions,'' eprint hep-th/9810041;
Class. Quant. Grav. {\bf 15} (1998) 2937 [hep-th/9710214];
P.M. Cowdall and P.K. Townsend,
Phys. Lett. {\bf B429} (1998) 281 [hep-th/9801165];
H. Singh, Phys. Lett. {\bf B429} (1998) 304 [hep-th/9801038];
Phys. Lett. {\bf B419} (1998) 195 [hep-th/9710189];
I.V. Lavrinenko, H. Lu and C.N. Pope, Class. Quant. Grav. {\bf 15} (1998)
2239 [hep-th/9710243].

\bibitem{berg} E. Bergshoeff, M. de Roo and E. Eyras,
Phys. Lett. {\bf B413} (1997) 70 [hep-th/9707130].

\bibitem{flux} I.V. Lavrinenko, H. Lu and  C.N. Pope, Nucl. Phys. {\bf B492}
(1997) 278 [hep-th/9611134].

\bibitem{schs} J. Scherk and  John H. Schwarz,
Phys. Lett. {\bf B82} (1979) 60; Nucl. Phys. {\bf B153} (1979) 61.

\bibitem{kkm} N. Kaloper, R.R. Khuri and R.C. Myers,
Phys. Lett. {\bf B428} (1998) 297 [hep-th/9803066].

\bibitem{cosmo} A. Lukas, B.A. Ovrut, K.S. Stelle and D. Waldram, ``Heterotic
M Theory in Five Dimensions,'' eprint hep-th/9806051;
``The Universe as a Domain Wall,'' eprint hep-th/9803235;
A. Lukas, B.A. Ovrut and D. Waldram, ``Cosmological Solutions of Horava-Witten
Theory,'' eprint hep-th/9806022;
N. Kaloper, I.I. Kogan and K.A. Olive,
Phys. Rev. {\bf D57} (1998) 7340 [hep-th/9711027];
H. Lu, S. Mukherji and
C.N. Pope, ``From P-branes to Cosmology,'' eprint hep-th/9612224;
H. Lu and C.N. Pope,
Mod. Phys. Lett. {\bf A12} (1997) 1087 [hep-th/9611079].


\bibitem{hull}{C.M. Hull, ``Massive String Theories From M-Theory and F-Theory,''
eprint  hep-th/9811021.}

\bibitem{ortin}{P. Meessen and  T. Ortin, ``An SL(2,Z) Multiplet of Nine-Dimensional
Type II Supergravity Theories,'' eprint hep-th/9806120.}

\bibitem{cmhull}{C.M. Hull, Phys. Rev. {\bf D30} (1984) 760;
Phys. Lett. {\bf 148B} (1984) 230; Physica {\bf 15D} (1985) 230.}

\bibitem{boucher}{W. Boucher, Nucl. Phys. {\bf B253} (1985) 541.}

\bibitem{witten} C.M. Hull and P.K. Townsend, Nucl. Phys. {\bf B438}
(1995) 109 [hep-th/9410167];
E. Witten, Nucl. Phys. {\bf B443} (1995) 85 [hep-th/9503124].

\bibitem{Uthe}see for example:\\
A. Sen, ``Developments in Superstring Theory,'' eprint
hep-ph/9810356;
``An Introduction to Nonperturbative String Theory,''
eprint hep-th/9802051;\\
C. Vafa, ``Lectures on Strings and Dualities,''
hep-th/9702201;\\
J. Polchinski, Rev. Mod. Phys. {\bf 68} (1996) 1245
[hep-th/9607050];\\
M.J. Duff, Int. J. Mod. Phys. {\bf A11} (1996) 5623
[hep-th/9608117];\\
J.H. Schwarz, Nucl. Phys. Proc. Suppl. {\bf 55B} (1997) 1
[hep-th/9607201].


\bibitem{MS} J. Maharana and J.H. Schwarz,
Nucl. Phys. {\bf B390} (1993) 3 [hep-th/9207016].

\bibitem{crjul} J. Scherk and J.H. Schwarz, Nucl. Phys. {\bf B153}
(1979) 61; E. Cremmer, in {\it Supergravity '81}, ed. S. Ferrara
and J.G. Taylor, (Cambridge Univ. Press, Cambridge 1982).

\bibitem{sen} K.A. Meissner and G. Veneziano, Phys. Lett.
{\bf B267} (1991) 33; Mod. Phys. Lett. {\bf A6} (1991) 3397;
M.J. Duff, Nucl. Phys. {\bf B335} (1990) 610;
A. Sen, Phys. Lett. {\bf B271} (1991) 295; Phys. Lett. {\bf B274}
(1992) 34. S.F. Hassan and A. Sen, Nucl. Phys. {\bf B375} (1992)
103;

\bibitem{buscher} T.H. Buscher, Phys. Lett. {\bf B194} (1987) 59;
Phys. Lett. {\bf B201} (1988) 466.

\bibitem{km} N. Kaloper and K.A. Meissner, Phys. Rev.
{\bf D56} (1997) 7940 [hep-th/9705193].

\bibitem{kaloper} N. Kaloper, Phys. Rev. {\bf D55} (1997) 3394-3402 [hep-th/9609087].

\bibitem{GPR} A. Giveon and M. Poratti,
Nucl. Phys. {\bf B355} (1991) 422; A. Giveon, M. Porrati and E.
Rabinovici, Phys. Rep. {\bf 244} (1994) 77.

\bibitem{more} M. de Roo, Nucl. Phys. {\bf 255} (1985) 515;
Phys. Lett. {\bf B156} (1985) 331; M. de Roo and P. Wagemans, Nucl.
Phys. {\bf 262} (1985) 644; P. Wagemans, Phys. Lett. {\bf 206}
(1988) 241; E. Bersghoeff, I.G. Koh and E. Sezgin, Phys. Lett. {\bf
155} (1985) 151.

\bibitem{wilczlars} F. Larsen and F. Wilczek, Nucl. Phys. {\bf B475} (1996) 627
[hep-th/9604134].

\bibitem{coset}{A. Giveon, E. Rabinovici and G. Veneziano,
Nucl. Phys. {\bf B322} (1989) 167.}

\bibitem{actor}{A. Giveon, N. Malkin and E. Rabinovici,
Phys. Lett. {\bf 238} (1990) 57.}

\bibitem{nont}{A. Sen, Nucl. Phys. {\bf B440} (1995) 421 [hep-th/9411187].}

\bibitem{modi}K.S. Narain, Phys. Lett. {\bf 169B} (1986) 41; K.S. Narain, M.H.
Sarmadi and E. Witten, Nucl. Phys. {\bf B279} (1987) 369.

\bibitem{fr}{P.G.O. Freund and M.A. Rubin, Phys. Lett. {\bf B97} (1980) 233.}

\bibitem{pseudo}{C.M.~Hull and A.~van Proeyen,
Phys. Lett. {\bf B351} (1995) 188 [hep-th/9503022].}

\bibitem{exact}{M. Ro\v cek and E. Verlinde, Nucl. Phys. {\bf B373} (1992) 630;
A. Giveon and M. Ro\v cek, Nucl. Phys. {\bf 380} (1992) 128.}

\bibitem{rlindil}{R.C. Myers, Phys. Lett. {\bf B199} (1987) 371.}

\bibitem{comsic}{B.R. Greene, A. Shapere, C. Vafa and S.-T. Yau,
Nucl. Phys. {\bf B337} (1990) 1.}

\bibitem{three}{R. Rohm and E. Witten, Annals Phys. {\bf 170} (1986) 454.}

\bibitem{kkna} B. de Witt, {\it Dynamical Theory of Groups and Fields},
Gordon and Breach Science Publishers, New York 1965;
R. Kerner, Ann. Inst. H. Poincare, {\bf 9} (1968) 143;
Y.M. Cho and P.G.O. Freund, Phys. Rev. {\bf D12} (1975) 1711;
A. Trautman, Rep. Math. Phys. {\bf 16} (1975) 2029;
A. Salam and J. Strathdee, Ann. Phys. {\bf 141} (1982) 316;
Y.M. Cho, Phys. Lett. {\bf B186} (1987) 38.

\bibitem{duffpope} M.J. Duff and C.N. Pope, Nucl. Phys. {\bf B255} (1985) 355;
M.J. Duff, B.E.W. Nilsson, C.N. Pope and N.P. Warner, Phys. Lett.
{\bf B149} (1984) 90.

\bibitem{ryanshepley} M.P. Ryan, Jr. and L.C. Shepley,
{\it Homogeneous Relativistic Cosmologies}, Princeton
University Press, Princeton, NJ 1975.

\bibitem{chams} A.H. Chamseddine and
M.S. Volkov, Phys. Rev. {\bf D57} (1998) 6242 [hep-th/9711181].

\bibitem{englert} F. Englert, Phys. Lett. {\bf B119} (1982) 339.
M.J. Duff, B.E.W. Nilsson and C.N. Pope,
Phys. Rev. Lett. {\bf 50} (1983) 2043.

\end{thebibliography}
\end{document}